%% file: main.tex
\documentclass[hidelinks,12pt]{article}
\input{preamble.tex}

\begin{document}

\newgeometry{top=1.5in, bottom=0.5in, left=1.0in, right=1.0in} 
\begin{titlepage}
 \begin{center}

\begin{spacing}{1.5}
{\fontsize{18}{18}\selectfont \textbf{Theory-guided relational learning for data product pricing: An interpretable meta-path-guided graph neural network approach}}
\\[0.5cm]
\end{spacing}

 {\large
 \begin{minipage*}{\textwidth} \begin{center}
Ruize Gao\symbolfootnote[1]{Beijing Institute of Mathematical Sciences and Applications, Beijing, China.},\
Feng Xiao\symbolfootnote[2]{Beijing Institute of Mathematical Sciences and Applications, Beijing, China.},\
Jinpu Li\symbolfootnote[3]{Tsinghua University, Beijing, China.},\
Ke Tang\symbolfootnote[4]{Tsinghua University, Beijing, China.},\ and\
Shaoze Cui\symbolfootnote[5]{Beijing Institute of Technology, Beijing, China. Corresponding author. Email: shaoze-cui@foxmail.com}
\end{center}
 	\end{minipage*}
  }

~ \\[0.5cm]
 {\today}\\[1.0cm]

\begin{spacing}{1.3}
\begin{abstract}

 {\small \noindent
 Data products are non-standardized and context-dependent, so their prices are jointly shaped by product attributes and contextual factors such as provider reputation, product comparability, and industry structure.
We formulate data product pricing as a learning task over a heterogeneous information network spanning data products, providers, marketplaces, industries, and pricing modes, and design theory-grounded meta-paths, brand-based, similarity-based, industry-based, and hybrid, to characterize distinct pricing mechanisms. Building on this network, we propose a Meta-Path-Guided Heterogeneous Graph Neural Network (MPGNN), which learns path-specific representations, fuses them through an attention-based gate, and is further augmented with large language model-generated textual features.
Empirical evaluation shows that MPGNN achieves statistically significant improvements over baseline models in both regression and classification across different pricing modes, and that provider-involving meta-paths dominate the learned weights, yielding predictive evidence consistent with a brand-equity-related pricing mechanism.
The proposed framework combines predictive accuracy with theory-driven interpretability, offering a transferable methodological template for pricing other non-standardized digital assets.
 }

\vspace{0.5cm}

\noindent \textit{\textbf{Keywords}: data product pricing, heterogeneous information network, meta-path, graph neural network, interpretable machine learning} \\

\vspace{1.0cm}

\end{abstract}
\end{spacing}

\vspace{1.0cm}

\end{center}

\end{titlepage}
\restoregeometry

\clearpage

\onehalfspacing
\clearpage
\section{Introduction} \label{sec:introduction}

With the continuous development of data element markets, data products have increasingly become important tradable objects in the digital economy. Various data exchanges, cloud marketplaces, and third-party data service platforms have emerged, and firms and institutions have begun to price and sell datasets, APIs, data services, and data analytics outputs as tradable products \citep{peiSurveyDataPricing2022,congDataPricingMachine2022}. This trend is reflected in the rapid expansion of data trading at scale: China's data market transaction volume exceeded 160 billion yuan (about 22.3 billion U.S. dollars) in 2024, an increase of more than 30\% year on year,\footnote{Source: Xinhua News Agency, ``China's 2024 data market transaction volume tops 160 billion yuan,'' based on figures released by the National Data Administration of China. \url{https://english.news.cn/20250111/51a5f4689ba541c5ae44f73cb8b510d0/c.html}.} while the global data marketplace platform market reached about 1.49 billion U.S. dollars in 2024 and is projected to grow to 5.73 billion U.S. dollars by 2030 at a compound annual growth rate of 25.2\%.\footnote{Source: Grand View Research, ``Data Marketplace Platform Market Size, Share \& Trends Analysis Report.'' \url{https://www.grandviewresearch.com/industry-analysis/data-marketplace-market-report}.} However, compared with traditional goods, data products are characterized by non-standardization, replicability, reusability, strong context dependence, and uncertain value, making their price formation mechanisms more complex \citep{liConceptsAccountingTreatment2024,renKG2026}. How to accurately, transparently, and interpretably evaluate the prices of data products has therefore become a key issue in data asset management, data trading, and the development of data element markets \citep{azcoitiaInterpretableMarketbasedData2025,zhaoResearchAnalyzingInfluencing2024}.

Unlike physical goods, which have observable production costs and abundant reference prices, data products carry near-zero marginal reproduction cost, no widely accepted cost basis, and highly context-dependent value, so in practice their prices are set with very little common reference. The resulting pricing opacity has tangible and costly consequences. On real trading platforms, data products with broadly comparable content are frequently listed at order-of-magnitude different prices, mirroring the large and persistent price dispersion documented for other online information goods \citep{ghosePriceDispersionTransaction2011,clemonsPriceDispersion2002,chellappaPriceFormats2011,liPriceDispersionLossLeader2013}. Such mispricing is costly on both sides of the market: over-pricing suppresses transactions and slows the circulation of data elements, whereas under-pricing leads to a systematic undervaluation of data assets, which in turn distorts the recognition and reporting of data assets on the balance sheet \citep{liConceptsAccountingTreatment2024} and weakens providers' incentives to supply high-quality data. Moreover, prices produced by opaque models are difficult to justify to counterparties and regulators, eroding trust in data trading. These observations indicate that data product pricing is not merely a prediction-accuracy exercise but a question with direct economic stakes, and that it calls for methods that are at once accurate, transparent, and economically interpretable.

Existing studies on data product pricing can be broadly divided into two categories. The first category develops pricing models based on economic theories, such as game theory, auction theory, and mechanism design \citep{hanDataValuationVertical2026,markovichCompetingCookiesPlatforms2026,suiOptimalDataPricing2025}. These studies can explain strategic interactions among participants in data transactions and market equilibria, but they typically rely on strong theoretical assumptions and provide insufficient characterization of complex product attributes, platform structures, and multi-agent relationships in real-world data trading markets. The second category adopts machine learning or intelligent computation methods to predict data prices based on product features, such as linear regression, ensemble learning, and deep learning models \citep{heckmanPricingModelData2015,haoHCEGHeterogeneousClustering2024,haoHowPriceDataset2025}. Although these methods can improve prediction accuracy, most of them mainly focus on the direct mapping between data product attributes and prices, while paying limited attention to the relational structures among data products, providers, industries, markets, and pricing modes. As a result, existing methods have difficulty fully revealing the relational mechanisms and economic meanings underlying data product price formation \citep{renKG2026}.

Therefore, to further understand the price formation mechanism of data products, it is insufficient to consider product attributes in isolation. Instead, data products need to be examined within the multi-actor and multi-layer relational networks in which they are embedded. In fact, data products do not exist independently: their prices are determined not only by their own attributes, but also by their relationships with data providers, similar products, industries, marketplace platforms, and other related entities.
Specifically, products offered by the same data provider often share similar brand reputation, service capabilities, and pricing strategies, and may therefore exhibit correlated pricing patterns. According to brand equity theory, a provider's brand reputation can influence consumers' value perceptions and willingness to pay \citep{kellerConceptualizingMeasuringManaging1993}, providing a theoretical explanation for such provider-level price relationships. Second, data products with similar content, application scenarios, or other product characteristics may serve as price references for one another. Product similarity theory suggests that consumers evaluate products by comparing them with alternatives that share similar characteristics, implying that the prices of similar data products may be interrelated \citep{lindenAmazoncomRecommendationsItemtoitem2003,kwonHybridRecommendationSystem2025}. Third, data products within the same industry typically face similar demand conditions, competitive environments, and regulatory constraints, and may therefore exhibit similar pricing patterns, consistent with industry structure theory regarding the influence of industry environments on pricing behavior \citep{singhPricingMarketConcentration2008}. In addition, marketplace environments and cross-industry business linkages may affect data product prices through different relational paths.
Taken together, these observations suggest that data product pricing is not merely a mapping from product attributes to prices, but is inherently characterized by complex relational structures. Therefore, an effective data product pricing method should jointly model product attributes, heterogeneous entity relationships, and different relational paths, while capturing the economic mechanisms embedded in these relationships \citep{renKG2026}.

To address these problems, we propose an interpretable Meta-Path-Guided Graph Neural Network (MPGNN) for data product pricing. Specifically, we first represent the data product market as a heterogeneous information network, in which nodes represent data products, data providers, data marketplaces, industries, and pricing modes, while edges capture product similarity, provision relationships, industry affiliation, market cooperation, and pricing-mode support. Based on this network, we design multiple economically meaningful meta-paths, grounded in brand equity theory, product similarity theory, and industry structure theory, to characterize price associations among data products through distinct relational mechanisms.
Building on this heterogeneous information network, MPGNN incorporates four interconnected methodological components. First, each meta-path is independently encoded using a Heterogeneous Graph Transformer (HGT), whose message-passing process is constrained layer by layer according to the edge sequence specified by the corresponding meta-path. 
Second, a self-attention gating mechanism is employed to integrate the representations learned from multiple meta-paths. This mechanism therefore not only improves the integration of heterogeneous relational information but also provides an interpretable view of how different economic relationships contribute to data product price formation.
Third, we introduce large language model-driven textual feature augmentation to enrich the representation of data products. We leverage the semantic understanding capabilities of LLMs to extract and generate additional informative features from textual descriptions, and integrate these augmented textual features with the original product attributes and relational representations. 
Fourth, for the price prediction task, MPGNN adopts a two-stage cascaded architecture inspired by the Mixture-of-Experts (MoE) paradigm. A dedicated price-tier classifier first estimates the probability distribution of each data product over $C$ price tiers. Subsequently, $C$ tier-specific regression experts generate price estimates for their respective price regions, and their outputs are aggregated through soft probabilistic routing according to the predicted tier probabilities. 
This design decomposes the challenging task of predicting data product prices spanning multiple orders of magnitude into two interconnected stages: coarse-grained price-tier identification and fine-grained within-tier price regression. 
By allowing each regression expert to specialize in a relatively narrow price region, the proposed architecture reduces the difficulty associated with substantial price dispersion and improves the model's ability to capture heterogeneous pricing patterns across different price levels.

To address these research gaps, this study makes three contributions.

\begin{itemize}

\item \textbf{Relational, theory-grounded formulation of data product pricing.}
We reconceptualize data product pricing as a relational learning problem over a heterogeneous information network that jointly models data products, providers, industries, marketplaces, and pricing modes, and design a set of interpretable meta-paths, grounded in brand equity theory, product similarity theory, and industry structure theory, to characterize distinct pricing mechanisms within this network.

\item \textbf{MPGNN: an integrated framework for price prediction and mechanism attribution.}
We propose MPGNN, a framework that integrates price prediction with path-level mechanism attribution through meta-path-specific heterogeneous graph transformer encoding, attention-gated fusion of path representations, large language model-enhanced textual features, and a two-stage cascaded mixture-of-experts architecture tailored to the long-tailed price distribution of data products.

\item \textbf{Empirical validation and economic interpretation across data markets.}
We evaluate MPGNN across three data markets and find that the supplier-mediated meta-path is the most influential pricing mechanism in both tasks, yielding predictive evidence consistent with a brand-equity-related mechanism in data product pricing. The proposed framework thus offers data markets an auditable and transferable tool for data product pricing.

\end{itemize}

The remainder of this paper is organized as follows. Section 2 reviews the related literature, including product pricing theory, data product pricing methods, and the interpretability of data product pricing. 
Section 3 introduces the proposed methodology.
Section 4 describes the experimental design and evaluation metrics. 
Section 5 reports the experimental results and discusses the main findings. Section 6 further discusses the implications of this study for information systems research and data product pricing practice. Finally, Section 7 concludes the paper and outlines directions for future research.

\section{Literature Review} \label{sec:literature}

The literature related to this study mainly covers three areas: product pricing theory, data product pricing methods, and the interpretability of data product pricing. Accordingly, we review the literature from these three perspectives.

\subsection{Product pricing theory}

Existing studies on product pricing can be broadly categorized into three theoretical perspectives, including brand equity theory, product similarity theory, and industry structure theory. 

From the perspective of brand equity, \cite{kellerConceptualizingMeasuringManaging1993} finds that customer-based brand equity arises when consumers are familiar with a brand and hold strong, favorable, and unique associations with it in memory, which enables firms to charge a price premium, a view echoed by subsequent research documenting the revenue and pricing outcomes associated with brand equity \citep{kellerBrandsBranding2006}. From the perspective of product similarity, \cite{lindenAmazoncomRecommendationsItemtoitem2003} propose an item-based collaborative filtering algorithm that generates recommendations by identifying items similar to those previously purchased or rated by users, thereby capturing the underlying similarity structure among products, a mechanism further supported by purchase-data-based and diversity-oriented recommendation studies \citep{bodapatiRecommendationSystems2008,flederBlockbusterCulture2009}. And then, \cite{kwonHybridRecommendationSystem2025} further incorporate price similarity into recommendation algorithms and demonstrates that integrating price similarity is an effective strategy for developing commercially viable recommender systems. From the perspective of industry structure, \cite{singhPricingMarketConcentration2008} show that industry structure significantly affects product pricing, with higher market concentration being associated with substantially higher prices. 
Overall, these perspectives suggest that product pricing is jointly determined by brand-driven value perception, similarity-based product relationships, and market-level competitive structures.

\subsection{Data product pricing method}

Regarding data product pricing methods, pricing strategies and pricing models constitute two essential components. Specifically, in terms of data pricing strategies, \cite{peiSurveyDataPricing2022} investigate several commonly adopted pricing mechanisms, including free data provision, usage-based pricing, pricing based on data volume or API calls under a fixed fee, flat-fee tariffs, and premium patterns.

Second, data pricing models can generally be categorized into two types of approaches: economics-theory-based models and intelligent computation-based models. On the one hand, economics-theory-based models primarily draw upon methodologies such as game theory, auction theory, and mechanism design \citep{hanDataValuationVertical2026}. For instance, \cite{markovichCompetingCookiesPlatforms2026} examine whether platforms should monetize user data under network effects. Similarly, \cite{suiOptimalDataPricing2025} analyze how competing two-sided platforms design data collection and pricing strategies to maximize profits under heterogeneous technological costs. Although these models provide important theoretical insights into data markets and participant behavior, they are typically built upon relatively strong assumptions and often rely on simplified representations of market structures. As a result, their direct applicability to real-world data product pricing scenarios is limited.
This stream connects to a broader economics-of-data and markets-for-information literature, which studies how a monopolistic seller should optimally design and price informative experiments \citep{bergemannDesignPriceInformation2018,babaioffOptimalMechanismsSelling2012,bergemannMarketsInformationIntroduction2019}, how information sales interact with downstream competition \citep{bimpikisInformationSaleCompetition2019}, including the direct sale of consumer data to marketers \citep{guoSellingDataMarketers2025}, the welfare and efficiency consequences of data as a nonrival good \citep{jonesTonettiNonrivalryEconomicsData2020,acemogluTooMuchData2022}, and the role of data in firms' competitive advantage \citep{hagiuDataEnabledLearning2023}. A related body of work examines the economics of privacy and digital markets, in which the value of personal data and the effects of privacy regulation are central concerns \citep{acquistiEconomicsPrivacy2016,goldfarbDigitalEconomics2019,goldfarbPrivacyRegulationOnline2011,montesValuePersonalInformation2019}. Classic information-goods studies further show that nonlinear and usage-based tariffs are essential instruments for monetizing digital goods \citep{sundararajanNonlinearPricingInformation2004,hinzPriceDiscriminationEcommerce2011}, complementing earlier work on the bundling of information goods \citep{bakosBundlingInformationGoods1999}, the role of search costs in electronic marketplaces \citep{bakosReducingBuyerSearch1997}, the choice between selling and pay-per-use mechanisms \citep{balasubramanianPricingInformationGoods2015}, the comparison between perpetual-license (selling) and subscription (leasing) pricing for information goods under consumer valuation depreciation \citep{douSellingLeasingPricing2017}, and the pricing of information products distributed on online servers \citep{jainPricingInformationProducts2002}. These studies illuminate the economic forces underlying data value, yet they generally stop short of operationalizing these mechanisms into predictive pricing tools for heterogeneous, real-world data products.

On the other hand, intelligent computation-based models primarily leverage data characteristics and their performance in downstream applications for value assessment and price prediction\citep{gaoTextualSemanticsMachine}. Different pricing objects require different approaches to value measurement and computational modeling. \cite{congDataPricingMachine2022} systematically review various pricing objects in machine learning settings, including raw datasets, data labels, and trained machine learning models. For the pricing of datasets or data products, early studies commonly employ traditional statistical methods, such as linear regression, to estimate prices based on observable product attributes \citep{heckmanPricingModelData2015}. In real-world markets, however, the relationships between data characteristics and prices are often nonlinear and involve complex feature interactions. Linear models alone may therefore fail to adequately capture such complexities, potentially resulting in biased price estimates.
To address this limitation, an increasing number of studies have introduced machine learning methods to better capture complex nonlinear pricing relationships. For example, \cite{haoHCEGHeterogeneousClustering2024} propose an ensemble learning model based on a ``gravity-weighting strategy,'' integrating fifteen different models to improve the accuracy and robustness of data product price prediction. Extending this line of research, \cite{haoHowPriceDataset2025} employ BERT to learn representations from the unstructured textual attributes of datasets and combine these representations with LightGBM for price prediction, thereby extending conventional structured product attributes to incorporate the semantic information embedded in data product descriptions. This text-based valuation strategy echoes earlier evidence that mining consumer-review text can reveal the pricing power of individual product features \citep{archakDerivingPricingPower2011}.

Beyond directly predicting market prices, another stream of research assesses the intrinsic value of data from the perspective of its utility for specific tasks. Because data are heterogeneous, complementary, and reusable, their value often depends on the particular application task and on how they are combined with other data. Accordingly, the machine learning literature has developed a range of task-based data valuation methods, among which Data Shapley \citep{ghorbaniDataShapley2019} is one of the most representative. Drawing on the concept of the Shapley value, Data Shapley quantifies the task-specific value of training data by measuring the marginal contribution of an individual data point to model performance across different data coalitions. Such approaches extend data valuation from asking ``what attributes does the data possess?'' to asking ``how much utility can the data generate?'', thereby providing a new computational perspective for understanding data value. Building on such valuation techniques, recent work designs fairshare pricing mechanisms for large language model training data that translate data valuation scores directly into seller compensation \citep{zhangFairshareDataPricing2025}.
Meanwhile, some studies have begun to explicitly incorporate relational information among data products and between data products and other market entities. For example, \cite{renKG2026} organize data assets, related entities, and their relationships into a knowledge graph, extract relational structural information from the graph, and combine it with XGBoost to predict data product prices. Compared with pricing approaches that rely solely on the intrinsic attributes of individual data products, this line of research suggests that incorporating inter-entity relationships can provide additional information for data valuation. It also indicates that the price formation of data products may not be determined independently by their own attributes.

Despite these advances in both the information used for data valuation and the computational methods employed, existing intelligent computation-based approaches leave considerable room for further development. On the one hand, attribute-based machine learning models primarily focus on the structured or unstructured characteristics of individual data products. On the other hand, task-oriented valuation methods such as Data Shapley mainly quantify the marginal contribution of data to specific predictive tasks. Although knowledge graph-based approaches have begun to incorporate relational information, limited attention has been paid to systematically distinguishing different types of relationships and, more importantly, to characterizing the economic mechanisms represented by these relationships. In practice, data products are embedded in complex relational networks involving providers, industries, trading markets, pricing schemes, and other data products. Their prices may therefore be shaped not only by their intrinsic attributes but also by relational contexts such as brand affiliation, product similarity, industry linkages, and market environments. It is thus necessary to further model the price formation of data products from a relational perspective.

\subsection{Interpretability of data product pricing}

Enhancing the interpretability of data product pricing can improve the transparency and trustworthiness of the pricing process. 
In this context, explainable machine learning methods provide an important technical pathway for improving the transparency of data product pricing.
From the perspective of model structure, interpretable models can generally be categorized into intrinsically interpretable models and post-hoc interpretable models.

Intrinsically interpretable models are designed with transparent structures during the modeling stage, where model parameters and functional forms carry clear statistical or economic meanings. 
Representative approaches include traditional statistical models such as linear models, generalized linear models, generalized additive models, and decision trees, as well as more recent methods such as Explainable Boosting Machines, GAMI-Net, and RuleFit. 
These methods can balance predictive performance and model transparency, thereby providing a basis for interpreting model decisions, an approach strongly advocated for high-stakes decision settings where intrinsically interpretable models are preferred over post-hoc explanations of black boxes \citep{rudinStopExplainingBlackBox2019}.
However, they typically rely on relatively simple functional forms or rule-based structures, making it difficult to effectively capture complex relational structures and interaction mechanisms in multi-entity systems.

In contrast, graph neural networks (GNNs) provide a powerful tool for modeling relationships among entities through graph structures, enabling the characterization of complex interactions among multiple entities. Representative architectures include graph convolutional networks \citep{kipfSemiSupervisedClassification2017}, relational GCNs designed for multi-relational data \citep{schlichtkrullModelingRelationalData2018}, attention-based transformer encoders \citep{vaswaniAttentionAllYouNeed2017}, and heterogeneous graph transformers that introduce node- and edge-type dependent parameters to model heterogeneous attention \citep{huHeterogeneousGraphTransformer2020}, while heterogeneous information network analysis offers the broader methodological foundation for reasoning over multi-typed nodes and edges \citep{shiSurveyHeterogeneousInformation2017}.
However, while improving model representation capacity, traditional GNNs often suffer from limited interpretability, as the learned relational structures and information propagation paths are difficult to interpret intuitively. 
In this context, meta-path-based methods offer a promising direction for enhancing structural interpretability. 

Meta-paths describe sequences of relationships among different types of entities in heterogeneous networks, capturing the semantic associations between them. 
In recent years, meta-path-based methods have been applied in various domains. For example, \cite{duHowRottenApple2025} apply meta-path techniques to corporate fraud detection by constructing a heterogeneous business network. Empirical results show that the top five meta-paths related to connected transactions exhibit strong predictive power for fraud detection. 
Similarly, \cite{maCustomerEngagementPrediction2025} apply meta-path methods to customer engagement prediction in social media contexts. By constructing a heterogeneous network of users and consumers and designing meta-paths based on user-related and brand-related factors, their analysis reveals that different types of meta-paths dominate under different scenarios.
Despite their demonstrated potential in areas such as fraud detection and social media analytics, the application of meta-path-based interpretable methods to data product pricing remains limited. In particular, there is still a lack of systematic research on how to integrate economic theories of data pricing with meta-path design.

In addition to intrinsically interpretable methods, another important line of research focuses on post-hoc interpretability. These methods aim to explain model predictions without altering the structure of complex models. Typical approaches include partial dependence plots (PDP), accumulated local effects (ALE), LIME~\citep{ribeiroWhyShouldTrust2016}, and SHAP~\citep{lundbergUnifiedApproach2017}.
In the context of interpretable data product pricing, such methods have been applied to identify key factors influencing prices across industries. 
For instance, \cite{zhaoResearchAnalyzingInfluencing2024} finds, based on SHAP analysis, that data volume is the most significant determinant of data prices in most industries, while the sensitivity to data update frequency exhibits clear industry heterogeneity, being particularly high in finance and technology sectors, but relatively low in industrial and healthcare domains. 
Moreover, \cite{azcoitiaInterpretableMarketbasedData2025} show that data content, data volume, and delivery methods are the primary drivers of data product pricing, whereas factors such as time span, historical records, and geographic coverage have relatively limited influence.

\subsection{Research gaps to be filled in this study}

Building on the three streams of literature reviewed above, three research
gaps remain to be addressed.

First, although classical pricing theories (brand equity, product
similarity, and industry structure) collectively imply that data product
prices are jointly shaped by providers, peer products, industries, markets,
and pricing modes, existing studies largely reduce pricing to an
attribute-price mapping and fail to systematically capture these
multi-entity relational structures that actually govern price formation in
real data marketplaces.

Second, although meta-path-based methods have proven effective for
encoding semantic relations in heterogeneous networks, their application to
data product pricing remains scarce; existing knowledge-graph-based
approaches to data pricing, in particular, do not adequately distinguish
among relation types or operationalize pricing theories, brand equity,
product similarity, and industry structure, into economically
interpretable, computationally tractable structures.

Third, existing interpretable approaches to data product pricing rely
mainly on post-hoc tools such as SHAP and partial dependence plots, and
lack intrinsically interpretable structural models that can reveal which
relational mechanism, brand, similarity, industry, market, or pricing mode,
drives the price and to what extent.

Filling up these three gaps requires a modeling framework that
(i) represents the data product market as a heterogeneous information
network covering products, providers, industries, markets, and pricing
modes; (ii) embeds established product pricing theories into the design of
economically meaningful meta-paths; and (iii) couples meta-path-guided
graph learning with a transparent path-level weighting mechanism so that
the relative importance of different pricing mechanisms can be read
directly from the model.

\section{Problem formulation}

In this study, we model the data trading market as a heterogeneous information network (HIN), also commonly referred to as a \emph{heterogeneous graph}, for a network that contains multiple types of nodes and/or multiple types of edges \citep{sunPathSim2011,wangHeterogeneousGraphAttention2019,dongMetapath2vec2017}, denoted as $\mathcal{G} = (\mathcal{V}, \mathcal{E})$, where $\mathcal{V}$ represents the set of nodes and $\mathcal{E}$ denotes the set of edges. Specifically, each node $v \in \mathcal{V}$ corresponds to either a data product or a related entity (e.g., data providers and industries), and each edge $e \in \mathcal{E}$ captures the relationship between entities. Let $\mathbf{X} \in \mathbb{R}^{N \times d}$ denote the feature matrix of data products, where $N$ is the number of products and $d$ is the feature dimension, and let $\mathbf{y} \in \mathbb{R}^{N}$ denote the corresponding price vector. To integrate heterogeneous information from multiple entity types and capture their joint influence on data product value, we construct a unified representation framework that jointly models node attributes and structural dependencies in the network.

To further capture high-order and multi-level relationships among heterogeneous entities, we design a set of meta-paths $\{\mathcal{P}_1, \mathcal{P}_2, \dots, \mathcal{P}_H\}$ tailored for data product pricing. Each meta-path encodes a specific semantic relationship. The design of these meta-paths is grounded in established pricing theories, such as brand equity theory \citep{kellerConceptualizingMeasuringManaging1993} and product similarity theory \citep{lindenAmazoncomRecommendationsItemtoitem2003}, ensuring their economic interpretability. Based on the feature matrix $\mathbf{X}$ and the meta-path-based relational structures $\{\mathbf{A}^{(h)}\}_{h=1}^H$, we aim to learn a pricing function $f(\cdot)$ such that
\[
\hat{\mathbf{y}} = f\left(\mathbf{X}, \{\mathbf{A}^{(h)}\}_{h=1}^H \right),
\]
where $\hat{\mathbf{y}}$ denotes the predicted price vector. The model is trained by minimizing a predefined loss function, enabling accurate prediction of target data product prices.

Table~\ref{tab:notation} summarizes the core mathematical notation used in this study.

\begin{table}[htbp]
\centering
\TableSpaced
\caption{Summary of key notation.}
\label{tab:notation}
\begin{tabular}{p{3.3cm}p{12.0cm}}
\toprule
Symbol & Description \\
\midrule
$\mathcal{G}=(\mathcal{V},\mathcal{E})$ & Heterogeneous information network: node set $\mathcal{V}$ and edge set $\mathcal{E}$ \\
$D$, $S$, $M$, $I$, $P$ & Five node types: data product, provider, market, industry, pricing mode \\
$\mathbf{X}$, $\mathbf{y}$, $\hat{\mathbf{y}}$ & Product feature matrix; true and predicted price vectors \\
$\mathcal{P}_k$, $K$ & The $k$-th meta-path; total number of meta-paths \\
$\mathcal{A}_{\mathcal{P}_k}$ & Relational structure (adjacency) restricted to meta-path $\mathcal{P}_k$ \\
$\mathbf{h}_{\mathcal{P}_k}$ & Node representation produced by the HGT encoder for meta-path $\mathcal{P}_k$ \\
$w_k$, $\alpha_k$ & Final fused weight and static learnable logit of meta-path $k$ \\
$\mathbf{z}$, $\mathbf{z}_{\text{repr}}$, $\mathbf{z}_{\text{price}}$ & Multi-path fused representation and its structural / pricing subspaces \\
$C$ & Number of price tiers (and tier-specific experts); $C=5$ \\
$f_{\text{cls}}$, $\mathbf{p}=(p_1,\dots,p_C)$ & Price-tier classifier; predicted tier-probability vector \\
$g_k(\cdot)$ & $k$-th tier-specific regression expert \\
$\hat{p}_i$, $p_{ik}$ & Final soft-routed price prediction for product $i$; probability that $i$ belongs to tier $k$ \\
$\mathcal{L}_{cls}$, $\mathcal{L}_{reg}$, $\mathcal{L}_{CL}$, $\mathcal{L}_{local}$, $\mathcal{L}_{path}$ & Classification, regression, contrastive, local expert-supervision, and path-regularization losses \\
\bottomrule
\end{tabular}
\end{table}

\section{Methodology} \label{sec:methodology}

\begin{figure}[!htbp]
    \centering
    \includegraphics[width=1\linewidth]{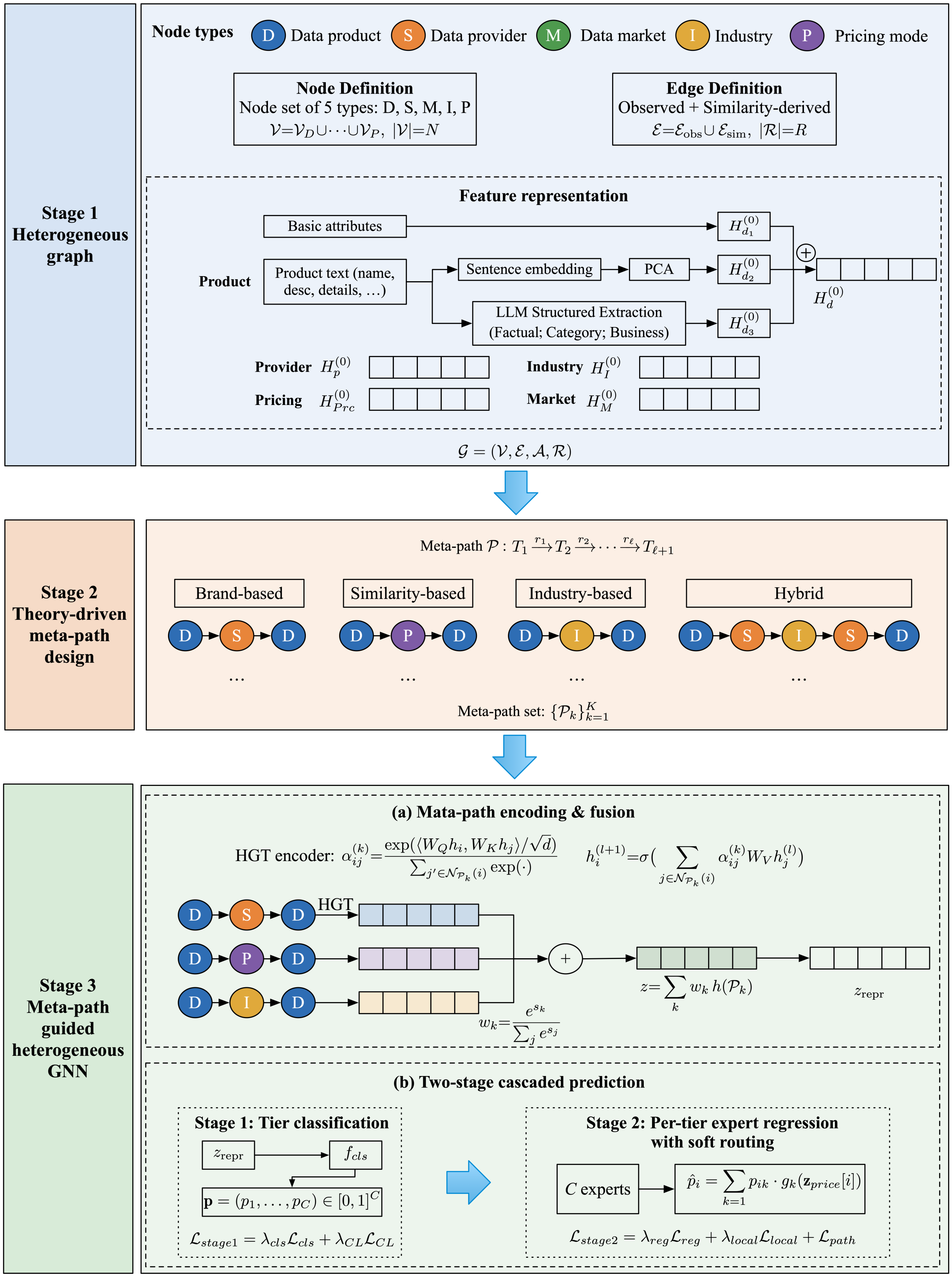}
    \caption{Framework of the proposed model.}
    \label{fig:MPGNN}
\end{figure}

The framework of the proposed model is shown in Figure \ref{fig:MPGNN}. The proposed framework consists of three core components. First, Heterogeneous Graph Construction models data products, data providers, data markets, industries, and pricing modes as different types of nodes and constructs a heterogeneous information network by integrating observed and similarity-derived relations. Meanwhile, the basic attributes and textual information of data products are encoded as node features to provide initial representations for subsequent graph learning. Second, Theory-driven Meta-path Design constructs brand-based, similarity-based, industry-based, and hybrid meta-paths based on economic theories related to data product price formation, thereby capturing higher-order associations among data products through different economic relationships. Third, the Meta-path-guided Heterogeneous GNN performs message passing and representation learning along different meta-paths and adaptively aggregates the resulting meta-path-specific representations. The learned representations are then used for price-tier classification and price regression to determine the final prices of data products.

\subsection{Design principles}
\label{sec:design-principles}

Before presenting the implementation details of each module, we make explicit the design logic that connects them to the characteristics of data product pricing. Non-standardized data products raise four distinct computational-design requirements, each of which is only partially addressed by conventional attribute-based or generic graph-based pricing methods; following the design science research paradigm in information systems \citep{hevnerDesignScience2004}, we state each requirement as a design principle (DP), identify the limitation of prior approaches it responds to, and note the cost or boundary condition under which it is expected to help.

\textbf{DP1: Explicit representation of heterogeneous relations.} Data product prices are jointly shaped by entities of fundamentally different semantic types, namely providers, marketplaces, industries, similar products, and pricing modes. Attribute-based pricing models (e.g., linear regression, tree ensembles, or feed-forward networks over product features) ignore these relations entirely, while a homogeneous graph that collapses all relation types into a single similarity network keeps the connections but discards their semantic identity, so neither can attribute predictive power to a specific relational mechanism. DP1 responds to this by encoding products, providers, markets, industries, and pricing modes as distinct node types connected by typed edges, so that each relation keeps its semantic identity throughout learning. The cost of this design is additional modeling and bookkeeping complexity relative to a single homogeneous graph, and its payoff is conditional on the relation types being genuinely separable in the data.

\textbf{DP2: Alignment between relational semantics and economic mechanisms.} Even once relations are typed, a generic heterogeneous GNN still propagates information across the whole graph without constraining which relation sequence produced a given signal, so brand, similarity, and industry effects remain blended into one opaque representation. DP2 responds to this by constraining propagation to theory-grounded meta-paths, so that each economic mechanism maps onto a distinct, inspectable relational pathway rather than being merged into the aggregate message-passing signal. This is a stronger assumption than letting the network discover its own propagation pattern: it presumes that the specified meta-paths are a reasonably complete and correctly operationalized reading of brand equity, similarity, and industry structure theories.

\textbf{DP3: Context-dependent integration of relational information.} Different data products rely on different relational mechanisms in practice: some are priced mainly by provider reputation, others mainly by content similarity or industry convention. A single, globally fixed weighting over meta-paths (e.g., a simple unweighted average) cannot adapt to this heterogeneity across products. DP3 responds to this with the attention-based meta-path gate, which learns a data-dependent weighting over paths instead of a fixed one. We note that, in the current implementation, the gate's attention scores are computed from path-level summary vectors obtained via global mean pooling, so the resulting weights vary across splits and tasks but are shared across all products within a given split rather than individually conditioned on each product.

\textbf{DP4: Tier-aware decomposition for a long-tailed price distribution.} Data product prices span several orders of magnitude, so a single regressor must simultaneously fit a dense mass of low- and mid-priced products and a sparse tail of high-value ones, which tends to under-fit the tail. DP4 responds to this by first identifying a coarse price tier and then routing to a tier-specific regression expert. Relative to a standard mixture-of-experts layer, in which the router is learned end-to-end purely from the downstream loss without any ground-truth signal on expert assignment, our router is instead a classifier supervised directly by observable price-tier labels and frozen before the experts are trained.

DP1--DP4 state the chain of design requirements this study responds to: heterogeneity must first be made explicit (DP1) before it can be aligned with economic mechanisms (DP2); that alignment must then be integrated in a way that is sensitive to product-level context (DP3); and the resulting representation must still be fitted to a price distribution that is itself heterogeneous across tiers (DP4). 

\subsection{Heterogeneous graph construction}
\label{sec:heterogeneous-graph}

\subsubsection{Node and edge definition}
The nodes primarily consist of data product nodes, data provider nodes, industry nodes, market nodes, and pricing mode nodes.
The heterogeneous data trading network includes the following types of edges:
(Data market, \textit{contains}, Data product), 
(Data market, \textit{collaborates with}, Data provider), 
(Data provider, \textit{provides}, Data product), 
(Data provider, \textit{belongs to}, industry),
(Data product, \textit{applies to}, industry),
(Data product, \textit{supports}, Pricing mode),
(Data product, \textit{is similar to}, Data product),
(Data provider, \textit{is similar to}, Data provider).

Among the above relationships, the edges
(Data product, \textit{is similar to}, Data product) and
(Data provider, \textit{is similar to}, Data provider)
are derived through similarity computation, while the others are directly obtained from observed data.

Specifically, the similarity between data products is computed using cosine similarity based on product descriptions:

\begin{equation}
\text{Sim}(d_i, d_j) = \frac{\mathbf{x}_i \cdot \mathbf{x}_j}{\|\mathbf{x}_i\| \|\mathbf{x}_j\|}
\end{equation}

The similarity between data providers is computed analogously, using cosine similarity over a feature vector that concatenates provider-level attributes, namely provider type (one-hot encoded) and public listing status:

\begin{equation}
\text{Sim}(p_a, p_b) = \frac{\mathbf{s}_a \cdot \mathbf{s}_b}{\|\mathbf{s}_a\| \|\mathbf{s}_b\|}
\end{equation}
where $\mathbf{s}_a$ and $\mathbf{s}_b$ denote the standardized attribute vectors of providers $p_a$ and $p_b$, respectively.

\subsubsection{Feature representation}
The data product nodes include the following attributes: product name, text features, label\_volume, label\_history, label\_future, label\_sensitive, labels\_coverage, label\_data\ sample, support Email, support URL, and labels\_format. The detailed descriptions of these attributes are presented in Table \ref{tab:A1}.
Data provider nodes include provider name, provider type and public listing status. Detailed descriptions are presented in Table \ref{tab:A2}.
Industry node attributes include Market scale 1, Market scale 2, Industry growth rate, and Data Sensitivity. Detailed descriptions are presented in Table \ref{tab:A3}.
Market node features include market type, market scale, and partner scale, as detailed in Table \ref{tab:A4}.
Pricing modes mainly include subscription-based pricing, request-based pricing, and one-time purchase.

\subsubsection{Feature Preprocessing}

During node feature preprocessing, price-related fields of data products are first transformed logarithmically to mitigate the skewness of price distributions. We use an uppercase $P$ to denote the original (untransformed) price and a lowercase $p$ for its log-transformed counterpart. For the price $P_{ij}$ of data product $i$ under pricing mode $j$, the transformed price is defined as:

\begin{equation}
p_{ij} = \log(1 + P_{ij}).
\end{equation}

For product textual information, text embeddings are used to represent product descriptions. PCA is then applied to reduce the text embedding dimension to 128:

\begin{equation}
\mathbf{x}_{text} = \text{PCA}_{128}(\mathbf{x}_{emb}).
\end{equation}

We adopt PCA rather than uncentered SVD because PCA first centers the embeddings around their mean before extracting principal directions, which is the appropriate choice for dense vectors such as sentence embeddings; uncentered SVD is instead motivated by preserving the sparsity of term-frequency matrices, a consideration that does not apply here since the input embeddings are already dense.

For numerical features, standard normalization is applied:

\begin{equation}
\mathbf{x}_{norm} = \frac{\mathbf{x} - \mu}{\sigma}.
\end{equation}

Categorical variables, such as data format and provider entity type, are encoded using one-hot encoding, while missing values in Boolean features are filled. Finally, features of all node types are mapped into the heterogeneous graph structure.

For supervised labels, the continuous price of each data product is used as the regression target. To support classification, valid price samples are further divided into several price intervals according to quantile thresholds, producing price-tier labels. In this implementation, prices are divided into five intervals for price-tier classification. Samples with missing or invalid prices are excluded from training and evaluation through a mask mechanism.

LLM-structured text features.
In addition to the PCA-reduced text embeddings, we employ a locally deployed
large language model to extract structured metadata from each
product's textual description (name, product\_description, and product\_detail).
Following a three-layer schema, the LLM extracts:
(i)~factual attributes such as record count, update frequency, geographic scope,
and number of data fields;
(ii)~category labels including data category, PII presence, subject type, and
source type;
(iii)~business-level inferences such as primary use case, target customer segment,
exclusivity rating (1--5), and pricing unit hint.
These 12 structured fields are encoded into a 54-dimensional feature vector via
log1p-normalization, z-score standardization, and one-hot encoding (with missing
indicators), and are appended to the product node features before graph construction.
To prevent information leakage, the LLM prompt is restricted to the product's
textual fields only; all price-related tokens are programmatically removed from
the input.
This design transforms the unstructured textual description into explicit,
machine-readable pricing signals that complement the latent semantics captured
by PCA embeddings. Because these fields are machine-extracted rather than directly observed, they may carry measurement error or misclassification inherited from the extraction process \citep{yangMindGapAccounting2018}.

\subsection{Theory-driven meta-path design}
\label{sec:meta-path-design}

\subsubsection{Meta-path schema}

To characterize the multi-relational dependence among data products in the pricing process, we construct a heterogeneous information network composed of five types of nodes, namely, data product nodes $D$, data provider (brand) nodes $S$, market nodes $M$, industry nodes $I$, and pricing mode nodes $P$. Based on product pricing theories, we further design a set of meta-paths to capture the potential pricing dependence between two data products.

Specifically, the design of meta-paths in this study is grounded in three major pricing theories, including brand equity theory, product similarity theory, and industry structure theory. In addition, some meta-paths reflect the joint effects of multiple theories and are therefore categorized as hybrid meta-paths. The meta-paths considered in this study can be grouped as follows.

\begin{itemize}
    \item \textbf{Brand-based meta-paths.} These meta-paths are motivated by brand equity theory, which suggests that products associated with the same brand may exhibit similar pricing patterns. Representative meta-paths include $D_1 \rightarrow S_1 \rightarrow D_2$.

    \item \textbf{Similarity-based meta-paths.} These meta-paths are motivated by product similarity theory. When two data products are similar in product content or share the same pricing mode, their value realization mechanisms and market comparability are likely to be close, and thus one product may reference the price of the other. Representative meta-paths include $D_1 \rightarrow D_2$ and $D_1 \rightarrow P_1 \rightarrow D_2$. We group the pricing-mode path $D_1 \rightarrow P_1 \rightarrow D_2$ under similarity for parsimony, although it is more precisely a market-segment or comparability constraint than a measure of content similarity.

    \item \textbf{Industry-based meta-paths.} These meta-paths are motivated by industry structure theory, which emphasizes that products within the same industry are subject to similar competitive environments, demand characteristics, and market structures. Therefore, prices may be formed with reference to other products in the same industry. A representative meta-path is $D_1 \rightarrow I_1 \rightarrow D_2$.

    \item \textbf{Hybrid meta-paths.} In practice, the pricing of data products is often jointly shaped by multiple factors, such as brand association, industry linkage, and market constraints. Therefore, we further construct several hybrid meta-paths to capture these composite transmission mechanisms, including $D_1 \rightarrow S_1 \rightarrow S_2 \rightarrow D_2$, 
    $D_1 \rightarrow S_1 \rightarrow I_1 \rightarrow S_2 \rightarrow D_2$,
    $D_1 \rightarrow S_1 \rightarrow M_1 \rightarrow S_2 \rightarrow D_2$,
    and $D_1 \rightarrow I_1 \rightarrow S_1 \rightarrow S_2 \rightarrow I_2 \rightarrow D_2$.
\end{itemize}

\subsubsection{Economic interpretation of meta-paths}
\label{sec:meta-path-interpretation}

To better explain the economic meaning of the proposed meta-paths, we provide an interpretation for each meta-path from the perspective of pricing theories and market mechanisms. 

\begin{enumerate}
\item[$(1)$] \textbf{$D_1 \rightarrow D_2$.}  
    If the product contents of $D_1$ and $D_2$ are similar, then $D_1$ may refer to the price of $D_2$ when determining its price. This meta-path reflects the idea of product similarity theory, that is, similar products tend to exhibit comparable value and price levels.
    This path operationalizes similarity through the cosine similarity of text-embedding-based product descriptions, not through a direct measure of substitutability, functional overlap, or buyer-perceived comparability. 

\item[$(2)$] \textbf{$D_1 \rightarrow S_1 \rightarrow D_2$.}  
    If both $D_1$ and $D_2$ are provided by the same brand $S_1$, then the price of $D_1$ is likely to reference the price of $D_2$. This meta-path reflects the role of brand equity theory, since products under the same brand usually share similar brand positioning, reputation, and pricing strategies.
    This path is measured purely from the categorical fact of sharing the same provider node, without an independent measure of brand reputation, perceived quality, or willingness-to-pay premium. 

\item[$(3)$] \textbf{$D_1 \rightarrow I_1 \rightarrow D_2$.}  
    If $D_1$ belongs to industry $I_1$ and $D_2$ also belongs to industry $I_1$, then the price of $D_1$ may refer to the price of $D_2$ within the same industry. This meta-path reflects industry structure theory, which suggests that products in the same industry are affected by similar demand conditions and competitive structures.
    This path is measured purely from shared industry-category membership, not from any direct measure of competitive intensity, market concentration, substitutability, or entry barriers, the constructs that industry structure theory is actually about.

\item[$(4)$] \textbf{$D_1 \rightarrow P_1 \rightarrow D_2$.}  
    This meta-path indicates that products $D_1$ and $D_2$ adopt the same pricing mode $P_1$, such as subscription-based pricing, one-off payment, or usage-based pricing. This path reflects a pricing mode constraint mechanism. Different pricing modes imply different value realization logics and price structures. Products sharing the same pricing mode are therefore highly comparable, and the price of $D_1$ may refer to that of $D_2$.
    This path is measured from a shared categorical pricing-mode label (e.g., both subscription-based), not from a direct measure of content or functional similarity.

\item[$(5)$] \textbf{$D_1 \rightarrow S_1 \rightarrow S_2 \rightarrow D_2$.}  
    If $D_1$ is provided by brand $S_1$, while $D_2$ is provided by brand $S_2$, and $S_1$ is associated with $S_2$, then the price of $D_1$ may be correlated with that of $D_2$. This meta-path captures a hybrid effect, in which brand association further extends pricing dependence across related brands.
    The $S_1$--$S_2$ association is measured by the provider-similarity score defined in Section~\ref{sec:heterogeneous-graph} (cosine similarity over provider type and listing status), not by an independent measure of inter-brand affiliation, partnership, or market overlap. 

\item[$(6)$] \textbf{$D_1 \rightarrow S_1 \rightarrow I_1 \rightarrow S_2 \rightarrow D_2$.}  
    This meta-path indicates that data product $D_1$ is provided by brand $S_1$, brand $S_1$ belongs to industry $I_1$, and within the same industry $I_1$, another brand $S_2$ provides product $D_2$. This path reflects an intra-industry cross-brand competition mechanism. Within the same industry, different brands face similar market conditions and competitive pressures, so the price of $D_1$ may reference that of $D_2$ to maintain market competitiveness and pricing rationality.
    This path is measured purely from shared industry-category membership between two different providers, inheriting the same coarse-proxy limitation noted for the $D$--$I$--$D$ path above; it does not measure competitive intensity or market-share overlap between $S_1$ and $S_2$ directly. 

\item[$(7)$] \textbf{$D_1 \rightarrow S_1 \rightarrow M_1 \rightarrow S_2 \rightarrow D_2$.}  
    This meta-path indicates that brands $S_1$ and $S_2$ are both active in the same market $M_1$, and provide products $D_1$ and $D_2$, respectively. This path reflects a market-level pricing constraint mechanism. Within the same data trading platform or marketplace, product pricing is shaped by common market rules, user structure, and competitive environment. Therefore, the price of $D_1$ may refer to the price distribution of other products such as $D_2$ in the same market.
    This path is measured from provider-market affiliation edges, i.e., both providers being active on the same marketplace, rather than from a direct measure of shared pricing rules or user-base overlap.

\item[$(8)$] \textbf{$D_1 \rightarrow I_1 \rightarrow S_1 \rightarrow S_2 \rightarrow I_2 \rightarrow D_2$.}  
    This meta-path indicates that product $D_1$ belongs to industry $I_1$, is connected to brand $S_2$ through brand $S_1$, and brand $S_2$ belongs to industry $I_2$ and provides product $D_2$. This path reflects a composite transmission mechanism across brands and industries. On the one hand, brands may be related through similarity; on the other hand, structural linkages may exist across industries. Therefore, the pricing of $D_1$ may be indirectly associated with that of $D_2$ through both brand and industry relations.
    This is the longest and most composite path considered, chaining the provider-similarity and industry-membership constructs discussed above across two industries and two providers; it therefore inherits both constructs' measurement limitations and compounds them.

\end{enumerate}

Overall, the above meta-paths reveal that data product pricing is shaped by a multi-layer relational structure rather than by product attributes alone. By embedding these meta-paths into the heterogeneous graph model, this study is able to capture the economic mechanisms underlying price formation and improve both the predictive performance and interpretability of data product pricing.

\begin{figure}[!ht]
    \centering
    \includegraphics[width=\linewidth]{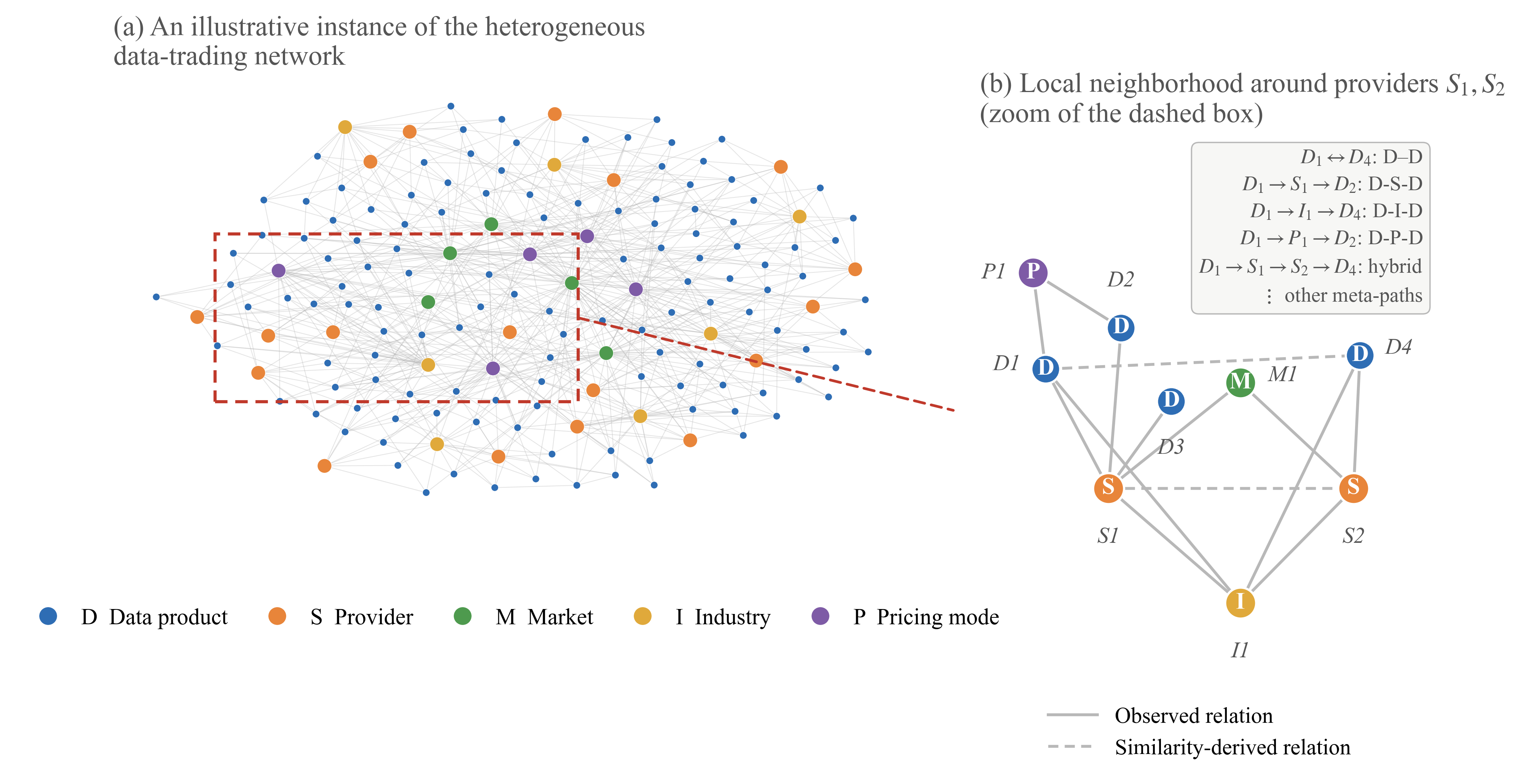}
    \caption{An illustrative example of the heterogeneous data-trading network. (a) A schematic network instance in which node color denotes node type (data product $D$, provider $S$, market $M$, industry $I$, pricing mode $P$); solid edges denote directly observed relations and dashed edges denote similarity-derived relations. (b) The local neighborhood enclosed by the dashed box, centered on two providers $S_1$ and $S_2$, magnified and labeled: this single neighborhood concretely instantiates five of the meta-paths discussed above.}
    \label{fig:hin_illustration}
\end{figure}

Figure~\ref{fig:hin_illustration} provides a visual illustration of the resulting heterogeneous network and of how the meta-paths introduced above are concretely instantiated among real node types.

\subsection{Meta-path-based data product pricing method}

Based on designed multiple meta-paths, we employ a heterogeneous graph transformer (HGT) to encode each meta-path, and introduce a meta-path weight gating mechanism to fuse representations from different semantic paths. The classification and regression tasks are trained in a cascaded, two-stage manner: the price-tier classifier is first trained to full convergence and frozen, after which the per-tier regression experts are trained using the frozen classifier's outputs as a fixed routing signal.
The pseudocode for the two stages is shown in Algorithm~\ref{alg:stage1_training} (Stage 1: classification training) and Algorithm~\ref{alg:stage2_training} (Stage 2: regression training).

\begin{algorithm}[htbp]
\caption{Stage-1 Classification Training}
\label{alg:stage1_training}
\begingroup
\small
\SingleSpacedXI
\begin{algorithmic}[1]
\item[] \textbf{Input:} Graph database, predefined meta-path set $\{\mathcal{P}_k\}_{k=1}^{K}$, price column $p$, maximum epochs $E_1$
\item[] \textbf{Output:} Trained HGT encoders $\{\text{HGT}_k\}_{k=1}^{K}$, meta-path gate, and classification head $f_{cls}$; stored tier probabilities $\mathbf{p}$ for all products
\STATE Extract node sets $\{V_D,V_S,V_M,V_I,V_P\}$ and relation set $\mathcal{E}$ from graph database
\STATE Construct the heterogeneous graph $\mathcal{G}=(\mathcal{V},\mathcal{E},\mathcal{A},\mathcal{R})$
\STATE Apply price log transformation, text PCA reduction, categorical encoding, and numerical normalization
\STATE Generate regression labels and bucketized price-tier labels; construct valid-sample masks
\STATE Split valid data product nodes into training and testing sets
\FOR{each meta-path $\mathcal{P}_k$}
    \STATE Generate reachable product pairs $\text{Pairs}_{\mathcal{P}_k}$ by path-constrained neighbor expansion
    \STATE Initialize an independent HGT encoder $\text{HGT}_k$ for $\mathcal{P}_k$
\ENDFOR
\STATE Initialize self-attention meta-path gate, classification head $f_{cls}$, and optimizer
\STATE $best\_metric_{cls} \leftarrow -\infty$
\FOR{$epoch = 1$ to $E_1$}
    \FOR{each meta-path $\mathcal{P}_k$}
        \STATE Compute path-specific representation $\mathbf{h}_{\mathcal{P}_k}=\text{HGT}_k(\mathbf{X},\mathcal{A}_{\mathcal{P}_k})$
    \ENDFOR
    \STATE Compute path summary vectors $\mathbf{E}$ via global mean pooling
    \STATE Compute dynamic attention scores via $\mathbf{Q}=\mathbf{E}\mathbf{W}_Q$, $\mathbf{K}=\mathbf{E}\mathbf{W}_K$, scaled dot-product attention
    \STATE Fuse path representations: $\mathbf{z}=\sum_{k=1}^{K}w_k\mathbf{h}_{\mathcal{P}_k}$
    \STATE Split $\mathbf{z}$ into $\mathbf{z}_{repr}$ (structural) and $\mathbf{z}_{price}$ (pricing) subspaces
    \STATE Compute meta-path contrastive loss $\mathcal{L}_{CL}$ using $\text{Pairs}_{\mathcal{P}_k}$ on $\mathbf{z}_{repr}$
    \STATE Predict tier probabilities $\mathbf{p} = \text{softmax}(f_{cls}(\mathbf{z}_{price}))$
    \STATE Compute classification loss $\mathcal{L}_{cls}$ (weighted cross-entropy)
    \STATE Compute Stage-1 loss $\mathcal{L}_{\text{stage1}} = \lambda_{cls}\mathcal{L}_{cls} + \lambda_{CL}\mathcal{L}_{CL}$
    \STATE Update the encoder, meta-path gate, and classification head by backpropagation
    \IF{validation classification metric improves}
        \STATE Save best Stage-1 model checkpoint
    \ENDIF
\ENDFOR
\STATE Restore the best Stage-1 model parameters
\STATE Freeze the classification head $f_{cls}$; using the frozen classifier, compute the tier probabilities $\mathbf{p}$ once for every product and store them as fixed values for Stage 2 (the encoder and meta-path gate remain trainable and carry over into Stage 2)
\end{algorithmic}
\endgroup
\end{algorithm}

\begin{algorithm}[htbp]
\caption{Stage-2 Regression Training with the Frozen Classifier}
\label{alg:stage2_training}
\begingroup
\small
\SingleSpacedXI
\begin{algorithmic}[1]
\item[] \textbf{Input:} HGT encoders $\{\text{HGT}_k\}_{k=1}^{K}$, meta-path gate, and stored tier probabilities $\mathbf{p}$ from Algorithm~\ref{alg:stage1_training}; price column $p$; maximum epochs $E_2$
\item[] \textbf{Output:} Trained per-tier expert regressors $\{g_k\}_{k=1}^{C}$, final meta-path weights, evaluation metrics
\STATE Initialize per-tier expert regressors $\{g_k\}_{k=1}^{C}$ and an optimizer over the encoder, meta-path gate, and expert parameters (the classification head is excluded)
\STATE $best\_metric_{reg} \leftarrow -\infty$
\FOR{$epoch = 1$ to $E_2$}
    \FOR{each meta-path $\mathcal{P}_k$}
        \STATE Compute path-specific representation $\mathbf{h}_{\mathcal{P}_k}=\text{HGT}_k(\mathbf{X},\mathcal{A}_{\mathcal{P}_k})$
    \ENDFOR
    \STATE Compute path summary vectors $\mathbf{E}$ via global mean pooling and fuse $\mathbf{z}=\sum_{k=1}^{K}w_k\mathbf{h}_{\mathcal{P}_k}$; take its pricing subspace $\mathbf{z}_{price}$
    \STATE Retrieve the frozen tier probabilities $p_{ik}$ stored from Stage 1 (the classification head is not re-evaluated)
    \STATE Compute per-tier expert outputs $\{g_k(\mathbf{z}_{price})\}_{k=1}^{C}$
    \STATE Aggregate via soft routing: $\hat{p} = \sum_{k} p_k \cdot g_k(\mathbf{z}_{price})$
    \STATE Compute global regression loss $\mathcal{L}_{reg}$ (MSE) and local expert supervision loss $\mathcal{L}_{local}$
    \STATE Compute path regularization loss $\mathcal{L}_{path}$ (LASSO + entropy)
    \STATE Compute Stage-2 loss $\mathcal{L}_{\text{stage2}}=\lambda_{reg}\mathcal{L}_{reg}+\lambda_{local}\mathcal{L}_{local}+\mathcal{L}_{path}$
    \STATE Update the encoder, meta-path gate, and per-tier experts by backpropagation (the classification head is excluded from the optimizer)
    \IF{validation regression metric improves}
        \STATE Save best Stage-2 model checkpoint
    \ENDIF
\ENDFOR
\STATE Restore the best Stage-2 model parameters
\STATE Save the trained model, meta-path weights, and evaluation results
\end{algorithmic}
\endgroup
\end{algorithm}

\subsubsection{Pair Generation}
For a given meta-path, $\mathcal{P} = (v_1, r_1, v_2, r_2, \ldots, r_{k-1}, v_k)$, we generate pairs of data product nodes that are reachable along the meta-path:

\begin{equation}
\text{Pairs}_{\mathcal{P}} = \{(v_i, v_j) \mid v_i \xrightarrow{\mathcal{P}} v_j\}.
\end{equation}

The pair generation process starts from each source product node and expands neighbors layer by layer according to the edge types specified by the meta-path. To avoid combinatorial explosion in high-order neighborhoods, at most a fixed number of neighbors are randomly sampled at each hop. A limited number of final target nodes are then retained, and source-target node pairs are constructed as positive samples. These pairs are used for subsequent meta-path contrastive learning.

\subsubsection{Heterogeneous Graph Transformer Encoder}

An independent HGT encoder is constructed for each meta-path. First, raw features of different node types are linearly projected into a unified hidden space:
\begin{equation}
\mathbf{h}^{(0)}_v = \mathbf{W}^{(0)}_{\tau(v)} \mathbf{x}_v,
\end{equation}
where $\tau(v)$ denotes the node type of node $v$, and $\mathbf{W}^{(0)}_{\tau(v)}$ is the type-specific linear projection matrix.
In a standard HGT, message passing at layer $l$ can be expressed as:

\begin{equation}
\mathbf{h}^{(l+1)}_v
=
\text{HGTConv}^{(l)}
\left(
\{\mathbf{h}^{(l)}_u : u \in \mathcal{N}(v)\},
\mathcal{E}
\right).
\end{equation}

We further introduce meta-path constraints. For the $k$-th meta-path, the HGT encoder does not perform unconstrained propagation over the full heterogeneous graph. Instead, it conducts message passing layer by layer according to the sequence of edge types in the meta-path:

\begin{equation}
\mathbf{h}_{\mathcal{P}_k}
=
\text{HGT}_k(\mathbf{X}, \mathcal{A}_{\mathcal{P}_k}),
\end{equation}
where $\mathcal{A}_{\mathcal{P}_k}$ denotes the relation structure constrained by meta-path $\mathcal{P}_k$. Specifically, at layer $l$, only the relation type corresponding to the $l$-th edge in the meta-path is retained, so that node representation updates are strictly constrained by the semantic structure of the meta-path. In this way, each meta-path produces a data product representation under a specific semantic context.

\subsubsection{Meta-Path Weight Gating Mechanism}
\label{sec:gating-mechanism}

Let $\mathbf{h}_{\mathcal{P}_k}$ denote the node representation generated by the $k$-th meta-path. The set of all meta-path-specific representations is $\{\mathbf{h}_{\mathcal{P}_1}, \mathbf{h}_{\mathcal{P}_2}, \ldots, \mathbf{h}_{\mathcal{P}_K}\}$. To fuse semantic information from different meta-paths, a meta-path weight gating module is introduced. Rather than treating all paths as equally important, this module learns to assign an adaptive weight to each path, so that more pricing-relevant relational mechanisms receive higher weights while redundant or noisy paths are suppressed.

Concretely, the gating module combines a \emph{static learnable bias} with a \emph{data-dependent dynamic attention} mechanism. The number of meta-paths is denoted $K$. First, global mean pooling is applied to the node representations of each path to obtain path-level summary vectors:

\begin{equation}
\mathbf{e}_k = \frac{1}{N} \sum_{i=1}^{N} \mathbf{h}_{\mathcal{P}_k}[i], \quad k = 1, \dots, K,
\end{equation}
where $\mathbf{h}_{\mathcal{P}_k}[i] \in \mathbb{R}^{D_h}$ is the representation of the $i$-th product node under the $k$-th meta-path, and $N$ is the number of product nodes. Stacking these summaries yields $\mathbf{E} = [\mathbf{e}_1; \dots; \mathbf{e}_K] \in \mathbb{R}^{K \times D_h}$.

A scaled dot-product self-attention is then applied to capture inter-path interactions:

\begin{equation}
\mathbf{Q} = \mathbf{E} \mathbf{W}_Q, \quad \mathbf{K} = \mathbf{E} \mathbf{W}_K,
\end{equation}
where $\mathbf{W}_Q, \mathbf{W}_K \in \mathbb{R}^{D_h \times d_{\text{attn}}}$ are learnable projection matrices and $d_{\text{attn}}$ is the attention dimension. The attention matrix is computed as:

\begin{equation}
\mathbf{A} = \text{softmax}\!\left(\frac{\mathbf{Q} \mathbf{K}^{\top}}{\sqrt{d_{\text{attn}}}}\right) \in \mathbb{R}^{K \times K},
\end{equation}
where $\mathbf{A}_{ij}$ quantifies how much path $i$ attends to path $j$. The dynamic importance of each path is then obtained by averaging the attention it receives from all other paths:

\begin{equation}
s_k^{\text{dyn}} = \frac{1}{K} \sum_{i=1}^{K} \mathbf{A}_{ik}, \quad k = 1, \dots, K.
\end{equation}

This formulation implements the intuition that a path that is consistently ``referenced'' by many other paths is likely to carry more pricing-relevant information.

The final path weight combines the static learnable logit $\alpha_k$ with the log-transformed dynamic score, followed by a temperature-scaled softmax:

\begin{equation}
w_k = \text{softmax}\!\left(\frac{\alpha_k + \log(s_k^{\text{dyn}} + \epsilon)}{\tau}\right)_k,
\end{equation}
where $\epsilon = 10^{-8}$ prevents numerical instability and the temperature $\tau = 5.0$ smooths the distribution to prevent any single path from dominating prematurely.

The multi-path fused representation is then:

\begin{equation}
\mathbf{z} = \sum_{k=1}^{K} w_k \, \mathbf{h}_{\mathcal{P}_k}.
\end{equation}

To encourage the model to identify a sparse yet diverse set of important paths, two complementary regularization terms are applied. First, an $\ell_1$ (LASSO) penalty on the static logits promotes sparsity:

\begin{equation}
\mathcal{L}_{\text{LASSO}} = \lambda_{\text{LASSO}} \sum_{k=1}^{K} |\alpha_k|.
\end{equation}

Second, an entropy regularization term rewards weight dispersion and prevents pathological collapse onto a single path:

\begin{equation}
\mathcal{L}_{\text{entropy}} = -\sum_{k=1}^{K} w_k \log(w_k + \epsilon).
\end{equation}

The combined path regularization loss is:

\begin{equation}
\mathcal{L}_{\text{path}} = \mathcal{L}_{\text{LASSO}} - \lambda_H \, \mathcal{L}_{\text{entropy}},
\end{equation}
where $\lambda_{\text{LASSO}} = 10^{-3}$ and $\lambda_H = 0.05$ control the strength of the two regularizers. The LASSO term drives unimportant paths toward zero weight, while the entropy term ensures that at least a few paths retain non-negligible weight; together, they produce a sparse but non-degenerate weighting that is both interpretable and stable. The learned weights are directly inspected after training (Section~\ref{sec:result}) to quantify the relative importance of each economic pricing mechanism.

\subsubsection{Meta-Path Contrastive Learning Loss}

Structural supervision is provided by meta-path contrastive learning. For positive node pairs generated by each meta-path, the model maximizes the similarity between the source and target node representations. The similarity between nodes $u$ and $v$ is defined as:

\begin{equation}
\text{sim}(u,v)
=
\frac{
\mathbf{z}_{repr}[u] \cdot \mathbf{z}_{repr}[v]
}{
||\mathbf{z}_{repr}[u]||_2 \cdot ||\mathbf{z}_{repr}[v]||_2
}.
\end{equation}

When negative samples are used, the InfoNCE loss~\citep{oordRepresentationLearningContrastive2018}
for meta-path $\mathcal{P}_k$ can be written as:

\begin{equation}
\mathcal{L}_{CL}^{\mathcal{P}_k}
=
-
\frac{1}{|\text{Pairs}_{\mathcal{P}_k}|}
\sum_{(u,v)}
\log
\frac{
\exp(\text{sim}(u,v)/\tau)
}{
\sum_{v'} \exp(\text{sim}(u,v')/\tau)
}.
\end{equation}

This study adopts the InfoNCE objective defined above, with explicitly constructed negative samples, rather than the simplified, negative-sample-free form. 

Under the cascaded training scheme adopted in this study, $\mathcal{L}_{CL}$ is used only in Stage 1 (classification training), where it is optimized jointly with the classification loss to inject structural-semantic supervision into the encoder. It is not used in Stage 2 (regression training); see the following two subsections for details.

\subsubsection{Two-Stage Cascaded Prediction with MoE-Style Soft Routing}
\label{sec:two-stage-moe}

In practice, the continuous price of a data product can vary by orders of magnitude across tiers (e.g., a few dollars for a simple CSV file versus tens of thousands of dollars for a real-time financial data feed), and fitting a single global regressor to such a wide range is inherently difficult.
To address this, we introduce a two-stage cascaded prediction architecture that explicitly decomposes the pricing task into (i)~coarse-grained tier identification and (ii)~fine-grained within-tier regression. The architecture is inspired by the Mixture-of-Experts (MoE) paradigm \citep{jacobsAdaptiveMixturesLocal1991,shazeerOutrageouslyLargeNeural2017,puigcerverSparseSoftMixtures2024}, where each expert specialises in a narrower price region and the final prediction is a probabilistically weighted combination of expert outputs. These two subtasks are not trained simultaneously; instead, they are trained sequentially in a cascade, with classification performed first and regression performed second, as detailed below. Unlike standard MoE architectures, in which the gating network is trained jointly and end-to-end with the experts purely from the downstream task loss, without any external supervision on expert assignment, the routing signal in our architecture is a classifier supervised by externally defined, economically meaningful ground-truth price-tier labels, and is frozen before the experts are trained. We adopt this supervised, decoupled design because, unlike in settings such as language modeling where no ground-truth expert assignment exists, our setting provides a natural and reliable partition of the output space (price tiers) that is more effectively learned through direct supervision than through latent self-organization. 

\paragraph{Stage 1: Tier Classification.}
In the first stage, a dedicated classifier $f_{\text{cls}}: \mathbb{R}^{D_{\text{price}}} \to \Delta^{C-1}$ maps the pricing subspace $\mathbf{z}_{\text{price}}$ to a probability distribution over $C$ price tiers:

\begin{equation}
\mathbf{p} = \text{softmax}\big(f_{\text{cls}}(\mathbf{z}_{\text{price}})\big) \in [0,1]^{C},
\end{equation}
where $C = 5$ in our implementation and $p_k = P(\text{tier}=k \mid \mathbf{z}_{\text{price}})$. The classifier is trained with a weighted cross-entropy loss. Specifically, this stage jointly trains the encoder, the meta-path gate, and the classifier $f_{\text{cls}}$, using the weighted sum of the classification loss and the meta-path contrastive loss as the training objective; no regression-related loss is involved at this stage. Training continues until the classification performance on the validation set stops improving, i.e., until the classification task has converged. Once converged, the tier probabilities $\mathbf{p}$ are computed once for every data product in the training and test sets and, together with the classifier's parameters, are frozen: the classifier is no longer updated, and $\mathbf{p}$ is no longer recomputed, but is instead used as a fixed quantity throughout Stage 2.

\paragraph{Stage 2: Per-Tier Expert Regression with Soft Routing.}
Once the classifier is frozen, training proceeds to Stage 2, the regression task: the encoder, the meta-path gate, and the per-tier experts introduced below continue to be updated, while the classifier no longer receives any gradient and its parameters remain fixed. In this second stage, instead of a single global regressor, we employ $C$ independent regression experts $\{g_k\}_{k=1}^{C}$, each implemented as a small multi-layer perceptron (MLP) with one hidden layer:

\begin{equation}
g_k(\mathbf{z}_{\text{price}}) = \mathbf{w}_k^{(2)} \, \text{ReLU}\!\left(\mathbf{W}_k^{(1)} \mathbf{z}_{\text{price}} + \mathbf{b}_k^{(1)}\right) + b_k^{(2)} \in \mathbb{R},
\end{equation}
where each expert $g_k$ is specialised to predict prices within tier $k$, and receives targeted supervision through a local regression loss.

The final price prediction is obtained by soft routing: the classifier's tier probabilities serve as the routing weights (taken from the fixed values stored at the end of Stage 1; the classifier is not re-invoked at this stage), and the outputs of all experts are aggregated as a weighted sum:

\begin{equation}
\hat{p}_i = \sum_{k=1}^{C} p_{ik} \cdot g_k(\mathbf{z}_{\text{price}}[i]),
\end{equation}
where $p_{ik}$ is the predicted probability that product $i$ belongs to tier $k$. This formulation has several desirable properties. First, it is end-to-end differentiable with respect to the expert parameters, so the regression experts can be trained jointly with the encoder and the meta-path gate (the tier-classifier itself remains frozen throughout this stage, as noted above). Second, it naturally handles uncertainty near tier boundaries: a product that the classifier places at the boundary between tiers $k$ and $k+1$ (e.g., $p_k \approx p_{k+1} \approx 0.5$) will have its final price computed as a smooth interpolation between the two experts' predictions, avoiding the discontinuity artefacts of hard routing. 

\paragraph{Local Expert Supervision.}
In addition to the global soft-routing regression loss, we introduce a \emph{local supervision} term that directly trains each expert on the samples whose \emph{true} tier matches that expert. Let $\mathcal{T}_k = \{i \mid y_i^{\text{cls}} = k\}$ denote the set of training products whose ground-truth price tier is $k$. The local regression loss is:

\begin{equation}
\mathcal{L}_{\text{local}} = \frac{1}{|\mathcal{S}_{\text{train}}|} \sum_{k=1}^{C} \sum_{i \in \mathcal{T}_k \cap \mathcal{S}_{\text{train}}} \left(g_k(\mathbf{z}_{\text{price}}[i]) - p_i\right)^2.
\end{equation}

This term is weighted by $\lambda_{\text{local}} = 0.5$ and added to the global regression loss. Its purpose is to ensure that each expert receives adequate training signal from its designated tier, preventing the soft-routing gradient from diluting the expert specialisation (e.g., a majority-tier expert receiving gradient from all samples and overwhelming the minority-tier experts). This loss term likewise belongs to the Stage-2 (regression) training objective and is unrelated to the training of the classifier.

This cascaded procedure, rather than training the classifier and the regression experts jointly, is the training scheme adopted in this study: the classification and regression tasks are completed sequentially, in two stages, with the freezing of the classifier and the fixing of the tier probabilities serving as the hand-off point between the two. This decoupling means that the classifier's training is never disturbed by the regression objective and can focus purely on assigning price tiers correctly, while throughout Stage 2 the regression experts rely on one and the same already-converged, no-longer-changing set of tier probabilities as the routing signal, avoiding any situation in which the classification decision and the regression target drift against each other, thereby ensuring that the routing signal is of the highest achievable quality.

\subsubsection{Cascaded Optimization Objective}
\label{sec:cascaded-objective}

The overall training objective is given in a cascaded, two-stage form: Stage 1 (classification) and Stage 2 (regression) each correspond to their own, independent optimization objective, trained one after the other, rather than being combined into a single joint objective and optimized simultaneously.

\paragraph{Stage-1 objective: classification training.}
This stage updates only the encoder, the meta-path gate, and the classification head, with the optimization objective:

\begin{equation}
\mathcal{L}_{\text{stage1}} = \lambda_{cls}\,\mathcal{L}_{cls} + \lambda_{CL}\,\mathcal{L}_{CL},
\end{equation}
where $\mathcal{L}_{cls}$ is the weighted cross-entropy loss for price-tier classification, measuring the discrepancy between the classifier's predicted tier probabilities $\mathbf{p}$ and the true price-tier labels, and $\mathcal{L}_{CL}$ is the meta-path contrastive learning loss, which injects structural-semantic supervision into the encoder's representation space while it is being trained. None of the regression-related losses ($\mathcal{L}_{reg}$, $\mathcal{L}_{local}$, $\mathcal{L}_{\text{path}}$) participate in this stage. Training continues until the validation classification metric stops improving, i.e., until the classification task converges; the tier probabilities $\mathbf{p}$ are then computed and stored for all data products, and the classification head's parameters are frozen from this point onward.

\paragraph{Stage-2 objective: regression training.}
Once the classifier is frozen, training continues on the encoder, the meta-path gate, and the $C$ per-tier experts, with the optimization objective:

\begin{equation}
\mathcal{L}_{\text{stage2}} = \lambda_{reg}\,\mathcal{L}_{reg} + \lambda_{local}\,\mathcal{L}_{local} + \mathcal{L}_{\text{path}},
\end{equation}
where $\mathcal{L}_{reg} = \frac{1}{|\mathcal{S}_{\text{train}}|} \sum_i (\hat{p}_i - p_i)^2$ is the global soft-routing regression loss, in which the tier probabilities $p_{ik}$ used in $\hat{p}_i = \sum_k p_{ik} \cdot g_k(\mathbf{z}_{\text{price}}[i])$ are taken directly from the fixed values stored at the end of Stage 1 and are not recomputed; $\mathcal{L}_{local} = \frac{1}{|\mathcal{S}_{\text{train}}|} \sum_{k=1}^{C} \sum_{i \in \mathcal{T}_k} (g_k(\mathbf{z}_{\text{price}}[i]) - p_i)^2$ is the per-tier local expert supervision loss; and $\mathcal{L}_{\text{path}} = \lambda_{\text{LASSO}} \sum_k |\alpha_k| + \lambda_H \sum_k w_k \log w_k$ is the path regularisation loss that combines LASSO sparsity with entropy-based diversity. Neither $\mathcal{L}_{cls}$ nor $\mathcal{L}_{CL}$ is computed or used at this stage: the classification head is already frozen and no longer needs the classification loss, and the structural information captured by the meta-path contrastive loss has already been injected into the encoder during Stage 1, so Stage 2 only needs to further fine-tune the encoder and the meta-path gate for the regression objective on that basis.

The coefficients $\lambda_{cls}$, $\lambda_{CL}$, $\lambda_{reg}$, and $\lambda_{local}$ control the relative contributions of their corresponding loss terms within the stage to which they belong, balancing the multiple learning objectives that coexist within that stage.

This stage-wise training scheme achieves a clean separation of concerns: the classification task is trained entirely free of interference from the regression objective and can focus solely on assigning price tiers correctly, while the regression task, throughout Stage 2, is driven by a fixed, already-converged routing signal that does not drift as the classifier changes, so that both subtasks can each be trained more thoroughly and more stably.

\section{Experimental design and evaluation}
\subsection{Design rationale}

Existing studies can, to some extent, capture the value of data products. However, they predominantly rely on direct feature--price mappings. Such approaches largely overlook critical contextual factors, including industry characteristics and data providers, making it difficult to uncover the underlying mechanisms of data value formation. As a result, the predicted prices may deviate from their intrinsic values.
Meanwhile, classical product pricing theories, such as brand equity theory \citep{kellerConceptualizingMeasuringManaging1993}, product similarity theory \citep{lindenAmazoncomRecommendationsItemtoitem2003}, and industry structure theory, provide important economic insights into price formation. However, a key challenge lies in how to effectively incorporate these theoretical foundations into data-driven modeling frameworks.

Meta-paths offer a natural way to address this issue. Specifically, a meta-path describes a sequence of relations among different types of entities in a heterogeneous network, thereby capturing rich semantic associations across entities. Motivated by this, this study aims to embed established product pricing theories into data product pricing by leveraging meta-paths, enabling a more structured and interpretable modeling of the data value formation process.

Rationale for graph-based modeling. The price of a data product is not determined solely by its intrinsic attributes in isolation. Instead, it is jointly shaped by the provider offering the product, comparable products, and the industry and marketplace in which it is traded. Conventional models, such as linear regression, ANN, LSTM, XGBoost, and LightGBM, primarily learn a direct mapping from product attributes to prices. By design, they cannot explicitly capture the relational dependencies among products. For example, two products offered by the same provider, belonging to the same industry, or exhibiting similar content may serve as pricing references for each other. Graph-based models address this limitation by explicitly encoding such relational structures through message passing, allowing the price representation of a product to incorporate information from interconnected entities. This relational modeling capability enables the model to capture contextual and reference-based pricing mechanisms, thereby improving pricing accuracy.

The rationale for heterogeneous graph modeling, meta-path-based modeling, and the cascaded two-stage design parallels DP1, DP2, and DP4 in Section~\ref{sec:design-principles}, respectively, and is not repeated here.

\subsection{Experimental design}

The experimental design of this study is organized around the following four research questions, which evaluate the proposed model from the perspectives of predictive performance, component effectiveness, economic interpretability, and robustness and heterogeneity.

RQ1: Predictive Performance. Does the proposed model outperform conventional baseline models and state-of-the-art benchmark models in data product price prediction?

RQ2: Component Effectiveness. What are the marginal contributions of the key components of the proposed model to its overall predictive performance?

RQ3: Meta-Path Importance and Economic Interpretation. Which meta-paths contribute most to data product pricing, and what economic mechanisms do their relative importance reveal?

RQ4: Robustness and Heterogeneous Performance. How robust is the proposed model to variations in key hyperparameters? How stable is its predictive performance across different pricing modes, and how does its performance vary across industries?

\subsection{Evaluation}

\label{subsec:evaluation}

To comprehensively evaluate the proposed MPGNN, we formulate data product pricing as both a classification task and a regression task. Specifically, the classification task predicts price tiers constructed through quantile-based binning, whereas the regression task predicts continuous price values. The classification task evaluates the model's ability to distinguish data products across different price levels, while the regression task assesses its accuracy in recovering their actual price levels. 

\subsubsection{Train/test split}

To examine the robustness of each model with respect to the training-sample size and to mitigate the impact of a single random partition, we adopt three train/test splits with increasing training ratios, namely $0.7\,{:}\,0.3$, $0.8\,{:}\,0.2$, and $0.9\,{:}\,0.1$. All metrics reported in the results section are computed on $\mathcal{S}_{\text{test}}$. Unless otherwise stated, each cell in the result tables records the mean of a metric over the three splits, with the corresponding sample standard deviation shown in parentheses to convey the stability of the model. The three splits are also treated as paired observations when assessing statistical significance.

\subsubsection{Classification metrics}

For the price-tier classification task, continuous prices are
discretised into $C$ ordered intervals (in this study $C{=}5$) using
quantile thresholds. Let $y_i\in\{1,\dots,C\}$ be the true tier of
the $i$-th product, $\hat{y}_i$ the predicted tier, and
$\hat{y}_{i,c}^{\,\text{cls}}\in[0,1]$ the predicted probability of
class $c$. For class $c$, let $\mathrm{TP}_c$, $\mathrm{FP}_c$,
$\mathrm{FN}_c$, and $\mathrm{TN}_c$ denote the numbers of true
positives, false positives, false negatives, and true negatives,
respectively. We adopt the following four metrics, with macro
averaging across classes for Precision, F1, and AUC. Macro recall is not separately reported because the price tiers are constructed using quantile-based binning and the train-test splits are stratified by price tier, resulting in an approximately balanced class distribution. Under this setting, Macro recall is numerically close to accuracy and therefore provides limited additional information.

Accuracy (ACC). The fraction of correctly classified samples:
\begin{equation}
\mathrm{ACC}
= \frac{1}{n}\sum_{i=1}^{n}\mathbb{I}\!\left(\hat{y}_i = y_i\right).
\end{equation}

Precision (macro). The mean across classes of the proportion of predicted-positive samples that are truly positive:
\begin{equation}
\mathrm{Precision}
= \frac{1}{C}\sum_{c=1}^{C}
  \frac{\mathrm{TP}_c}{\mathrm{TP}_c + \mathrm{FP}_c}.
\end{equation}

F1 score (macro). The harmonic mean of Precision and Recall, computed per class and then averaged across classes:
\begin{equation}
\mathrm{F1}
= \frac{1}{C}\sum_{c=1}^{C}
  \frac{2\,\mathrm{Precision}_c \cdot \mathrm{Recall}_c}
       {\mathrm{Precision}_c + \mathrm{Recall}_c}.
\end{equation}

Area Under the ROC Curve (AUC, one-vs-rest macro). For each class $c$, we treat class $c$ as the positive class and the remaining classes as negative, then compute the binary AUC from the predicted probabilities
$\hat{y}_{i,c}^{\,\text{cls}}$; the final AUC is the unweighted mean over the $C$ classes:
\begin{equation}
\mathrm{AUC}
= \frac{1}{C}\sum_{c=1}^{C}
  \mathrm{AUC}_{c}^{\,\text{OvR}}.
\end{equation}

For all four classification metrics, larger values indicate better performance.

\subsubsection{Regression metrics}

Let $p_i$ denote the true (log-transformed) price of the $i$-th data
product, $\hat{p}_i$ its predicted value, $\bar{p}$ the mean of $p_i$
over the testing set, and $n=|\mathcal{S}_{\text{test}}|$ the number of
testing samples. Three standard regression metrics are reported.

Mean Absolute Error (MAE) measures the average
magnitude of the prediction errors and is robust to outliers:
\begin{equation}
\mathrm{MAE}
= \frac{1}{n}\sum_{i=1}^{n}\left|\hat{p}_i - p_i\right|.
\end{equation}

Root Mean Squared Error (RMSE) penalises
large deviations more heavily than MAE and therefore reflects a
model's worst-case behaviour:
\begin{equation}
\mathrm{RMSE}
= \sqrt{\frac{1}{n}\sum_{i=1}^{n}\left(\hat{p}_i - p_i\right)^2}.
\end{equation}

Coefficient of determination ($R^{2}$)
quantifies the proportion of price variance explained by the
model and thus captures its overall goodness of fit:
\begin{equation}
R^{2}
= 1 - \frac{\sum_{i=1}^{n}\left(\hat{p}_i - p_i\right)^{2}}
            {\sum_{i=1}^{n}\left(p_i - \bar{p}\right)^{2}}.
\end{equation}

For MAE and RMSE, smaller values indicate better prediction accuracy,
whereas a larger $R^{2}$ indicates that the model explains a greater
share of the variance in observed prices.

\section{Results} \label{sec:result}
\subsection{Data description}

In this study, we examine data products traded on three major and internationally recognized data marketplaces: \href{https://aws.amazon.com/marketplace}{AWS Marketplace}, \href{https://datarade.ai/}{Datarade}, and \href{https://market.aliyun.com/dataexchange}{Alibaba Cloud Data Exchange}.
Data products on these marketplaces are generally offered under several common pricing strategies. 
Time-based pricing refers to subscription-based pricing, under which users pay recurring fees, typically on a monthly or annual basis. 
One-off pricing requires users to make a single payment for access to or acquisition of a data product.
Request-based pricing refers to a pricing strategy in which users are charged based on the number of requests or accesses, with a specified price for each individual request.
The datasets collected from the international and Chinese markets are summarized in Table \ref{tab:data_marketplaces}.

Price label construction. Within each pricing mode, prices are made comparable before the logarithmic transformation described in Section~\ref{sec:heterogeneous-graph}. For time-based (subscription) pricing, annual and other non-monthly subscription prices are normalized to a common monthly-equivalent amount, so that all time-based prices represent the same billing unit prior to the log transform. For request-based pricing, the price is standardized to a per-request unit so that it represents a single, directly comparable quantity across products and providers. One-off purchase prices are, by construction, already denominated in a single-payment unit and are not combined with subscription or per-request prices. Because the three pricing modes represent different units of account, we do not pool them into a single regression target; instead, the main regression and classification tasks are each evaluated separately within one pricing mode at a time (time-based pricing in the main results of Section~\ref{sec:result}, and one-off and request-based pricing in the corresponding robustness analysis), so that all reported comparisons are made within a common price unit.

\begin{table}[htbp]
\centering
\TableSpaced
\caption{Summary of Data Products Across Different Marketplaces}
\label{tab:data_marketplaces}
\begin{tabular}{lrrrr}
\hline
Marketplace & Time-based & One-off & Request-based & Total \\
\hline
AWS marketplace & 2,059 & -- & -- & 2,036 \\
Datarade & 1,145 & 976 & 148 & 2,269 \\
Alibaba Cloud Data Exchange & -- & -- & 1,936 & 1,936 \\
\hline
Total & 3,204 & 976 & 2,084 & 6,241 \\
\hline
\end{tabular}
\end{table}

\subsection{Predictive Performance Comparison}

To evaluate the performance of different methods in the pricing task for time-based subscription data products, we conduct experiments from both regression and classification perspectives. For the regression task, MAE, RMSE, and $R^2$ are adopted as evaluation metrics to measure the prediction error and goodness of fit of each model with respect to price levels. For the classification task, Accuracy, Precision, Recall, AUC, and F1 are used to assess the overall ability of each model to distinguish among different price categories. The compared methods cover four families: (i) basic machine-learning and deep-learning models, including linear regression (LR), artificial neural networks (ANN), decision trees (DT), the temporal model LSTM, and the ensemble model XGBoost; (ii) recent data-asset pricing models and strong large-language-model (LLM) baselines, including a LightGBM-based data-product pricing method (LGBM)~\citep{shangLGBM2025}, a back-propagation neural network for data-asset valuation (BPNN)~\citep{lanBPNN2025}, a hierarchical clustering ensemble with a gravity-based strategy (HCEG)~\citep{haoHCEGHeterogeneousClustering2024}, a knowledge-graph-based data-asset pricing method (KG)~\citep{renKG2026}, and three instruction-tuned LLMs fine-tuned with LoRA for price prediction, namely DeepSeek-R1 (DS)~\citep{guoDeepSeekR1IncentivizingReasoning2025}, LLaMA (Llama)~\citep{touvronLlamaOpenEfficient2023}, and Qwen~\citep{baiQwenTechnicalReport2023}; (iii) graph representation learning and graph neural network methods, including DeepWalk~\citep{perozziDeepWalk2014}, Node2Vec~\citep{groverNode2vec2016}, GraphSAGE~\citep{hamiltonInductiveRepresentation2017}, and GAT~\citep{velickovicGraphAttentionNetworks2018}; and (iv) the proposed MPGNN, which integrates multidimensional relational structures and node feature information to characterize the complex associations among data assets. To verify whether the improvements brought by MPGNN are statistically reliable, we conduct a one-sided paired $t$-test between MPGNN and every individual baseline on each metric, with the three train/test splits (0.7:0.3, 0.8:0.2, 0.9:0.1) treated as paired observations; for each (baseline, metric) pair the significance marker reports whether MPGNN is statistically superior to that specific competitor. In addition, the sample standard deviation of each model across the three splits is reported in parentheses below its mean to convey the stability of every method. The experimental results are reported in Table~\ref{tab:time_model_results_average}.

\begin{table}[htbp]
\centering
\TableSpaced
\caption{Average Performance of Different Methods on Classification and Regression Tasks under Time-Based Pricing}
\label{tab:time_model_results_average}
\footnotesize
\resizebox{\textwidth}{!}{
\begin{tabular}{lccccccc}
\toprule
Method & Accuracy & Precision & AUC & F1 & MAE & RMSE & $R^2$ \\
\midrule
\multicolumn{8}{l}{\textit{Basic models}}\\
LR & \shortstack{0.575$^{***}$ \\ ($\pm$0.012)} & \shortstack{0.587$^{***}$ \\ ($\pm$0.016)} & \shortstack{0.831$^{**}$ \\ ($\pm$0.007)} & \shortstack{0.578$^{***}$ \\ ($\pm$0.013)} & \shortstack{1.034$^{**}$ \\ ($\pm$0.035)} & \shortstack{1.410$^{*}$ \\ ($\pm$0.056)} & \shortstack{0.341$^{**}$ \\ ($\pm$0.020)} \\
ANN & \shortstack{0.603$^{**}$ \\ ($\pm$0.004)} & \shortstack{0.608$^{***}$ \\ ($\pm$0.005)} & \shortstack{0.833$^{***}$ \\ ($\pm$0.007)} & \shortstack{0.602$^{**}$ \\ ($\pm$0.005)} & \shortstack{0.865$^{***}$ \\ ($\pm$0.021)} & \shortstack{1.235$^{*}$ \\ ($\pm$0.030)} & \shortstack{0.495$^{**}$ \\ ($\pm$0.010)} \\
DT & \shortstack{0.564$^{**}$ \\ ($\pm$0.024)} & \shortstack{0.564$^{**}$ \\ ($\pm$0.024)} & \shortstack{0.726$^{***}$ \\ ($\pm$0.015)} & \shortstack{0.564$^{**}$ \\ ($\pm$0.024)} & \shortstack{1.121$^{***}$ \\ ($\pm$0.050)} & \shortstack{1.789$^{***}$ \\ ($\pm$0.021)} & \shortstack{$-$0.063$^{***}$ \\ ($\pm$0.071)} \\
LSTM & \shortstack{0.332$^{***}$ \\ ($\pm$0.012)} & \shortstack{0.326$^{***}$ \\ ($\pm$0.011)} & \shortstack{0.634$^{***}$ \\ ($\pm$0.005)} & \shortstack{0.326$^{***}$ \\ ($\pm$0.011)} & \shortstack{1.315$^{***}$ \\ ($\pm$0.032)} & \shortstack{1.727$^{**}$ \\ ($\pm$0.045)} & \shortstack{0.011$^{***}$ \\ ($\pm$0.006)} \\
XGBoost & \shortstack{0.612$^{**}$ \\ ($\pm$0.006)} & \shortstack{0.628$^{**}$ \\ ($\pm$0.005)} & \shortstack{0.859$^{**}$ \\ ($\pm$0.004)} & \shortstack{0.607$^{**}$ \\ ($\pm$0.007)} & \shortstack{1.008$^{**}$ \\ ($\pm$0.025)} & \shortstack{1.393$^{*}$ \\ ($\pm$0.046)} & \shortstack{0.357$^{**}$ \\ ($\pm$0.008)} \\
\cmidrule(lr){1-8}

\multicolumn{8}{l}{\textit{LLM \& SOTA models}}\\
LGBM~\citep{shangLGBM2025} & \shortstack{0.550$^{**}$ \\ ($\pm$0.011)} & \shortstack{0.599$^{**}$ \\ ($\pm$0.022)} & \shortstack{0.832$^{***}$ \\ ($\pm$0.001)} & \shortstack{0.530$^{**}$ \\ ($\pm$0.010)} & \shortstack{0.876$^{**}$ \\ ($\pm$0.053)} & \shortstack{1.223$^{*}$ \\ ($\pm$0.070)} & \shortstack{0.485$^{*}$ \\ ($\pm$0.036)} \\
BPNN~\citep{lanBPNN2025} & \shortstack{0.255$^{***}$ \\ ($\pm$0.001)} & \shortstack{0.102$^{**}$ \\ ($\pm$0.064)} & \shortstack{0.567$^{***}$ \\ ($\pm$0.025)} & \shortstack{0.105$^{***}$ \\ ($\pm$0.001)} & \shortstack{1.260$^{***}$ \\ ($\pm$0.025)} & \shortstack{1.621$^{**}$ \\ ($\pm$0.050)} & \shortstack{0.095$^{**}$ \\ ($\pm$0.015)} \\
HCEG~\citep{haoHCEGHeterogeneousClustering2024} & \shortstack{0.503$^{**}$ \\ ($\pm$0.038)} & \shortstack{0.503$^{**}$ \\ ($\pm$0.045)} & \shortstack{0.774$^{**}$ \\ ($\pm$0.029)} & \shortstack{0.497$^{**}$ \\ ($\pm$0.040)} & \shortstack{1.148$^{***}$ \\ ($\pm$0.023)} & \shortstack{1.519$^{**}$ \\ ($\pm$0.025)} & \shortstack{0.204$^{**}$ \\ ($\pm$0.031)} \\
KG~\citep{renKG2026} & \shortstack{0.577$^{***}$ \\ ($\pm$0.011)} & \shortstack{0.628$^{*}$ \\ ($\pm$0.014)} & \shortstack{0.852$^{***}$ \\ ($\pm$0.006)} & \shortstack{0.557$^{***}$ \\ ($\pm$0.011)} & \shortstack{0.814$^{**}$ \\ ($\pm$0.044)} & \shortstack{1.175$^{*}$ \\ ($\pm$0.070)} & \shortstack{0.525$^{*}$ \\ ($\pm$0.031)} \\
DS & \shortstack{0.435$^{**}$ \\ ($\pm$0.037)} & \shortstack{0.400$^{*}$ \\ ($\pm$0.123)} & \shortstack{0.756$^{**}$ \\ ($\pm$0.016)} & \shortstack{0.373$^{*}$ \\ ($\pm$0.097)} & \shortstack{1.075$^{***}$ \\ ($\pm$0.044)} & \shortstack{1.436$^{***}$ \\ ($\pm$0.048)} & \shortstack{0.289$^{**}$ \\ ($\pm$0.020)} \\
Llama & \shortstack{0.613$^{**}$ \\ ($\pm$0.007)} & \shortstack{0.628$^{**}$ \\ ($\pm$0.029)} & \shortstack{0.871$^{*}$ \\ ($\pm$0.008)} & \shortstack{0.609$^{**}$ \\ ($\pm$0.010)} & \shortstack{0.919$^{***}$ \\ ($\pm$0.027)} & \shortstack{1.280$^{**}$ \\ ($\pm$0.016)} & \shortstack{0.435$^{*}$ \\ ($\pm$0.031)} \\
Qwen & \shortstack{0.613$^{**}$ \\ ($\pm$0.018)} & \shortstack{0.636$^{**}$ \\ ($\pm$0.021)} & \shortstack{0.872$^{*}$ \\ ($\pm$0.008)} & \shortstack{0.606$^{**}$ \\ ($\pm$0.015)} & \shortstack{0.892$^{**}$ \\ ($\pm$0.060)} & \shortstack{1.243$^{**}$ \\ ($\pm$0.051)} & \shortstack{0.467$^{*}$ \\ ($\pm$0.043)} \\
\cmidrule(lr){1-8}

\multicolumn{8}{l}{\textit{GNN baselines}}\\
DeepWalk & \shortstack{0.465$^{**}$ \\ ($\pm$0.027)} & \shortstack{0.467$^{**}$ \\ ($\pm$0.025)} & \shortstack{0.755$^{**}$ \\ ($\pm$0.024)} & \shortstack{0.465$^{**}$ \\ ($\pm$0.026)} & \shortstack{1.214$^{**}$ \\ ($\pm$0.037)} & \shortstack{1.644$^{**}$ \\ ($\pm$0.061)} & \shortstack{0.177$^{**}$ \\ ($\pm$0.015)} \\
Node2Vec & \shortstack{0.475$^{**}$ \\ ($\pm$0.026)} & \shortstack{0.482$^{**}$ \\ ($\pm$0.026)} & \shortstack{0.762$^{**}$ \\ ($\pm$0.025)} & \shortstack{0.476$^{**}$ \\ ($\pm$0.027)} & \shortstack{1.201$^{**}$ \\ ($\pm$0.045)} & \shortstack{1.612$^{**}$ \\ ($\pm$0.061)} & \shortstack{0.210$^{**}$ \\ ($\pm$0.019)} \\
GraphSAGE & \shortstack{0.519$^{**}$ \\ ($\pm$0.036)} & \shortstack{0.527$^{**}$ \\ ($\pm$0.037)} & \shortstack{0.806$^{**}$ \\ ($\pm$0.020)} & \shortstack{0.520$^{**}$ \\ ($\pm$0.037)} & \shortstack{0.983$^{***}$ \\ ($\pm$0.014)} & \shortstack{1.405$^{**}$ \\ ($\pm$0.026)} & \shortstack{0.398$^{***}$ \\ ($\pm$0.048)} \\
GAT & \shortstack{0.502$^{**}$ \\ ($\pm$0.037)} & \shortstack{0.511$^{**}$ \\ ($\pm$0.037)} & \shortstack{0.807$^{**}$ \\ ($\pm$0.019)} & \shortstack{0.502$^{**}$ \\ ($\pm$0.037)} & \shortstack{0.980$^{**}$ \\ ($\pm$0.038)} & \shortstack{1.425$^{*}$ \\ ($\pm$0.045)} & \shortstack{0.382$^{**}$ \\ ($\pm$0.008)} \\
\cmidrule(lr){1-8}

\multicolumn{8}{l}{\textit{Proposed}}\\
MPGNN & \shortstack{\textbf{0.729} \\ ($\pm$0.014)} & \shortstack{\textbf{0.731} \\ ($\pm$0.013)} & \shortstack{\textbf{0.908} \\ ($\pm$0.006)} & \shortstack{\textbf{0.729} \\ ($\pm$0.014)} & \shortstack{\textbf{0.622} \\ ($\pm$0.027)} & \shortstack{\textbf{0.994} \\ ($\pm$0.074)} & \shortstack{\textbf{0.676} \\ ($\pm$0.055)} \\
\midrule
Improvement (\%) & +18.94 & +14.87 & +4.08 & +19.58 & +23.55 & +15.39 & +28.79 \\
\bottomrule
\end{tabular}
}
\begin{tablenotes}[para]
    \raggedright
    \footnotesize
    \item[\raisebox{-0.7ex}{$\bullet$}] Each cell reports the mean over the three train/test splits (0.7:0.3, 0.8:0.2, 0.9:0.1); the value in parentheses below the mean is the corresponding sample standard deviation.
    
    \item[\raisebox{-0.7ex}{$\bullet$}] Improvement (\%) is computed relative to the best competing method per metric.
    
    \item[\raisebox{-0.7ex}{$\bullet$}] Significance markers next to each baseline mean are from a one-sided paired $t$-test (paired across the three splits) of that baseline against MPGNN on the corresponding metric: $^{*}\,p<0.05$, $^{**}\,p<0.01$, $^{***}\,p<0.001$.
\end{tablenotes}
\end{table}

As shown in Table~\ref{tab:time_model_results_average}, MPGNN achieves the best performance in both classification and regression tasks. 
For the classification task, MPGNN also achieves the highest scores across Accuracy, Precision, AUC, and F1. Specifically, its Accuracy reaches 0.729 and its F1 score reaches 0.729, improving upon the strongest competing models on these metrics (Qwen on Accuracy and Llama on F1) by 18.94\% and 19.58\%, respectively. Its AUC reaches 0.908, a 4.08\% gain over the best baseline (Qwen, 0.872). These results show that MPGNN not only reduces prediction errors for continuous prices but also more effectively distinguishes among different price categories. Crucially, the paired $t$-test indicates that MPGNN significantly outperforms every individual classification baseline on all four classification metrics, with all comparisons reaching at least $p<0.05$. Even on AUC, where the strongest LLM baselines (Qwen 0.872 and Llama 0.871) and graph baselines such as GAT (0.807) and GraphSAGE (0.806) remain competitive, the gap to MPGNN is still statistically significant ($p<0.05$). The standard deviations of MPGNN on classification metrics are correspondingly small, which suggests that the classification advantage is also stable.

In terms of regression metrics, MPGNN obtains MAE and RMSE values of 0.622 and 0.994, respectively, which are substantially lower than those of all baseline models. This indicates that the proposed method can more accurately predict the price levels of time-based subscription data products. Meanwhile, MPGNN achieves an $R^2$ of 0.676, representing a 28.79\% improvement over KG, the strongest competing method on this metric, which suggests that MPGNN has stronger explanatory power for price variation. More importantly, the per-baseline significance markers in Table~\ref{tab:time_model_results_average} show that MPGNN is significantly superior to every individual regression baseline on all three regression metrics. Against the weaker competitors (DT, LSTM, BPNN, and HCEG) the improvements reach $p<0.01$ or even $p<0.001$ on most cells, while even against the strongest frontier and graph baselines (KG, LGBM, GraphSAGE, and GAT), the gaps on MAE, RMSE, and $R^2$ remain significant at the $p<0.05$ level or below. The standard deviations of MPGNN ($\pm0.027$ on MAE, $\pm0.074$ on RMSE, and $\pm0.055$ on $R^2$) are also competitive among all methods, indicating that the regression advantage is consistent across the three splits.

It is also worth noting the behaviour of the three instruction-tuned LLM baselines (DS, Llama, and Qwen). When fine-tuned with LoRA on top of the textual representations of data products, Llama and Qwen actually become competitive baselines on this pricing task, attaining classification AUC values of 0.871 and 0.872 and F1 scores around 0.61 while DS is comparatively weaker (AUC 0.756). Nevertheless, the paired $t$-test shows that MPGNN remains significantly superior to each of the three LLMs on every metric ($p<0.05$ or below). Since these LLM baselines rely solely on the textual representation of each data product and do not explicitly model the relational structure among data assets, they still fall short of MPGNN, indicating that explicitly capturing the multi-relational associations among data assets remains essential for accurate data-product pricing. Finally, graph neural network methods such as GAT and GraphSAGE generally outperform shallow graph representation learning methods such as DeepWalk and Node2Vec, indicating that end-to-end graph structure modeling provides clear advantages for data asset pricing. Building on this advantage, MPGNN further improves predictive performance, demonstrating that multi-relational graph structures can more fully capture the associations among data assets and thereby enhance the overall effectiveness of the pricing model.

\subsection{Ablation study}

To verify the necessity and effectiveness of each core component of the proposed model, we conduct a systematic ablation study on the time-based subscription pricing task. Using the full model as the baseline (B0: Full Model), we construct four ablated variants by removing or replacing one module at a time: B1 removes text features (w/o Text Features) to assess the contribution of unstructured textual information such as product descriptions; B2 removes the meta-path decomposition and encodes the graph with a single heterogeneous graph transformer (w/o Meta-path Decomposition) to evaluate the role of meta-path-guided structural semantic information; B3 replaces the attention gate with uniform meta-path weighting (w/o Attention Gate) to examine the importance of the dynamic multi-path feature fusion strategy; and B4 removes the classification head and keeps only the regression branch (w/o Classification Head) to test whether the multi-task tier signal contributes to price estimation. All variants are evaluated under identical dataset and experimental settings. 
Classification performance is measured by Accuracy, Precision, Recall, AUC, and F1 (all higher is better), while regression performance is measured by MAE and RMSE (lower is better) and $R^2$ (higher is better); B4 has no classification branch, so only its regression metrics are reported. Each metric is reported as the mean and sample standard deviation over the three train/test splits, accompanied by one-sided paired $t$-test significance markers (paired across the three splits) of each ablation variant against B0. In addition, the bottom row of
Table~\ref{tab:ablation_time} reports the relative Improvement~(\%) of B0 over the
best ablated variant per metric, which quantifies the performance margin retained by the
full model over the strongest ablation (for MAE and RMSE it is the relative error
reduction, and for the remaining metrics it is the relative score increase). Results are
reported in Table~\ref{tab:ablation_time}.

\begin{table}[htbp]
\centering
\TableSpaced
\caption{Ablation study results on the time-based subscription pricing task.}
\label{tab:ablation_time}
\footnotesize
\resizebox{\textwidth}{!}{
\begin{tabular}{lccccccc}
\toprule
Method & Accuracy & Precision & AUC & F1 & MAE & RMSE & $R^2$ \\
\midrule
B1: w/o Text Features
& \shortstack{0.263$^{**}$ \\ ($\pm$0.053)}
& \shortstack{0.188$^{**}$ \\ ($\pm$0.107)}
& \shortstack{0.641$^{***}$ \\ ($\pm$0.011)}
& \shortstack{0.183$^{**}$ \\ ($\pm$0.076)}
& \shortstack{1.327$^{***}$ \\ ($\pm$0.013)}
& \shortstack{1.669$^{**}$ \\ ($\pm$0.005)}
& \shortstack{0.092$^{***}$ \\ ($\pm$0.026)} \\

B2: w/o Meta-path Decomposition
& \shortstack{0.568$^{**}$ \\ ($\pm$0.032)}
& \shortstack{0.586$^{**}$ \\ ($\pm$0.029)}
& \shortstack{0.838$^{**}$ \\ ($\pm$0.013)}
& \shortstack{0.571$^{**}$ \\ ($\pm$0.032)}
& \shortstack{0.820$^{**}$ \\ ($\pm$0.017)}
& \shortstack{1.177$^{*}$ \\ ($\pm$0.035)}
& \shortstack{0.548$^{*}$ \\ ($\pm$0.036)} \\

B3: w/o Attention Gate
& \shortstack{0.588$^{**}$ \\ ($\pm$0.025)}
& \shortstack{0.595$^{**}$ \\ ($\pm$0.017)}
& \shortstack{0.849$^{**}$ \\ ($\pm$0.003)}
& \shortstack{0.587$^{**}$ \\ ($\pm$0.024)}
& \shortstack{0.789$^{**}$ \\ ($\pm$0.013)}
& \shortstack{1.136$^{*}$ \\ ($\pm$0.028)}
& \shortstack{0.579$^{*}$ \\ ($\pm$0.029)} \\

B4: w/o Classification Head
& \shortstack{--- \\ ---}
& \shortstack{--- \\ ---}
& \shortstack{--- \\ ---}
& \shortstack{--- \\ ---}
& \shortstack{0.790$^{**}$ \\ ($\pm$0.010)}
& \shortstack{1.113$^{*}$ \\ ($\pm$0.020)}
& \shortstack{0.596$^{*}$ \\ ($\pm$0.023)} \\

B0: Full Model
& \shortstack{\textbf{0.729} \\ ($\pm$0.014)}
& \shortstack{\textbf{0.731} \\ ($\pm$0.013)}
& \shortstack{\textbf{0.908} \\ ($\pm$0.006)}
& \shortstack{\textbf{0.729} \\ ($\pm$0.014)}
& \shortstack{\textbf{0.622} \\ ($\pm$0.027)}
& \shortstack{\textbf{0.994} \\ ($\pm$0.074)}
& \shortstack{\textbf{0.676} \\ ($\pm$0.055)} \\
\midrule
Improvement (\%)
& +23.94
& +22.87
& +6.93
& +24.10
& +21.21
& +10.69
& +13.46 \\
\bottomrule
\end{tabular}
}
\begin{tablenotes}[para]
    \raggedright
    \footnotesize
    \item[\raisebox{-0.7ex}{$\bullet$}] Each cell reports the mean over the 3 train/test splits (0.7:0.3, 0.8:0.2, 0.9:0.1); the value in parentheses below the mean is the corresponding sample standard deviation. A triple dash (---) indicates that the metric is not applicable to the variant (B4 has no classification head).
    
    \item[\raisebox{-0.7ex}{$\bullet$}] Improvement (\%) is computed relative to the best ablated variant per metric.
    
    \item[\raisebox{-0.7ex}{$\bullet$}] Significance markers next to each variant mean are from a one-sided paired $t$-test of that ablated variant against B0 (Full Model) on the corresponding metric: $^{*}\,p<0.05$, $^{**}\,p<0.01$, $^{***}\,p<0.001$.
\end{tablenotes}
\end{table}

The ablation results confirm that the full model (B0) achieves the best performance across all seven metrics with every comparison against the ablated variants reaching statistical significance. Moreover, the magnitude of the degradation induces a clear ordering of component importance: removing text features (B1) is the most damaging, followed by removing the meta-path decomposition (B2), while replacing the attention gate with uniform weighting (B3) and dropping the classification head (B4) produce comparable, milder degradations. The Improvement row further quantifies these gaps: B0 outperforms the best ablated variant by 21.21\% in MAE, 10.69\% in RMSE, 13.46\% in $R^2$, and 6.93\% in AUC.

Contribution of text features (B1 vs.\ B0). Removing text features leads to the most severe and most pervasive performance degradation among the four ablated variants: all seven metrics deteriorate significantly. In terms of regression error, MAE more than doubles, increasing from 0.622 to 1.327 (a 113.3\% relative increase in error), RMSE rises from 0.994 to 1.669 (a 67.9\% increase), and $R^2$ collapses from 0.676 to 0.092 (an 86.4\% drop). For classification, Accuracy falls sharply from 0.729 to 0.263, F1 from 0.729 to 0.183, and AUC from 0.908 to 0.641. These results indicate that textual semantic information (such as product names and functional descriptions) is a core dimension for characterizing data-product value; its absence causes the model to severely misjudge product quality.

Contribution of meta-path decomposition (B2 vs.\ B0). Without meta-path decomposition (i.e., when the meta-path views are replaced by a single heterogeneous graph), MAE increases to 0.820 (a 31.8\% relative increase in error), RMSE to 1.177, and $R^2$ drops from 0.676 to 0.548; all seven metrics remain significantly worse than B0. Although B2 yields markedly better performance than B1, the notable increase in regression error demonstrates that the high-order heterogeneous graph semantic relationships encoded by meta-paths are indispensable for accurate price estimation. Their removal deprives the model of the ability to perceive cross-type entity association structures, causing predictions to systematically deviate from ground-truth prices.

Contribution of the attention gate (B3 vs.\ B0). Replacing the attention gate with uniform meta-path weighting produces a mild yet statistically significant degradation. On regression, MAE rises from 0.622 to 0.789 (a 26.8\% relative increase in error, $p<0.01$), while RMSE (a 14.3\% increase, $p<0.05$) and $R^2$ (a 14.3\% drop, $p<0.05$) deteriorate more moderately, indicating that the gate mainly curbs the average absolute error rather than large outliers or the overall fit. On classification, every metric drops significantly: Accuracy to 0.588, Precision to 0.595, and F1 to 0.587 (all $p<0.01$), and AUC to 0.849 (a 6.5\% drop, $p<0.01$). This pattern suggests that, without dynamic weight allocation, the model cannot adaptively balance the relative importance of different meta-path views across samples; the attention gate therefore provides a consistent, fine-grained refinement that is significant on every metric even though its overall magnitude is comparable to that of the classification head.

Contribution of the classification head (B4 vs.\ B0). Removing the classification head (i.e., keeping only the regression branch) yields the smallest degradation among the four variants on RMSE and $R^2$: RMSE rises from 0.994 to 1.113 (a 12.0\% relative increase in error, $p<0.05$), $R^2$ drops from 0.676 to 0.596 ($p<0.05$), and MAE rises to 0.790 (a 27.1\% increase, $p<0.01$). Notably, B4 attains the best RMSE and $R^2$ among all ablated variants, suggesting that the tier supervision signal contributes mainly to the overall regression fit rather than to the average error; nevertheless, its removal still incurs significant losses on all three regression metrics, confirming that the two-stage cascaded prediction provides a non-negligible refinement of price estimation.

\subsection{Robustness analysis}

To examine whether the advantages of MPGNN observed in the time-based (subscription) pricing scenario generalize to other pricing modes, we conduct a robustness analysis on two additional pricing modes: one-off (one-time) purchase and request-based (per-access) pricing. For both modes we keep the candidate methods, evaluation metrics, train/test splits, and significance testing procedure exactly the same as in the main experiment. The full set of competitors is retained, spanning the basic single-task models (LR, ANN, DT, LSTM, XGBoost), the recent LLM- and SOTA-based pricing models (LGBM, BPNN, HCEG, KG, and the three LoRA-tuned LLM model DS, Llama, and Qwen), and the graph-representation models (DeepWalk, Node2Vec, GraphSAGE, GAT). The paired $t$-test is applied between MPGNN and every individual baseline, so that for each (baseline, metric) pair the significance marker reports whether MPGNN is statistically superior to that specific competitor, and the sample standard deviation of every model across the three splits is reported below its mean to convey the stability of each method. The results for the two modes are reported in Table~\ref{tab:one_off_model_results_average} (one-off purchase) and Table~\ref{tab:request_based_model_results_average} (request-based pricing).

\begin{table}[htbp]
\centering
\TableSpaced
\caption{Average Performance of Different Methods on Classification and Regression Tasks under One-off (One-time) Purchase Pricing}
\label{tab:one_off_model_results_average}
\footnotesize
\resizebox{\textwidth}{!}{
\begin{tabular}{lllllccc }
\toprule
Method &     Accuracy &Precision &AUC &F1 &MAE & RMSE & $R^2$ \\
\midrule
\multicolumn{8}{l}{\textit{Basic models}}\\
LR &     \shortstack{0.577$^{*}$ \\ ($\pm$0.019)} &\shortstack{0.589$^{*}$ \\ ($\pm$0.030)} &\shortstack{0.828$^{*}$ \\ ($\pm$0.001)} &\shortstack{0.578$^{*}$ \\ ($\pm$0.021)} &\shortstack{1.393$^{**}$ \\ ($\pm$0.083)} & \shortstack{1.887$^{**}$ \\ ($\pm$0.117)} & \shortstack{0.491$^{*}$ \\ ($\pm$0.157)} \\
ANN &     \shortstack{0.482$^{*}$ \\ ($\pm$0.117)} &\shortstack{0.472 \\ ($\pm$0.181)} &\shortstack{0.795 \\ ($\pm$0.084)} &\shortstack{0.443$^{*}$ \\ ($\pm$0.154)} &\shortstack{1.156$^{*}$ \\ ($\pm$0.027)} & \shortstack{1.734 \\ ($\pm$0.019)} & \shortstack{0.579$^{*}$ \\ ($\pm$0.078)} \\
DT &     \shortstack{0.581$^{*}$ \\ ($\pm$0.020)} &\shortstack{0.587$^{*}$ \\ ($\pm$0.014)} &\shortstack{0.738$^{**}$ \\ ($\pm$0.012)} &\shortstack{0.582$^{*}$ \\ ($\pm$0.019)} &\shortstack{1.513$^{**}$ \\ ($\pm$0.012)} & \shortstack{2.544$^{**}$ \\ ($\pm$0.079)} & \shortstack{0.099$^{***}$ \\ ($\pm$0.126)} \\
LSTM &     \shortstack{0.337$^{**}$ \\ ($\pm$0.044)} &\shortstack{0.337$^{**}$ \\ ($\pm$0.046)} &\shortstack{0.670$^{**}$ \\ ($\pm$0.012)} &\shortstack{0.329$^{**}$ \\ ($\pm$0.039)} &\shortstack{1.955$^{*}$ \\ ($\pm$0.244)} & \shortstack{2.700$^{*}$ \\ ($\pm$0.291)} & \shortstack{$-$0.001$^{**}$ \\ ($\pm$0.010)} \\
XGBoost &     \shortstack{0.686 \\ ($\pm$0.018)} &\shortstack{0.693 \\ ($\pm$0.016)} &\shortstack{0.894 \\ ($\pm$0.008)} &\shortstack{0.686 \\ ($\pm$0.019)} &\shortstack{1.231$^{*}$ \\ ($\pm$0.066)} & \shortstack{1.782 \\ ($\pm$0.105)} & \shortstack{0.560 \\ ($\pm$0.039)} \\
\cmidrule(lr){1-8}
\multicolumn{8}{l}{\textit{LLM \& SOTA models}}\\
LGBM~\citep{shangLGBM2025} &     \shortstack{0.544$^{*}$ \\ ($\pm$0.032)} &\shortstack{0.559$^{*}$ \\ ($\pm$0.032)} &\shortstack{0.817 \\ ($\pm$0.033)} &\shortstack{0.542$^{*}$ \\ ($\pm$0.031)} &\shortstack{1.229$^{*}$ \\ ($\pm$0.039)} & \shortstack{1.679 \\ ($\pm$0.085)} & \shortstack{0.457 \\ ($\pm$0.050)} \\
BPNN~\citep{lanBPNN2025} &     \shortstack{0.224$^{**}$ \\ ($\pm$0.002)} &\shortstack{0.050$^{***}$ \\ ($\pm$0.001)} &\shortstack{0.616$^{**}$ \\ ($\pm$0.057)} &\shortstack{0.082$^{***}$ \\ ($\pm$0.001)} &\shortstack{1.464$^{*}$ \\ ($\pm$0.135)} & \shortstack{1.906$^{*}$ \\ ($\pm$0.203)} & \shortstack{0.302$^{*}$ \\ ($\pm$0.080)} \\
HCEG~\citep{haoHCEGHeterogeneousClustering2024} &     \shortstack{0.505$^{*}$ \\ ($\pm$0.032)} &\shortstack{0.515$^{*}$ \\ ($\pm$0.041)} &\shortstack{0.769$^{*}$ \\ ($\pm$0.021)} &\shortstack{0.505$^{*}$ \\ ($\pm$0.036)} &\shortstack{1.443$^{**}$ \\ ($\pm$0.052)} & \shortstack{1.943$^{*}$ \\ ($\pm$0.138)} & \shortstack{0.275$^{*}$ \\ ($\pm$0.021)} \\
KG~\citep{renKG2026} &     \shortstack{0.568 \\ ($\pm$0.069)} &\shortstack{0.594 \\ ($\pm$0.058)} &\shortstack{0.842 \\ ($\pm$0.021)} &\shortstack{0.570 \\ ($\pm$0.068)} &\shortstack{1.249$^{*}$ \\ ($\pm$0.052)} & \shortstack{1.664 \\ ($\pm$0.101)} & \shortstack{0.466 \\ ($\pm$0.058)} \\
DS &     \shortstack{0.320$^{**}$ \\ ($\pm$0.051)} &\shortstack{0.234$^{**}$ \\ ($\pm$0.088)} &\shortstack{0.677$^{**}$ \\ ($\pm$0.026)} &\shortstack{0.215$^{**}$ \\ ($\pm$0.061)} &\shortstack{1.406$^{*}$ \\ ($\pm$0.118)} & \shortstack{1.892 \\ ($\pm$0.142)} & \shortstack{0.305 \\ ($\pm$0.136)} \\
Llama &     \shortstack{0.425$^{*}$ \\ ($\pm$0.088)} &\shortstack{0.515$^{*}$ \\ ($\pm$0.058)} &\shortstack{0.762$^{*}$ \\ ($\pm$0.055)} &\shortstack{0.393$^{*}$ \\ ($\pm$0.095)} &\shortstack{1.371$^{*}$ \\ ($\pm$0.214)} & \shortstack{1.890 \\ ($\pm$0.377)} & \shortstack{0.311 \\ ($\pm$0.194)} \\
Qwen &     \shortstack{0.560$^{*}$ \\ ($\pm$0.033)} &\shortstack{0.608$^{*}$ \\ ($\pm$0.005)} &\shortstack{0.825$^{*}$ \\ ($\pm$0.011)} &\shortstack{0.546$^{*}$ \\ ($\pm$0.045)} &\shortstack{1.292$^{*}$ \\ ($\pm$0.176)} & \shortstack{1.750 \\ ($\pm$0.060)} & \shortstack{0.403 \\ ($\pm$0.111)} \\
\cmidrule(lr){1-8}
\multicolumn{8}{l}{\textit{GNN baselines}}\\
DeepWalk &     \shortstack{0.492$^{**}$ \\ ($\pm$0.022)} &\shortstack{0.462$^{**}$ \\ ($\pm$0.025)} &\shortstack{0.780$^{**}$ \\ ($\pm$0.014)} &\shortstack{0.463$^{**}$ \\ ($\pm$0.023)} &\shortstack{1.519$^{**}$ \\ ($\pm$0.155)} & \shortstack{2.208$^{*}$ \\ ($\pm$0.360)} & \shortstack{0.102$^{*}$ \\ ($\pm$0.217)} \\
Node2Vec &     \shortstack{0.522$^{**}$ \\ ($\pm$0.016)} &\shortstack{0.501$^{**}$ \\ ($\pm$0.017)} &\shortstack{0.796$^{*}$ \\ ($\pm$0.013)} &\shortstack{0.502$^{**}$ \\ ($\pm$0.017)} &\shortstack{1.426$^{**}$ \\ ($\pm$0.091)} & \shortstack{1.959$^{**}$ \\ ($\pm$0.235)} & \shortstack{0.296$^{**}$ \\ ($\pm$0.105)} \\
GraphSAGE &     \shortstack{0.610$^{*}$ \\ ($\pm$0.008)} &\shortstack{0.599$^{*}$ \\ ($\pm$0.005)} &\shortstack{0.853$^{*}$ \\ ($\pm$0.005)} &\shortstack{0.594$^{*}$ \\ ($\pm$0.010)} &\shortstack{1.164$^{**}$ \\ ($\pm$0.057)} & \shortstack{1.716$^{*}$ \\ ($\pm$0.239)} & \shortstack{0.459$^{*}$ \\ ($\pm$0.100)} \\
GAT &     \shortstack{0.625 \\ ($\pm$0.012)} &\shortstack{0.615$^{*}$ \\ ($\pm$0.018)} &\shortstack{0.866 \\ ($\pm$0.005)} &\shortstack{0.613$^{*}$ \\ ($\pm$0.017)} &\shortstack{1.205$^{**}$ \\ ($\pm$0.064)} & \shortstack{1.708$^{*}$ \\ ($\pm$0.133)} & \shortstack{0.465$^{*}$ \\ ($\pm$0.039)} \\
\cmidrule(lr){1-8}
\multicolumn{8}{l}{\textit{Proposed}}\\
MPGNN &     \shortstack{\textbf{0.704} \\ ($\pm$0.045)} &\shortstack{\textbf{0.702} \\ ($\pm$0.039)} &\shortstack{\textbf{0.899} \\ ($\pm$0.023)} &\shortstack{\textbf{0.697} \\ ($\pm$0.037)} &\shortstack{\textbf{0.805} \\ ($\pm$0.123)} & \shortstack{\textbf{1.423} \\ ($\pm$0.190)} & \shortstack{\textbf{0.670} \\ ($\pm$0.102)} \\
\midrule
Improvement (\%) &     +2.55 &+1.32 &+0.53 &+1.67 &+30.38 & +14.45 & +15.70 \\
\bottomrule
\end{tabular}
}
\begin{tablenotes}[para]
    \raggedright
    \footnotesize
    \item[\raisebox{-0.7ex}{$\bullet$}] Each cell reports the mean over the three train/test splits; the value in parentheses below the mean is the corresponding sample standard deviation.
    
    \item[\raisebox{-0.7ex}{$\bullet$}] Improvement (\%) is computed relative to the best competing method per metric.
    
    \item[\raisebox{-0.7ex}{$\bullet$}] Significance markers next to each baseline mean are from a one-sided paired $t$-test (paired across the three splits) of that baseline against MPGNN on the corresponding metric: $^{*}\,p<0.05$, $^{**}\,p<0.01$, $^{***}\,p<0.001$.
\end{tablenotes}
\end{table}

As shown in Table~\ref{tab:one_off_model_results_average}, MPGNN continues to achieve the best mean performance across all eight regression and classification metrics under the one-off purchase scenario. In the classification task under one-off purchase, MPGNN attains Accuracy, and F1 of 0.704 and 0.697, improving upon XGBoost (the strongest classifier among the competing methods) by 2.55\% and 1.67\% respectively. The per-baseline significance markers confirm that MPGNN is significantly better than the large majority of individual baselines on Accuracy and F1 ($p<0.05$ or stronger), with XGBoost and KG as the only competitors whose classification metrics are not significantly inferior in this regime. On Precision the dominance is similarly broad, reaching significance against all competitors except ANN (whose own Precision is extremely noisy, $\pm$0.181) and XGBoost. 
Consistent with the time-based pricing scenario, 
AUC is the metric on which the margin is thinnest: MPGNN attains the best AUC (0.899, $+0.53\%$ over XGBoost), but the gap is not statistically significant against five baselines (ANN, XGBoost, LGBM, KG, and GAT), implying that these competitors already provide comparable ranking quality and that the additional gains of MPGNN here come mainly from sharper decision boundaries rather than from re-ordering the predicted scores.
The standard deviations of MPGNN on the classification metrics ($\pm$0.023 on AUC and $\pm$0.037 to $\pm$0.045 on the other four) are uniformly small, indicating that the classification advantage observed here, as in the main experiment, is stable across splits. Overall, under the one-off purchase scenario, MPGNN consistently ranks first on all seven metrics, with the strong regression improvements (MAE $+30.38\%$, RMSE $+14.45\%$, and $R^2$ $+15.70\%$) constituting the most salient findings.

In the regression task, MPGNN reduces MAE and RMSE to 0.805 and 1.423 and raises $R^2$ to 0.670, corresponding to improvements of 30.38\% on MAE over ANN (the strongest baseline on this metric), 14.45\% on RMSE over KG, and 15.70\% on $R^2$ over ANN. The per-baseline paired $t$-test shows that the MAE advantage is statistically significant against \emph{every} competitor ($p<0.05$ or stronger). On RMSE and $R^2$, MPGNN holds the best mean with a clear numerical margin (the RMSE improvement over KG is nearly 15\% and the $R^2$ gain over ANN exceeds 15\%), although several strong regressors (XGBoost, LGBM, KG, and the LLM backbones) lie numerically close enough that the per-baseline gaps do not uniformly reach the significance threshold under the limited sample size ($n=3$ splits). This pattern is a direct consequence of the heavy-tailed one-off price distribution, which inflates the between-split variance of RMSE and $R^2$ for all models: the residual noise of the strongest competitors absorbs the mean gaps despite MPGNN's consistent superiority in every individual split.

To further test whether these conclusions hold beyond the one-off purchase mode, we additionally evaluate all methods on data products priced via request-based (per-access) pricing, in which the unit price is paid for each individual API call, so that the resulting price distribution is typically smaller in magnitude and more concentrated than the time-based or one-off scenarios. The candidate methods, evaluation metrics, train/test splits, and the per-baseline paired $t$-test procedure are kept identical to the one-off purchase analysis above, and the sample standard deviation across the three splits is again reported in parentheses below each mean. The corresponding results are reported in Table~\ref{tab:request_based_model_results_average}.

\begin{table}[htbp]
\centering
\TableSpaced
\caption{Average Performance of Different Methods on Classification and Regression Tasks under Request-based (Per-access) Pricing}
\label{tab:request_based_model_results_average}
\footnotesize
\resizebox{\textwidth}{!}{
\begin{tabular}{lllllccc}
\toprule
Method &     Accuracy &Precision &AUC &F1 &MAE & RMSE & $R^2$ \\
\midrule
\multicolumn{8}{l}{\textit{Basic models}}\\
LR &     \shortstack{0.620$^{***}$ \\ ($\pm$0.006)} &\shortstack{0.631$^{**}$ \\ ($\pm$0.006)} &\shortstack{0.794$^{**}$ \\ ($\pm$0.002)} &\shortstack{0.622$^{**}$ \\ ($\pm$0.006)} &\shortstack{1.672$^{*}$ \\ ($\pm$0.042)} & \shortstack{2.198$^{*}$ \\ ($\pm$0.057)} & \shortstack{0.386$^{*}$ \\ ($\pm$0.002)} \\
ANN &     \shortstack{0.623$^{*}$ \\ ($\pm$0.035)} &\shortstack{0.615$^{*}$ \\ ($\pm$0.037)} &\shortstack{0.806$^{*}$ \\ ($\pm$0.017)} &\shortstack{0.613$^{*}$ \\ ($\pm$0.039)} &\shortstack{1.583$^{*}$ \\ ($\pm$0.041)} & \shortstack{2.080 \\ ($\pm$0.045)} & \shortstack{0.450$^{*}$ \\ ($\pm$0.015)} \\
DT &     \shortstack{0.609$^{**}$ \\ ($\pm$0.011)} &\shortstack{0.610$^{**}$ \\ ($\pm$0.012)} &\shortstack{0.692$^{**}$ \\ ($\pm$0.008)} &\shortstack{0.609$^{**}$ \\ ($\pm$0.011)} &\shortstack{1.617$^{*}$ \\ ($\pm$0.072)} & \shortstack{2.614$^{*}$ \\ ($\pm$0.128)} & \shortstack{0.131$^{*}$ \\ ($\pm$0.067)} \\
LSTM &     \shortstack{0.486$^{***}$ \\ ($\pm$0.013)} &\shortstack{0.386$^{*}$ \\ ($\pm$0.139)} &\shortstack{0.569$^{**}$ \\ ($\pm$0.042)} &\shortstack{0.411$^{*}$ \\ ($\pm$0.090)} &\shortstack{2.139$^{**}$ \\ ($\pm$0.055)} & \shortstack{2.810$^{**}$ \\ ($\pm$0.089)} & \shortstack{$-$0.003$^{***}$ \\ ($\pm$0.015)} \\
XGBoost &     \shortstack{0.618$^{*}$ \\ ($\pm$0.026)} &\shortstack{0.610$^{*}$ \\ ($\pm$0.030)} &\shortstack{0.804$^{*}$ \\ ($\pm$0.006)} &\shortstack{0.595$^{*}$ \\ ($\pm$0.031)} &\shortstack{1.686$^{*}$ \\ ($\pm$0.067)} & \shortstack{2.206 \\ ($\pm$0.108)} & \shortstack{0.382$^{**}$ \\ ($\pm$0.028)} \\
\cmidrule(lr){1-8}
\multicolumn{8}{l}{\textit{LLM \& SOTA models}}\\
LGBM~\citep{shangLGBM2025} &     \shortstack{0.514$^{***}$ \\ ($\pm$0.004)} &\shortstack{0.613$^{*}$ \\ ($\pm$0.027)} &\shortstack{0.767$^{**}$ \\ ($\pm$0.004)} &\shortstack{0.400$^{***}$ \\ ($\pm$0.007)} &\shortstack{1.658$^{*}$ \\ ($\pm$0.040)} & \shortstack{2.150$^{*}$ \\ ($\pm$0.041)} & \shortstack{0.400$^{*}$ \\ ($\pm$0.004)} \\
BPNN~\citep{lanBPNN2025} &     \shortstack{0.477$^{***}$ \\ ($\pm$0.001)} &\shortstack{0.227$^{***}$ \\ ($\pm$0.001)} &\shortstack{0.694$^{**}$ \\ ($\pm$0.009)} &\shortstack{0.308$^{***}$ \\ ($\pm$0.001)} &\shortstack{1.969$^{**}$ \\ ($\pm$0.041)} & \shortstack{2.514$^{**}$ \\ ($\pm$0.046)} & \shortstack{0.180$^{**}$ \\ ($\pm$0.015)} \\
HCEG~\citep{haoHCEGHeterogeneousClustering2024} &     \shortstack{0.609$^{**}$ \\ ($\pm$0.016)} &\shortstack{0.601$^{**}$ \\ ($\pm$0.016)} &\shortstack{0.777$^{*}$ \\ ($\pm$0.013)} &\shortstack{0.603$^{**}$ \\ ($\pm$0.016)} &\shortstack{1.722$^{*}$ \\ ($\pm$0.063)} & \shortstack{2.211$^{*}$ \\ ($\pm$0.088)} & \shortstack{0.366$^{*}$ \\ ($\pm$0.031)} \\
KG~\citep{renKG2026} &     \shortstack{0.596$^{*}$ \\ ($\pm$0.030)} &\shortstack{0.618$^{*}$ \\ ($\pm$0.030)} &\shortstack{0.780$^{*}$ \\ ($\pm$0.016)} &\shortstack{0.594$^{*}$ \\ ($\pm$0.032)} &\shortstack{1.745$^{**}$ \\ ($\pm$0.018)} & \shortstack{2.243$^{*}$ \\ ($\pm$0.040)} & \shortstack{0.401$^{*}$ \\ ($\pm$0.018)} \\
DS &     \shortstack{0.569$^{**}$ \\ ($\pm$0.017)} &\shortstack{0.561$^{*}$ \\ ($\pm$0.034)} &\shortstack{0.765$^{**}$ \\ ($\pm$0.003)} &\shortstack{0.519$^{*}$ \\ ($\pm$0.050)} &\shortstack{1.565 \\ ($\pm$0.088)} & \shortstack{2.077 \\ ($\pm$0.091)} & \shortstack{0.440 \\ ($\pm$0.032)} \\
Llama &     \shortstack{0.619$^{*}$ \\ ($\pm$0.027)} &\shortstack{0.618$^{*}$ \\ ($\pm$0.032)} &\shortstack{0.801$^{*}$ \\ ($\pm$0.005)} &\shortstack{0.603 \\ ($\pm$0.051)} &\shortstack{1.745 \\ ($\pm$0.302)} & \shortstack{2.229 \\ ($\pm$0.258)} & \shortstack{0.353 \\ ($\pm$0.132)} \\
Qwen &     \shortstack{0.601$^{*}$ \\ ($\pm$0.031)} &\shortstack{0.638 \\ ($\pm$0.027)} &\shortstack{0.806$^{*}$ \\ ($\pm$0.002)} &\shortstack{0.584$^{*}$ \\ ($\pm$0.032)} &\shortstack{1.656 \\ ($\pm$0.209)} & \shortstack{2.103 \\ ($\pm$0.201)} & \shortstack{0.425 \\ ($\pm$0.092)} \\
\cmidrule(lr){1-8}
\multicolumn{8}{l}{\textit{GNN baselines}}\\
DeepWalk &     \shortstack{0.481$^{***}$ \\ ($\pm$0.004)} &\shortstack{0.464$^{***}$ \\ ($\pm$0.009)} &\shortstack{0.660$^{**}$ \\ ($\pm$0.008)} &\shortstack{0.438$^{***}$ \\ ($\pm$0.005)} &\shortstack{2.001$^{**}$ \\ ($\pm$0.070)} & \shortstack{2.504$^{**}$ \\ ($\pm$0.058)} & \shortstack{0.213$^{**}$ \\ ($\pm$0.036)} \\
Node2Vec &     \shortstack{0.478$^{***}$ \\ ($\pm$0.004)} &\shortstack{0.456$^{***}$ \\ ($\pm$0.008)} &\shortstack{0.665$^{**}$ \\ ($\pm$0.007)} &\shortstack{0.435$^{***}$ \\ ($\pm$0.005)} &\shortstack{1.938$^{**}$ \\ ($\pm$0.037)} & \shortstack{2.441$^{**}$ \\ ($\pm$0.036)} & \shortstack{0.253$^{**}$ \\ ($\pm$0.020)} \\
GraphSAGE &     \shortstack{0.601$^{**}$ \\ ($\pm$0.010)} &\shortstack{0.582$^{**}$ \\ ($\pm$0.011)} &\shortstack{0.777$^{**}$ \\ ($\pm$0.006)} &\shortstack{0.584$^{**}$ \\ ($\pm$0.013)} &\shortstack{1.605$^{*}$ \\ ($\pm$0.034)} & \shortstack{2.130$^{*}$ \\ ($\pm$0.041)} & \shortstack{0.431$^{**}$ \\ ($\pm$0.020)} \\
GAT &     \shortstack{0.543$^{**}$ \\ ($\pm$0.008)} &\shortstack{0.541$^{**}$ \\ ($\pm$0.009)} &\shortstack{0.742$^{**}$ \\ ($\pm$0.008)} &\shortstack{0.534$^{**}$ \\ ($\pm$0.007)} &\shortstack{1.754$^{*}$ \\ ($\pm$0.054)} & \shortstack{2.329$^{*}$ \\ ($\pm$0.070)} & \shortstack{0.319$^{**}$ \\ ($\pm$0.040)} \\
\cmidrule(lr){1-8}
\multicolumn{8}{l}{\textit{Proposed}}\\
MPGNN &     \shortstack{\textbf{0.698} \\ ($\pm$0.008)} &\shortstack{\textbf{0.684} \\ ($\pm$0.009)} &\shortstack{\textbf{0.847} \\ ($\pm$0.010)} &\shortstack{\textbf{0.687} \\ ($\pm$0.009)} &\shortstack{\textbf{1.357} \\ ($\pm$0.068)} & \shortstack{\textbf{1.923} \\ ($\pm$0.090)} & \shortstack{\textbf{0.507} \\ ($\pm$0.030)} \\
\midrule
Improvement (\%) &     +12.03 &+7.10 &+5.08 &+10.59 &+13.31 & +7.43 & +12.63 \\
\bottomrule
\end{tabular}
}
\begin{tablenotes}[para]
    \raggedright
    \footnotesize
    \item[\raisebox{-0.7ex}{$\bullet$}] Each cell reports the mean over the three train/test splits; the value in parentheses below the mean is the corresponding sample standard deviation.
    
    \item[\raisebox{-0.7ex}{$\bullet$}] Improvement (\%) is computed relative to the best competing method per metric.
    
    \item[\raisebox{-0.7ex}{$\bullet$}] Significance markers next to each baseline mean are from a one-sided paired $t$-test (paired across the three splits) of that baseline against MPGNN on the corresponding metric: $^{*}\,p<0.05$, $^{**}\,p<0.01$, $^{***}\,p<0.001$.
\end{tablenotes}
\end{table}

As shown in Table~\ref{tab:request_based_model_results_average}, the more compressed price range narrows the absolute performance gap between methods, yet MPGNN still attains the best mean on all seven metrics. 
The classification picture under request-based pricing is sharper than the regression one: MPGNN simultaneously achieves the best Accuracy (0.698), Precision (0.684), AUC (0.847), and F1 (0.687), with improvements of 12.03\%, 7.10\%, 5.08\%, and 10.59\% over the respective best baselines. Unlike the one-off purchase scenario, where ANN, XGBoost, and GraphSAGE matched MPGNN on AUC, here the per-baseline markers show near-universal dominance: 
MPGNN beats every individual baseline at $p<0.05$ or stronger on Accuracy and AUC, including AUC, where every competitor is significantly outperformed, with significance ranging from $p<0.01$ for most competitors down to $p<0.05$ for ANN, XGBoost, HCEG, KG, Llama, and Qwen.
The only two cells that do not reach significance are Precision against Qwen and F1 against Llama, the single strongest competitor on each of those two metrics. The standard deviations of MPGNN on the classification metrics are again uniformly small ($\pm$0.010 on AUC and $\pm$0.008 to $\pm$0.016 on the other four), confirming that the classification advantage is stable across splits in this scenario as well. Taken together with the one-off purchase results, this indicates that the gains of MPGNN generalize across qualitatively different pricing modes, and that the proposed multi-relational graph modeling delivers its strongest and most uniformly significant advantage on the classification side of the data-product pricing task.

On the regression side it obtains the smallest MAE (1.357) and RMSE (1.923) and the highest $R^2$ (0.507), with improvements of 13.31\% on MAE and 7.43\% on RMSE over DS and 12.63\% on $R^2$ over ANN. The per-baseline paired $t$-test shows that, on MAE and $R^2$, MPGNN significantly outperforms every competitor except the three LLM backbones DS, Llama, and Qwen, whose per-access regression errors come close to MPGNN but whose large split-to-split variance ($\pm$0.302 on Llama's MAE, $\pm$0.258 on its RMSE, and $\pm$0.092 to $\pm$0.132 on the $R^2$ of Llama and Qwen) prevents the gap from reaching the threshold. On RMSE the same three backbones, together with the closest single-task and boosting regressors ANN and XGBoost, likewise stay just below significance, whereas all remaining baselines are beaten significantly. In other words, under request-based pricing the LoRA-tuned language models emerge as the most competitive regression baselines, but MPGNN still improves upon them on every regression metric while strictly and significantly dominating all remaining methods. The standard deviation of MPGNN in this scenario ($\pm$0.068 on MAE, $\pm$0.090 on RMSE, and $\pm$0.030 on $R^2$) is markedly tighter than these high-variance language-model baselines, indicating that the regression behavior of MPGNN remains consistent across splits even when the absolute price magnitudes shrink.

\subsection{Industry heterogeneity analysis}

Data products from different industries exhibit substantially different pricing characteristics: financial datasets tend to be priced against standardized benchmarks, gaming datasets are sold under highly diverse licensing schemes, and public-sector data products span a wide range from free open data to high-priced specialized feeds. A model that is reported to work well on average may therefore mask substantial variability across industries.
To examine whether the proposed model generalizes uniformly or instead performs unevenly across industry segments, we conduct a per-industry heterogeneity analysis on the time-based subscription pricing task.

We train the shared model on the complete heterogeneous graph and, to convey the stability of every per-industry estimate, repeat the analysis over the same three train/test splits used throughout the paper, reporting for each industry the mean and sample standard deviation of every metric across the three splits.
For each split the trained model is evaluated separately on the subset of test products that apply to each industry. Industries with fewer than 50 data products in the graph are excluded to avoid unstable estimates; this leaves all 11 industries in the dataset. Because a single data product can be applicable to multiple industries via the product$\to$industry ``apply'' relation, products are counted in every industry they apply to (replication strategy), and hence $\sum_i n_{\text{test},i}$ exceeds the size of the global test set.
The same seven metrics used in the main experiment (Accuracy, Precision, AUC, and F1 for the five-tier price classification and MAE, RMSE, and $R^2$ for regression) are reported alongside the average per-industry test size $\bar{n}_{\text{test}}$. Results are reported in Table~\ref{tab:industry_heterogeneity}.

\begin{table}[htbp]
\centering
\TableSpaced
\caption{Per-industry performance of the proposed model on the time-based subscription pricing task, reported as mean $\pm$ standard deviation over the three train/test splits. Rows are sorted by mean MAE in ascending order (best regression performance first).}
\label{tab:industry_heterogeneity}
\footnotesize
\resizebox{\textwidth}{!}{
\begin{tabular}{lllllcccc}
\toprule
Industry &     Accuracy &Precision &AUC &F1 &MAE & RMSE & $R^2$ & $\bar{n}_{\text{test}}$ \\
\midrule
Environmental &     \shortstack{0.802 \\ ($\pm$0.050)} &\shortstack{0.707 \\ ($\pm$0.027)} &\shortstack{0.875 \\ ($\pm$0.039)} &\shortstack{0.713 \\ ($\pm$0.021)} &\shortstack{0.446 \\ ($\pm$0.023)} & \shortstack{0.805 \\ ($\pm$0.036)} & \shortstack{0.691 \\ ($\pm$0.019)} & 52 \\
Financial Services &     \shortstack{0.810 \\ ($\pm$0.025)} &\shortstack{0.708 \\ ($\pm$0.057)} &\shortstack{0.940 \\ ($\pm$0.031)} &\shortstack{0.698 \\ ($\pm$0.072)} &\shortstack{0.545 \\ ($\pm$0.064)} & \shortstack{1.006 \\ ($\pm$0.122)} & \shortstack{0.557 \\ ($\pm$0.111)} & 62 \\
Telecommunications &     \shortstack{0.783 \\ ($\pm$0.063)} &\shortstack{0.773 \\ ($\pm$0.072)} &\shortstack{0.939 \\ ($\pm$0.026)} &\shortstack{0.741 \\ ($\pm$0.056)} &\shortstack{0.578 \\ ($\pm$0.050)} & \shortstack{0.921 \\ ($\pm$0.050)} & \shortstack{0.810 \\ ($\pm$0.050)} & 43 \\
E-commerce and Business &     \shortstack{0.758 \\ ($\pm$0.024)} &\shortstack{0.761 \\ ($\pm$0.018)} &\shortstack{0.931 \\ ($\pm$0.006)} &\shortstack{0.730 \\ ($\pm$0.014)} &\shortstack{0.585 \\ ($\pm$0.085)} & \shortstack{1.074 \\ ($\pm$0.220)} & \shortstack{0.537 \\ ($\pm$0.192)} & 68 \\
Cars and Automotive &     \shortstack{0.700 \\ ($\pm$0.026)} &\shortstack{0.646 \\ ($\pm$0.025)} &\shortstack{0.868 \\ ($\pm$0.023)} &\shortstack{0.625 \\ ($\pm$0.022)} &\shortstack{0.622 \\ ($\pm$0.012)} & \shortstack{0.967 \\ ($\pm$0.019)} & \shortstack{0.708 \\ ($\pm$0.027)} & 104 \\
Public Sector &     \shortstack{0.671 \\ ($\pm$0.006)} &\shortstack{0.639 \\ ($\pm$0.032)} &\shortstack{0.883 \\ ($\pm$0.019)} &\shortstack{0.630 \\ ($\pm$0.039)} &\shortstack{0.626 \\ ($\pm$0.036)} & \shortstack{0.959 \\ ($\pm$0.051)} & \shortstack{0.457 \\ ($\pm$0.093)} & 82 \\
Media and Entertainment &     \shortstack{0.757 \\ ($\pm$0.028)} &\shortstack{0.801 \\ ($\pm$0.028)} &\shortstack{0.892 \\ ($\pm$0.008)} &\shortstack{0.772 \\ ($\pm$0.023)} &\shortstack{0.693 \\ ($\pm$0.013)} & \shortstack{1.028 \\ ($\pm$0.032)} & \shortstack{0.471 \\ ($\pm$0.138)} & 56 \\
Healthcare and Life Sciences &     \shortstack{0.715 \\ ($\pm$0.019)} &\shortstack{0.641 \\ ($\pm$0.047)} &\shortstack{0.883 \\ ($\pm$0.007)} &\shortstack{0.605 \\ ($\pm$0.040)} &\shortstack{0.724 \\ ($\pm$0.093)} & \shortstack{1.064 \\ ($\pm$0.186)} & \shortstack{0.687 \\ ($\pm$0.105)} & 44 \\
Resources &     \shortstack{0.666 \\ ($\pm$0.037)} &\shortstack{0.621 \\ ($\pm$0.024)} &\shortstack{0.871 \\ ($\pm$0.017)} &\shortstack{0.591 \\ ($\pm$0.054)} &\shortstack{0.739 \\ ($\pm$0.045)} & \shortstack{1.034 \\ ($\pm$0.159)} & \shortstack{0.736 \\ ($\pm$0.091)} & 43 \\
Gaming &     \shortstack{0.726 \\ ($\pm$0.078)} &\shortstack{0.799 \\ ($\pm$0.090)} &\shortstack{0.945 \\ ($\pm$0.041)} &\shortstack{0.756 \\ ($\pm$0.103)} &\shortstack{0.761 \\ ($\pm$0.192)} & \shortstack{1.207 \\ ($\pm$0.327)} & \shortstack{0.704 \\ ($\pm$0.137)} & 23 \\
Manufacturing &     \shortstack{0.617 \\ ($\pm$0.024)} &\shortstack{0.593 \\ ($\pm$0.037)} &\shortstack{0.842 \\ ($\pm$0.020)} &\shortstack{0.558 \\ ($\pm$0.041)} &\shortstack{0.770 \\ ($\pm$0.109)} & \shortstack{1.275 \\ ($\pm$0.276)} & \shortstack{0.539 \\ ($\pm$0.180)} & 66 \\
\midrule
Overall (full test set) &     \shortstack{0.729 \\ ($\pm$0.014)} & \shortstack{0.731 \\ ($\pm$0.013)} & \shortstack{0.908 \\ ($\pm$0.006)} & \shortstack{0.729 \\ ($\pm$0.014)} & \shortstack{0.622 \\ ($\pm$0.027)} & \shortstack{0.994 \\ ($\pm$0.074)} & \shortstack{0.676 \\ ($\pm$0.055)} & 641\\
\bottomrule
\end{tabular}
}
\begin{tablenotes}[para]
    \raggedright
    \footnotesize
    \item[\raisebox{-0.7ex}{$\bullet$}] The shared model is trained once per split on the complete heterogeneous graph; each cell reports the mean over the three splits with the sample standard deviation below it. Per-industry metrics are computed on the subset of test products that apply to each industry; a product applying to multiple industries contributes to each of them (replication strategy), so $\sum_i \bar{n}_{\text{test},i} > \bar{n}_{\text{test}}^{\text{overall}}$.
    
    \item[\raisebox{-0.7ex}{$\bullet$}] Only industries with at least 50 products in the graph are reported (11 out of 11 industries pass the threshold). 
    
    \item[\raisebox{-0.7ex}{$\bullet$}] $\bar{n}_{\text{test}}$ is the average per-industry test-subset size across the three splits; for Gaming ($\bar{n}_{\text{test}}=23$) this subset is small, so its per-industry estimates carry larger variance. The bottom ``Overall'' row reports the global test-set metrics (mean $\pm$ std over the three splits) for reference.
\end{tablenotes}
\end{table}

First, across all eleven industries the model achieves usable and reasonably stable performance: the mean per-industry MAE ranges from 0.446 (Environmental) to 0.770 (Manufacturing), and the mean per-industry AUC from 0.842 (Manufacturing) to 0.945 (Gaming), both clustered around the global test-set values (MAE 0.622, AUC 0.908). 
Five industries, namely Environmental, Financial Services, Telecommunications, E-commerce and Business, and Cars and Automotive, exhibit MAE values below the overall level, with Cars and Automotive performing nearly on par with the overall benchmark. In contrast, the remaining six industries have MAE values above the overall level, although Public Sector exceeds it only marginally. Regarding AUC, four industries, namely Gaming, Financial Services, Telecommunications, and E-commerce and Business, outperform the overall benchmark, while the remaining industries exhibit slightly lower AUC values.
The cross-split standard deviations are small for the majority of segments (AUC std between 0.006 and 0.041, MAE std between 0.012 and 0.192), so these per-industry differences are stable rather than split artefacts; the main exceptions are the $R^2$ of E-commerce and Business ($\pm$0.192), Manufacturing ($\pm$0.180), Media and Entertainment ($\pm$0.138), and Gaming ($\pm$0.137), whose wide or heavy-tailed test subsets make the explained-variance estimate noisy. We therefore regard the spread across industries as moderate heterogeneity rather than a localized failure mode.

Second, Environmental, Financial Services, and Telecommunications are the model's strongest segments. Environmental posts the lowest regression errors across the board (MAE 0.446, RMSE 0.805), and Financial Services and Telecommunications combine low regression errors with strong classification: Financial Services attains the second-lowest MAE (0.545), the highest Accuracy (0.810), and the second-highest AUC (0.940), while Telecommunications attains the highest $R^2$ (0.810, the only industry above 0.8), together with a top-three AUC (0.939) and the third-lowest MAE (0.578). This is consistent with the fact that these segments contain products with relatively standardized pricing conventions: environmental data products tend to follow public-sector-style fixed-fee schedules, financial datasets are priced against transparent market benchmarks, and telecommunication subscription plans follow clearly tier-structured price lists. This makes both their price levels and their discrete tiers easier to recover from textual features and seller context.

Third, Manufacturing, Gaming, and Resources form the bottom three on MAE (0.770, 0.761, 0.739 respectively). 
Manufacturing is in fact the single weakest segment across both heads: it posts the highest MAE (0.770) and RMSE (1.275) together with the lowest Accuracy (0.617), F1 (0.558), and AUC (0.842).
For Gaming, the small test sample ($\bar{n}_{\text{test}}=23$) inflates the variance of the exact-price estimate (MAE $\pm$0.192, RMSE $\pm$0.327), yet its top-ranked AUC (0.945) and mid-table $R^2$ (0.704) show that the tier ordering is still learned well, so the difficulty is confined to precise value prediction rather than to ranking. Resources, by contrast, has a moderate sample size ($\bar{n}_{\text{test}}=43$) but covers heterogeneous sub-domains (satellite-derived commodity feeds, supply-chain telemetry), each with its own pricing logic; it posts both a bottom-three regression error and the second-lowest classification scores (Accuracy/F1 0.666/0.591, AUC 0.871), so the shared encoder is genuinely weaker on this segment on both heads. 
Public Sector is the only industry exhibiting a substantial decline in $R^2$ (0.457, compared with the median of 0.687 across the eleven industries). Its MAE (0.626) is also slightly higher than the overall benchmark (0.622), indicating relatively weak performance on both regression metrics.

\begin{figure}[htbp]
  \centering
  \includegraphics[width=\linewidth]{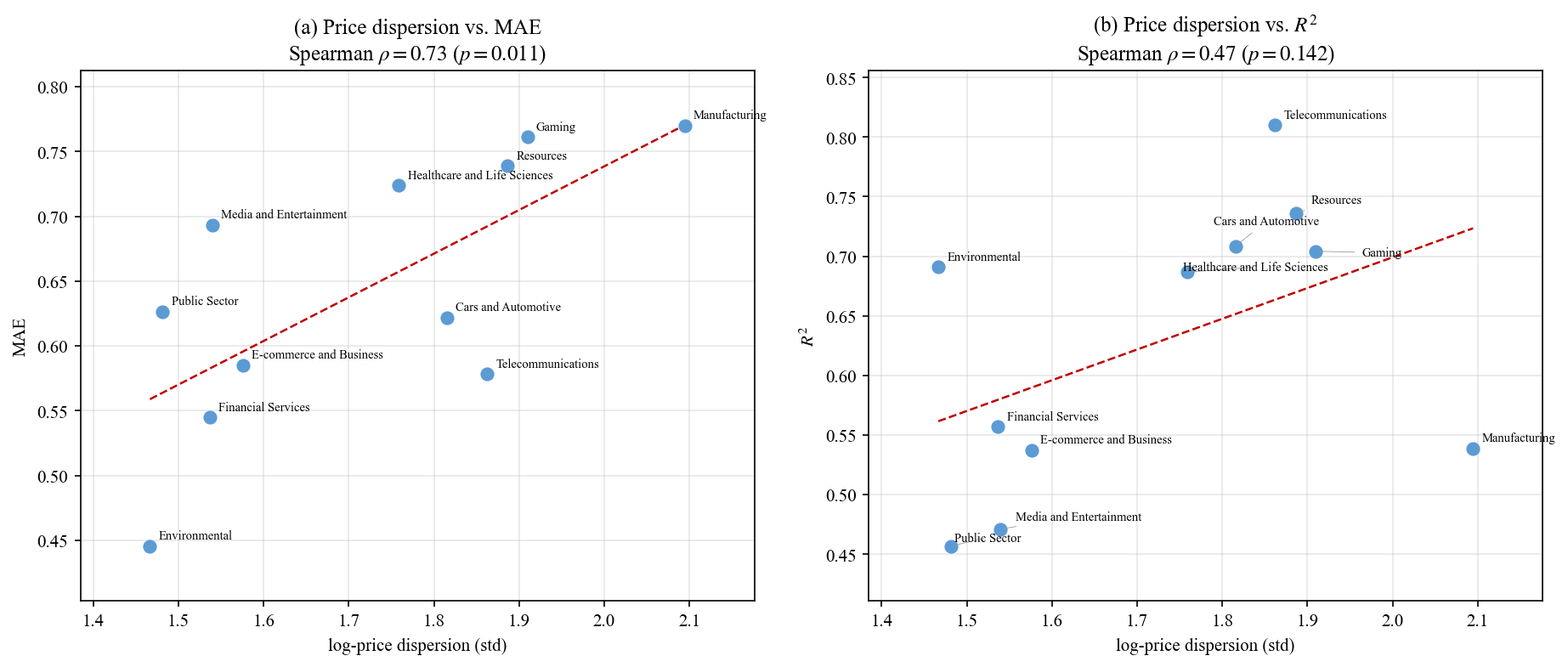}
  \caption{A data-level explanation of cross-industry differences. Each point represents one industry ($n = 11$); dashed lines show least-squares fits, and $\rho$ denotes Spearman's rank correlation coefficient.
  (a) An industry's log-price dispersion is significantly and positively correlated with its regression MAE ($\rho = 0.73$, $p = 0.011$): the model makes larger absolute pricing errors in industries with more dispersed prices.
  (b) Log-price dispersion shows a positive but non-significant association with $R^2$ ($\rho = 0.47$, $p = 0.142$). This trend may arise because the denominator of $R^2$, the target variance, grows with price dispersion.}
  \label{fig:industry_explain}
\end{figure}

To test whether the qualitative explanations above are supported by the data rather than merely plausible, we relate each industry's performance to a few simple, model-free statistics of its own price data: the dispersion of (log) prices, the price range, and the balance of the five price tiers (Shannon entropy), together with its average test-subset size, using Spearman rank correlations across the eleven industries (Table~\ref{tab:industry_explain} and Fig.~\ref{fig:industry_explain}).
Only one relationship is both sizeable and statistically significant: an industry's log-price dispersion correlates positively with its regression MAE ($\rho=0.73$, $p=0.01$), i.e.\ the model makes larger absolute pricing errors precisely in industries whose prices are more spread out, a direct, data-backed account of why Manufacturing, Gaming and Resources are the hardest to price exactly. The remaining intuitions receive at best weak or mixed support. 

\begin{table}[htbp]
\centering
\TableSpaced
\caption{Spearman rank correlations between simple per-industry price/sample statistics and per-industry performance ($n=11$ industries). A positive $\rho$ means the two quantities increase together. Only the price-dispersion vs.\ MAE relationship is significant at $p<0.05$.}
\label{tab:industry_explain}
\begin{tabular}{lcc}
\toprule
Industry-level relationship & Spearman $\rho$ & $p$ \\
\midrule
log-price dispersion vs.\ MAE & $+0.73$ & 0.011 \\
log-price dispersion vs.\ $R^2$ & $+0.47$ & 0.142 \\
price range vs.\ $R^2$ & $-0.13$ & 0.709 \\
tier-distribution entropy vs.\ AUC & $-0.45$ & 0.160 \\
average test-subset size vs.\ $R^2$ std & $+0.11$ & 0.749 \\
\bottomrule
\end{tabular}
\end{table}

The model generalizes reasonably across all eleven industries, and the per-industry metrics are stable across the three splits (small standard deviations on all classification metrics and on
most regression metrics). Measured by regression MAE, performance is strongest on industries with standardized pricing conventions (Environmental, Financial Services, Telecommunications) and weakest on industries with heterogeneous sub-domains or unusually wide price dispersion (Manufacturing, Gaming, Resources). This ordering is not uniform across metrics: Gaming attains the highest AUC (0.945) among all eleven industries, and Resources the second-highest $R^2$ (0.736), so their weak ranking on MAE reflects absolute pricing error alone rather than weak tier-ranking ability or model fit; Manufacturing is the only industry that performs poorly on every metric. These findings argue against using a single industry-agnostic baseline as the sole evaluation criterion: industry-stratified evaluation reveals heterogeneity that the global average alone conceals, and identifies concrete segments (chiefly Manufacturing, Resources, and Gaming) that would benefit from additional industry-specific modelling effort in future work.

\subsection{Identification and interpretation of key meta-paths}

\begin{figure}[htbp]
    \centering
    \includegraphics[width=\linewidth]{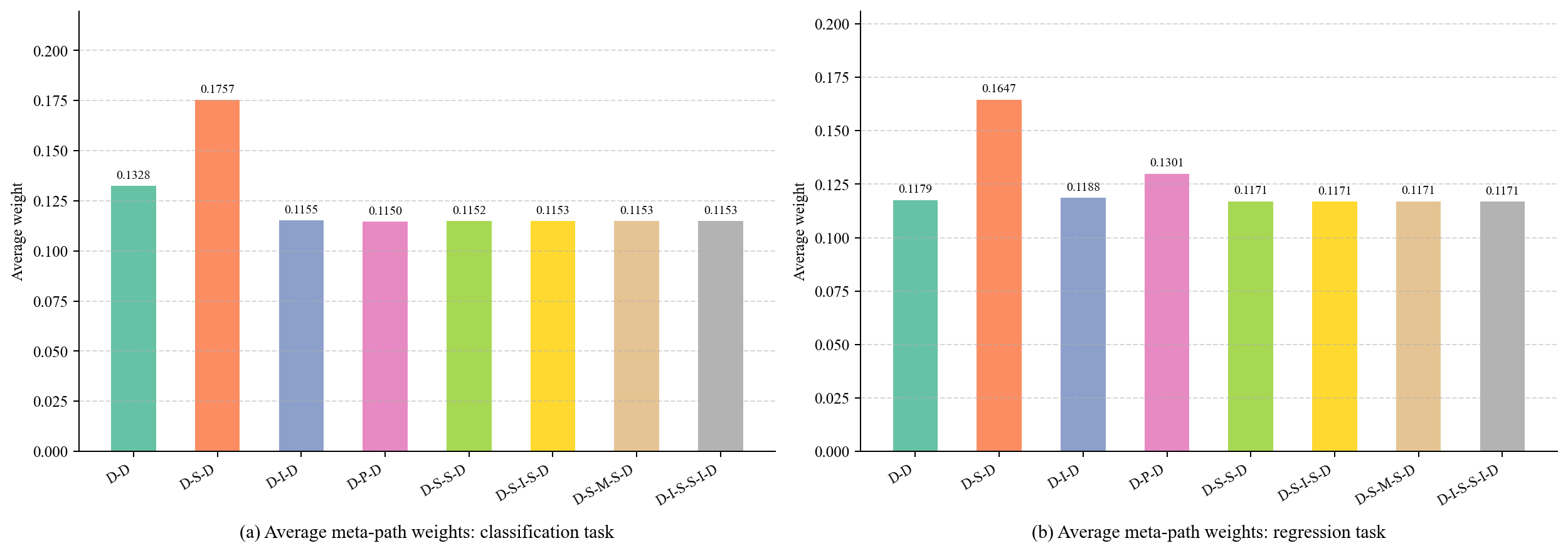}
    \caption{Average meta-path weights learned by the attention gate under the classification task~(a) and regression task~(b).}
    \label{fig:path_weights_full}
\end{figure}

To further understand the extent to which the model relies on different meta-paths for pricing tasks, we visualize the meta-path weights learned by the attention-based gating mechanism. Specifically, for each of the three dataset split ratios, we record the attention weight assigned to each meta-path upon model convergence and then average the weights across the three splits to mitigate the randomness arising from any single data partition. We consider eight meta-paths, each of which captures a distinct type of structural relationship between data product nodes through different intermediate entity types. Panels (a) and (b) of Figure \ref{fig:path_weights_full} report the resulting average weight distributions for the classification and regression tasks, respectively.

As shown in Fig.~\ref{fig:path_weights_full}, the D-S-D meta-path consistently dominates both the classification and regression tasks. For the classification task (Fig.~\ref{fig:path_weights_full}(a)), D-S-D achieves the highest average attention weight of 0.1757, approximately 1.32 times that of the second-ranked D-D meta-path (0.1328). Similarly, for the regression task (Fig.~\ref{fig:path_weights_full}(b)), D-S-D retains the highest weight of 0.1647, exceeding that of the second-ranked D-P-D meta-path (0.1301). Notably, the relative rankings of all meta-path weights remain identical across the three dataset split ratios for both tasks, indicating that the model's reliance on different meta-paths is highly stable. Furthermore, the attention weights of the remaining meta-paths exhibit a highly concentrated distribution. Specifically, in the classification task, the weights of the six meta-paths other than D-S-D and D-D fall within a narrow range of 0.1150--0.1155. Likewise, in the regression task, the weights of the six meta-paths other than D-S-D and D-P-D range from 0.1171 to 0.1188. These findings suggest that the model assigns greater importance to the most influential meta-paths while maintaining a relatively balanced allocation of attention across the remaining meta-paths.

\subsubsection{Why does the D-S-D path dominate? A data-level analysis}
\label{subsec:dsd_analysis}

The attention gate assigns the largest weight to the D-S-D path (i.e., the connectivity pattern ``data product $\to$ supplier $\to$ data product,'' meaning that two products are linked through a shared supplier). However, a weight learned by the model can only tell us that the supplier-linked path indeed carries useful information; it cannot tell us why: the magnitude of the weight is not, by itself, an explanation. To investigate this question, we conduct a model-independent analysis directly on the raw data to test the following hypothesis: a data product's price is largely determined by its supplier, so that two products sharing the same supplier (i.e., a D-S-D neighbor pair) should be far more similar in price than two randomly chosen products. All statistics reported below are computed on time-based pricing products (i.e., data products priced by duration/subscription), using the same sampling rule, valid-sample scope, and logarithmic transformation as the main model, and the same quintile-based price-tier binning as the classification task, so that the conclusions drawn here map directly onto the model's training setup. The supplier dimension is also well supported by the data: among the 373 suppliers that offer priced products, 237 (63.5\%) offer at least two priced products, covering 95.8\% of all product--supplier associations; this yields 66,257 same-supplier product pairs, precisely the edges that the D-S-D path can traverse on the graph to carry information.

The evidence below tightens the case along three independent angles in turn: how much price variance supplier identity alone can explain $\to$ how close in price two same-supplier products actually are $\to$ whether this closeness still holds at the coarser scale of price tiers.

Variance decomposition. We first treat the supplier as the sole grouping factor in a one-way ANOVA on the log price: the single variable ``which supplier the product belongs to'' already explains 81.2\% of the total price variance ($\eta^2=0.812$), that is, differences among products of the same supplier account for less than 19\% of the total price variance, while the remaining 81.2\% comes from differences between suppliers. The intra-class correlation, $\mathrm{ICC}=0.789$, likewise shows that prices of same-supplier products are highly correlated; the between-group effect is highly significant (one-way ANOVA, $F=32.83$, $df=(372,2835)$, $p<0.001$), ruling out the possibility that this clustering is merely due to random fluctuation. Put more intuitively, the median within-supplier coefficient of variation of the raw price is only $0.48$, whereas the global coefficient of variation across all products reaches $2.70$, about $5.6\times$ larger. Prices are thus tightly clustered within a given supplier but widely dispersed across suppliers, which is exactly the structure that the D-S-D path is positioned to capture (summarized in Table~\ref{tab:dsd_evidence}).

\begin{table}[htbp]
\centering
\TableSpaced
\caption{Data-level evidence for the impact of the D-S-D (shared-supplier) path on time-based (subscription) pricing.}
\label{tab:dsd_evidence}
\footnotesize
\resizebox{\textwidth}{!}{
\begin{tabular}{llcp{5.3cm}}
\toprule
Aspect & Indicator & Value & Interpretation \\
\midrule
\multirow{2}{*}{Variance decomposition}
 & Variance in price explained by supplier ($\eta^2$) & 0.812 & Supplier identity alone explains 81.2\% of price variance \\
 & Intra-class correlation (ICC) & 0.789 & Same-supplier product prices are highly correlated \\
\cmidrule(lr){1-4}
\multirow{2}{*}{Within-supplier dispersion}
 & Median within-supplier CV (raw price) & 0.48 & Same-supplier product prices are close to one another \\
 & Global CV (raw price) & 2.70 & Overall price dispersion is about $5.6\times$ the within-supplier level \\
\cmidrule(lr){1-4}
\multirow{4}{*}{Price gap $|\Delta\log p|$ (median)}
 & Same-supplier pairs (D-S-D) & 0.12 ($\approx 1.13\times$) & Typically differ by only $\approx$13\% \\
 & Similar-product pairs (D-D) & 0.18 ($\approx 1.20\times$) & Typically differ by only $\approx$20\% \\
 & Same-industry pairs (D-I-D) & 1.48 ($\approx 4.39\times$) & Can differ by more than $4\times$ \\
 & Random pairs (baseline) & 1.59 ($\approx 4.90\times$) & Differ by nearly $5\times$, over ten times the same-supplier gap \\
\cmidrule(lr){1-4}
\multirow{3}{*}{Price-tier view (classification)}
 & Same-tier rate, same-supplier pairs & 0.690 & About 69\% of same-supplier products fall in the same price tier \\
 & Same-tier rate, random pairs & 0.206 & Close to the 20\% chance rate under 5 equal-frequency tiers \\
 & Normalized mutual information (supplier, tier) & 0.34 & Knowing the supplier narrows tier uncertainty to under 2 effective candidates \\
\bottomrule
\end{tabular}
}
\vspace{2pt}

\begin{tablenotes}[para]
\raggedright
\item[\raisebox{-0.7ex}{$\bullet$}] {\footnotesize Note: all indicators are computed on time-based (subscription) pricing valid samples ($N=3{,}204$ products, $3{,}208$ product--supplier links, $373$ suppliers), using the same log transform and quintile-based tier binning as the MPGNN training pipeline.}
\end{tablenotes}
\end{table}

\begin{figure}[htbp]
    \centering
    \includegraphics[width=0.72\linewidth]{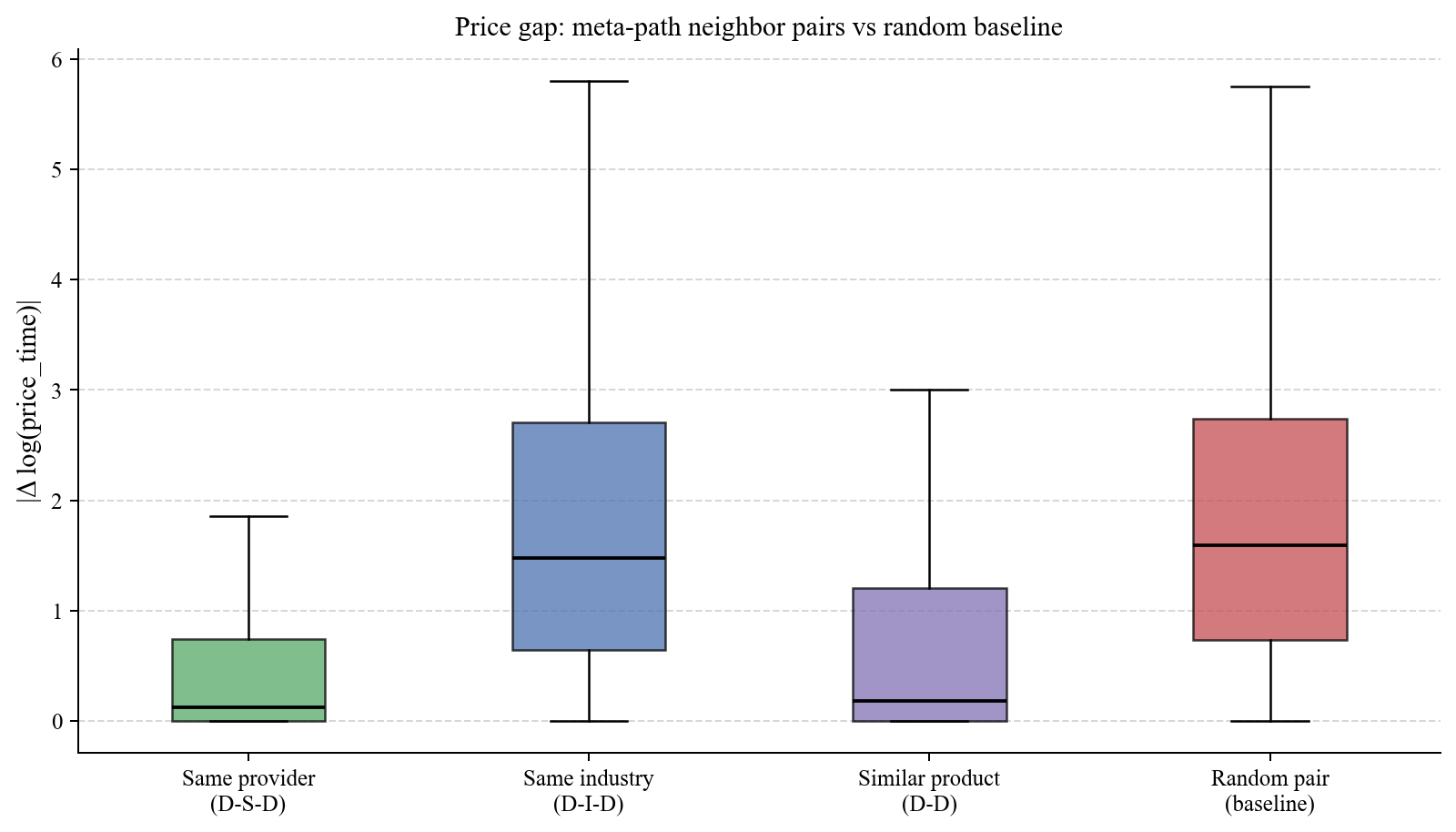}
    \caption{Distribution of the absolute log-price gap for meta-path neighbor pairs relative to a random-pair baseline. Shared-supplier (D-S-D) pairs and similar-product (D-D) pairs show much smaller price gaps (median differences of about 13\% and 20\%, respectively) than same-industry (D-I-D) pairs and random pairs (median differences exceeding $4\times$), corroborating the learned meta-path weights.}
    \label{fig:dsd_pricegap}
\end{figure}

Neighbor pairs vs.\ random pairs. We next compare the absolute log-price gap $|\Delta\log(\text{price})|$ between same-supplier pairs (the pairs that the D-S-D path actually samples on the graph) and an equal-sized set of random product pairs (distribution in Figure~\ref{fig:dsd_pricegap}; medians summarized in Table~\ref{tab:dsd_evidence}). This quantity is the absolute log of the price ratio between two products, so it translates directly into an intuitive ``price-multiple'' gap: same-supplier pairs (D-S-D) have a median gap of $0.12$, corresponding to a typical difference of about $1.13\times$ (i.e., $\approx$13\%); similar-product pairs (D-D, products with similar content that are not necessarily from the same supplier) have a median gap of $0.18$, or about $1.20\times$ ($\approx$20\%); same-industry pairs (D-I-D, products in the same industry that are not necessarily from the same supplier) have a median gap of $1.48$, or about $4.39\times$; and random pairs (baseline) have a median gap of $1.59$, or about $4.90\times$, more than an order of magnitude larger than the same-supplier gap. A one-sided Mann--Whitney $U$ test confirms that the difference between same-supplier pairs and random pairs is statistically significant ($p<0.001$, rank-biserial correlation $0.66$, a large effect size). This ordering is unambiguous: supplier $\approx$ content $\ll$ industry: knowing that two products share a supplier, or have similar content, is associated with a small price gap, whereas knowing only that they belong to the same industry still leaves the price gap nearly as large as for random pairs. This corroborates the learned meta-path weights and explains why the supplier-linked path (D-S-D), rather than the industry-linked path (D-I-D), carries the strongest pricing signal.

Classification view. Supplier homogeneity is even more pronounced at the coarser scale of price tiers, corroborating the fact that D-S-D receives its largest weight in the classification task. Specifically, same-supplier pairs fall into the same price tier with probability $0.690$, versus only $0.206$ for random pairs; the latter is close to the theoretical chance rate of 20\% under five equal-frequency price tiers, confirming that the tier partition itself is balanced, while same-supplier pairs hit the same tier at roughly $3.3\times$ the random rate (Table~\ref{tab:dsd_evidence}). From an information-theoretic perspective, the entropy of the price tier with no side information is $H(\text{tier})=2.30$ bits, already close to the theoretical maximum of $\log_2 5\approx 2.32$ bits for five equally likely tiers, i.e., one can barely guess which tier a product falls into. Once the supplier is known, this entropy drops to $H(\text{tier}\mid\text{supplier})=0.69$ bits, a 70\% reduction, corresponding to a normalized mutual information of $0.34$ (Table~\ref{tab:dsd_evidence}).

\subsubsection{Case studies}
\label{subsec:dsd_cases}

Concrete supplier cases further illustrate how this mechanism operates in practice, and where it breaks down (key figures for each case are summarized in Table~\ref{tab:dsd_cases}). Pattern 1: uniform brand-level pricing, that is, knowing the supplier is knowing the price. This is the clearest category of cases: a supplier applies essentially the same price to almost all of its products, regardless of differences in content. For example, 180byTwo splits a single ``eCHO\texttrademark\ Predictive B2B Intent Data'' product into 121 industry-specific packages, yet prices every one of them at exactly \$2{,}500 (tier 4, coefficient of variation $\mathrm{CV}=0$, i.e., zero dispersion); all 7 of Aviation Edge's aviation-data APIs are uniformly priced at \$123.25 (tier 0); and all 4 of Enetpulse's sports-data feeds are uniformly priced at \$793.1 (tier 2). For suppliers of this type, the within-group coefficient of variation is identically zero (Table~\ref{tab:dsd_cases}); the single fact of ``which supplier'' already fully determines the price.

Pattern 2: different suppliers anchor at different price levels. Pricing levels can differ enormously across suppliers, which is precisely why ``supplier identity'' is so informative. For instance, umlaut prices its nationwide mobile-coverage maps at \$12{,}000--20{,}000 (tier 4, the high end), while COMSPOC prices its ARGUS space-data services at \$0.083 (tier 0, the low end). It is exactly this large cross-supplier price gap that makes the structural information ``who the supplier is'' so predictive.

Pattern 3: noticeable within-supplier dispersion. The mechanism does not hold for every supplier. TRADExperts sells the same ``B2B advertising data'' product across different countries at prices ranging from \$83 to \$8{,}333 ($\mathrm{CV}=2.91$, spanning tiers 0 through 4, almost the entire price range); Nimbler's contact datasets mostly cluster around \$133--150, but include a \$12{,}500 identity-graph product as an outlier. For these suppliers, within-supplier price variation is driven mainly by differences in data volume or geographic coverage rather than by brand alone, so the signal carried by the D-S-D path is noisier and must be supplemented by content-similarity-based paths (D-D) and pricing-mode-based paths (D-P-D), consistent with the more even distribution of meta-path weights observed for the regression task.

\begin{table}[htbp]
\centering
\TableSpaced
\caption{Representative suppliers illustrating the D-S-D pricing mechanism under time-based (subscription) pricing.}
\label{tab:dsd_cases}
\resizebox{\textwidth}{!}{
\begin{tabular}{lllccl}
\toprule
Pattern & Supplier & \# Products & Price range (USD) & CV & Price tier(s) \\
\midrule
\multirow{3}{*}{Pattern 1: Uniform brand pricing (strong)}
 & 180byTwo      & 121 & 2{,}500 (flat)       & 0.00 & 4 (single) \\
 & Aviation Edge & 7   & 123.25 (flat)        & 0.00 & 0 (single) \\
 & Enetpulse     & 4   & 793.1 (flat)         & 0.00 & 2 (single) \\
\cmidrule(lr){1-6}
\multirow{2}{*}{Pattern 2: Distinct price anchors}
 & umlaut        & 16  & 12{,}000--20{,}000   & 0.18 & 4 (single) \\
 & COMSPOC       & 10  & 0.083 (flat)         & 0.00 & 0 (single) \\
\cmidrule(lr){1-6}
\multirow{2}{*}{Pattern 3: Within-supplier dispersion (weak, boundary case)}
 & TRADExperts   & 31  & 83--8{,}333          & 2.91 & 0--4 (spread) \\
 & Nimbler       & 11  & 133--12{,}500        & 2.90 & 0/1/4 \\
\bottomrule
\end{tabular}
}
\end{table}

Taken together, three independent angles, namely the variance decomposition, the comparison of price gaps between neighbor pairs and random pairs, and the mutual-information analysis under the price-tier view (Table~\ref{tab:dsd_evidence}), are mutually independent yet converge on the same conclusion: the supplier dimension is the strongest source of price homogeneity in the data, and this signal is carried by the supplier structure itself, without relying on any explicit supplier price feature. This directly explains why the attention gate assigns the largest weight to the D-S-D path, and why this dominance is most pronounced in the price-tier classification task.

\subsection{Parameter sensitivity analysis}
\label{subsec:parameter-sensitivity}

\begin{figure}[htbp]
  \centering
  \includegraphics[width=\linewidth]{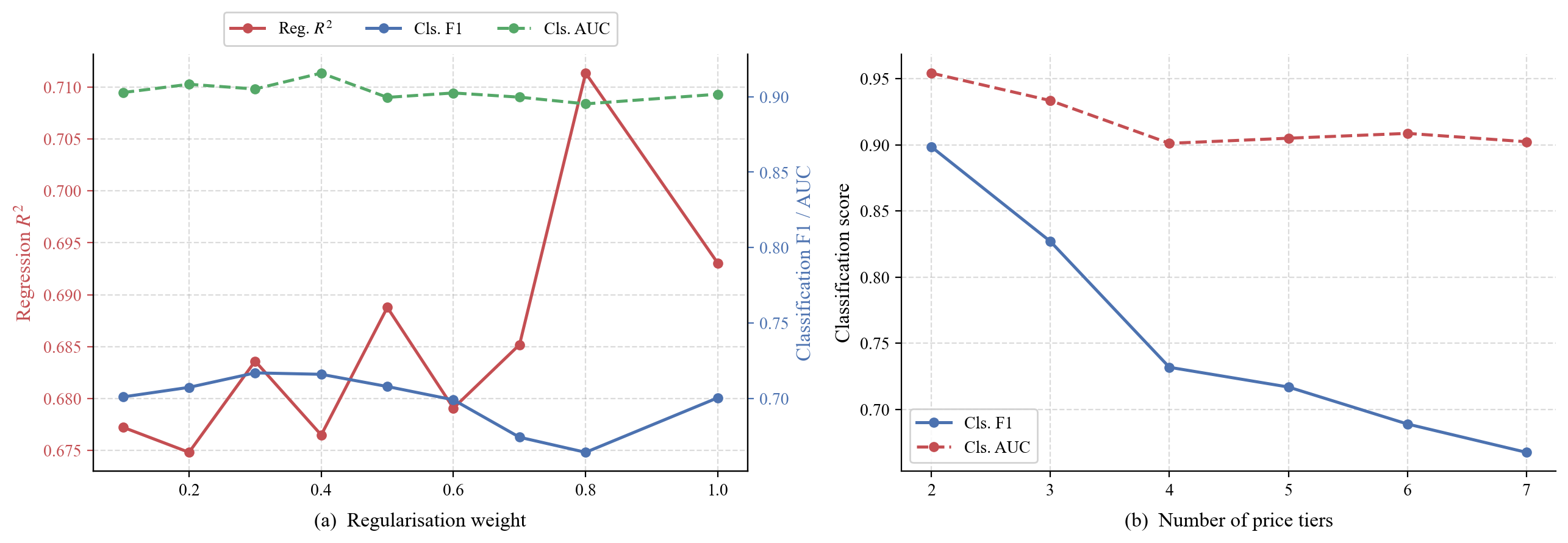}
  \caption{Robustness of MPGNN to two key parameters.
  (a) regularisation weight: regression $R^2$ (left axis)
  together with classification F1 and AUC (right axis);
  (b) number of price tiers: classification F1
  and AUC.}
  \label{fig:robustness-reg-tier}
\end{figure}

To assess the robustness of the proposed model with respect to its key
configuration choices, we conduct a sensitivity analysis on two
representative parameters: the regularisation weight, which controls the strength of the meta-path regularisation
term, and the number of price tiers,
which sets the granularity of the price-classification task. For each parameter we vary its value over a wide range while keeping all other settings fixed, and report performance on the classification task (and,
for regularisation weight, also on the regression task; the regression target
is the continuous price and is therefore independent of the number of price
tiers). The results are summarised in
Fig.~\ref{fig:robustness-reg-tier}. 

As shown in Fig.~\ref{fig:robustness-reg-tier}(a), the model is highly robust
to the regularisation weight over the entire scanned range $[0.1, 1.0]$: all
curves remain almost flat and fluctuate only within narrow bands. On the
classification task, AUC stays within $0.895$--$0.916$ (a variation of about
$2\%$), while F1 and Accuracy range over $0.664$--$0.718$. On the
regression task, $R^2$ remains within
$0.671$--$0.711$, RMSE within $0.936$--$0.999$ and MAE within
$0.611$--$0.637$. The only discernible pattern is a mild and somewhat
irregular upward drift in regression $R^2$ at larger weights: $R^2$ stays
close to $0.671$--$0.677$ for weights in $[0.1, 0.4]$, then fluctuates to higher values for weights in $[0.5, 1.0]$, peaking
at $R^2=0.711$ when the weight is $0.8$ before settling back to $0.693$ at a
weight of $1.0$. No setting causes any breakdown, confirming that the model
has a wide tolerance band for this weight.

The number of price tiers directly controls the granularity of the
price-classification task: a larger number of tiers turns the problem into a
finer-grained, intrinsically harder multi-class problem. Consistent with this
intuition, Fig.~\ref{fig:robustness-reg-tier}(b) shows that the threshold
metrics decline monotonically as the number of tiers grows, from F1
$=0.899$ and Accuracy $=0.899$ at two tiers down to
F1$\approx0.67$--$0.72$ and Accuracy$\approx0.66$--$0.72$ at five to seven
tiers. Crucially, however, the ranking-based
AUC stays high throughout, never dropping below $0.901$ and remaining within
a narrow $0.901$--$0.909$ band from four tiers upward. This contrast shows
that the decline in the threshold metrics reflects the increasing difficulty
of placing exact decision boundaries among many adjacent tiers rather than
any deterioration of the model's discriminative power, which stays strong
across all granularities. 

\section{Discussions} \label{sec:discussion}

The results in Section~\ref{sec:result} establish that MPGNN consistently outperforms strong attribute-based, ensemble, and graph-based baselines across two distinct pricing modes and eleven industries. We now discuss what these findings imply for information systems research and for data product pricing practice, and how the proposed framework can be extended to other downstream tasks beyond data products.

\subsection{Theoretical contribution to IS}

The first theoretical contribution of this study is to reformulate data product pricing as a learning problem on a multi-relational heterogeneous information network rather than as an attribute-to-price regression, situating it within the predictive-analytics tradition in IS research that treats prediction as a research goal in its own right, complementary to explanatory modeling \citep{shmueliPredictiveAnalytics2011}. Existing IS research on data valuation either approaches the problem from the perspective of economic mechanism design \citep{hanDataValuationVertical2026,suiOptimalDataPricing2025}, which yields rigorous but assumption-heavy models, or treats it as a pure prediction task that maps product features to prices \citep{heckmanPricingModelData2015,haoHowPriceDataset2025}, which delivers accuracy but ignores the relational context in which a price is formed. By explicitly modeling data products together with their providers, markets, industries, and pricing modes in a single graph, we provide empirical evidence that the relational context of a data product carries pricing-relevant information that attribute-only models systematically miss.

The second contribution is to operationalize classical product pricing theories as meta-path structures, thereby making them empirically testable within a deep-learning pipeline. Brand equity, product similarity, and industry structure are translated respectively into the paths $D{-}S{-}D$, $D{-}D$ (or $D{-}P{-}D$), and $D{-}I{-}D$, while hybrid paths such as $D{-}S{-}I{-}S{-}D$ and $D{-}S{-}M{-}S{-}D$ capture cross-brand-and-industry and cross-market mechanisms. The learned attention weights then act as a built-in mechanism-attribution device: the dominant weight on $D{-}S{-}D$ in both the regression and classification tasks yields predictive evidence consistent with a brand-equity-related mechanism in the data product context: two data products from the same provider are by far the most informative comparable for listed-price prediction. The more uniform spread across paths in the regression task indicates that continuous price level depends on a combination of mechanisms rather than a single dominant one. To our knowledge, this is the first study to embed economic pricing theories into meta-path design for data products and to recover the relative importance of each theoretical mechanism in a fully data-driven way.

The third contribution is methodological: MPGNN demonstrates that structural interpretability and predictive performance are not in tension for data product pricing. The ablation study shows that removing the meta-path decomposition leads to significant degradations on all seven metrics, and removing the attention gate sharply degrades classification quality; this indicates that the interpretability-oriented components are also the components that drive accuracy. This is a meaningful response to the longstanding tension in IS between ``interpretable but weak'' linear models and ``strong but opaque'' deep models, and offers a concrete design template for building interpretable graph models in other IS prediction problems.

\subsection{Practical implications}

For data marketplaces and data providers, the results carry several
actionable implications. 
First, the dominance of the provider-bridge path $D{-}S{-}D$ in both tasks
indicates that the most informative reference point when pricing a new
data product is the existing portfolio of the same provider; in practice
this argues for provider-level pricing consistency tooling on marketplaces
(e.g., flagging products whose model-implied price deviates sharply from
the provider's other listings).
Second, the per-industry analysis in
Table~\ref{tab:industry_heterogeneity} shows that industries with
standardized pricing conventions (Environmental, Financial Services, and
Telecommunications) are predicted well by a single industry-agnostic
model, whereas Manufacturing is the weakest segment across both tasks,
Gaming and Resources show higher regression error, and Public Sector
exhibits a sharply lower $R^2$.
Marketplaces operating in the latter segments may benefit from
industry-specific fine-tuning or stratified pricing guidelines rather
than a single global price recommendation.
Third, regression and classification difficulty can be decoupled: some
industries are hard to price exactly but easy to tier, and vice
versa. This decoupling suggests a tiered deployment strategy: where exact prediction is
unreliable, the price-tier classifier can still be used to provide
guardrails on listed prices, supporting downstream decisions such as
catalog organization, dynamic discounting, and pre-listing screening.

\subsection{Role to downstream tasks and generalization to other fields}

Although developed for data product pricing, the proposed framework is not specific to data as a commodity; rather, it is designed for any market whose tradable objects are non-standardized, context-dependent, and embedded in a multi-entity ecosystem. At least three classes of downstream tasks can directly reuse the framework. (i) Pricing of other digital and non-standardized assets. Algorithmic models, machine-learning APIs, datasets-as-a-service, intellectual-property licenses, and SaaS modules share the heterogeneity and re-usability that motivated this study; the same heterogeneous graph schema and the same theory-driven meta-paths can be applied with only schema-level adjustments.
(ii) Valuation tasks beyond pricing. Because the model fuses heterogeneous relational signals into product-level embeddings, the same backbone can be fine-tuned to predict adjacent targets such as transaction probability, customer retention on subscription products, or expected lifetime value, simply by swapping the regression and classification heads. (iii) Recommendation and matching in data marketplaces. The meta-path contrastive objective already produces an economically-meaningful similarity space among data products; this space can directly support search, product-to-buyer matching, and bundle recommendation, complementing the price prediction output, in the same spirit as machine-learning-based search personalization in online marketplaces\citep{yoganarasimhanSearchPersonalizationMachine2020}.

Two boundary conditions should be noted when transferring the framework. First, the economic interpretation of each meta-path is tied to the domain-specific reading of nodes and edges; deploying the framework in a new market requires re-mapping its entities to brand, similarity, industry, and market relations. Second, as the per-industry analysis shows, performance in segments with heterogeneous sub-domains or extremely wide price dispersion is more modest, indicating that industry-stratified or hierarchical extensions of MPGNN are a natural next step rather than a limitation of the modeling philosophy itself.

\section{Conclusion} \label{sec:conclusion}

This study addresses the problem of pricing non-standardized data products by combining classical product pricing theories with heterogeneous graph representation learning. We model the data trading market as a heterogeneous information network over data products, providers, markets, industries, and pricing modes, and design multiple meta-paths whose semantics are directly grounded in brand equity theory, product similarity theory, industry structure theory, and their hybrid combinations. On top of this network, we propose MPGNN, a meta-path-guided heterogeneous graph neural network that encodes each meta-path with an independent Heterogeneous Graph Transformer, fuses the resulting representations through a regularized attention gating mechanism, and is jointly trained for meta-path contrastive learning, price regression, and price-tier classification.

Empirical evaluation on data products collected from Datarade, AWS Marketplace, and Alibaba Cloud Data Exchange yields four main findings. 
First, MPGNN achieves the best performance on all seven regression and classification metrics under both time-based subscription and one-off purchase pricing modes, with paired $t$-tests confirming statistically significant improvements over nearly every individual baseline, including strong graph baselines such as GraphSAGE and GAT.
On the main time-based pricing task, MPGNN attains an RMSE of 0.994, $R^2$ of 0.676, and classification Accuracy of 0.729, with improvements of 15.39\%, 28.79\%, and 18.94\% over the strongest competing baselines, respectively.
Second, the ablation study shows that the three core components (text features, meta-path decomposition, and the attention gate) all contribute non-trivially and complementarily, with text features dominating overall, meta-path decomposition essential for regression accuracy, and the attention gate critical for classification sharpness.
Third, the learned meta-path weights identify the provider-bridge path $D{-}S{-}D$ as the most influential path for prediction in both tasks, while showing that regression-level prediction benefits from a more balanced combination of mechanisms than classification does.
Fourth, an industry-stratified analysis confirms that the model generalizes consistently across eleven industries with no catastrophic failure, and clarifies that residual errors concentrate in segments with heterogeneous sub-domains (Gaming, Resources, Manufacturing) or unusually wide price dispersion (Public Sector).

The study contributes to information systems research by recasting data product pricing as a relational learning problem, operationalizing established economic pricing theories as meta-paths, and showing that structural interpretability and predictive accuracy can be advanced simultaneously.
For practice, the framework provides marketplaces and data providers with both an accurate pricing tool and a transparent mechanism-level explanation, supporting pricing-consistency monitoring, tier-based recommendation, and industry-specific deployment strategies.

Several limitations point to future research.
First, the present study evaluates the framework on three large data marketplaces; extending it to additional platforms and to longitudinal price histories would enable the modeling of price dynamics rather than only cross-sectional levels.
Second, the meta-paths are currently designed manually from theory; an interesting next step is to combine theory-driven design with automated meta-path discovery so that mechanisms not yet covered by classical pricing theory can be uncovered.
Third, the residual heterogeneity observed across industries motivates industry-aware or hierarchical extensions of MPGNN, in which industry- or market-specific sub-modules are layered on top of the shared backbone.
Finally, the proposed framework can be transferred to the pricing and valuation of other non-standardized digital assets (machine-learning APIs, models, IP licenses, and SaaS modules), and we view these as promising directions for further validation of the approach.

\section*{Acknowledgments}
The authors acknowledge the financial support from the National Natural Science Foundation of China (No. 72601057, 72342008, 72401029 and 72473013).

\clearpage
\begin{spacing}{0.94}
\setlength{\bibsep}{0.0em}
\bibliographystyle{jf}
\bibliography{reference}

\newpage

\renewcommand{\thesection}{}
\renewcommand{\thesubsection}{\Alph{subsection}.}
\renewcommand{\thesubsubsection}{\Alph{subsection}.\arabic{subsubsection}}
\renewcommand{\thetable}{\Alph{subsection}\arabic{table}}
\renewcommand{\thefigure}{\Alph{subsection}\arabic{figure}}
\renewcommand{\theequation}{\Alph{subsection}.\arabic{equation}}
\setcounter{table}{0}
\setcounter{figure}{0}
\setcounter{equation}{0}
\setcounter{section}{0}
\setcounter{subsection}{0}

\section{Appendix}

\subsection{Node Feature Definitions}

\begin{table}[htbp]
\centering
\TableSpaced
\caption{Features of data product nodes}
\label{tab:A1}
\resizebox{\textwidth}{!}{
\begin{tabular}{llp{4cm}p{6cm}l}
\hline
    ID & Variable & Definition & Calculation Method & Feature Type \\
\midrule
    D & Product Name & Identifier of different data products & Name of the data product & Node identifier (non-feature) \\
    D1 & Text Features & Textual description of the product & Data product descriptions (including overview, details, refund policy, Q\&A, etc.) into a 128-dimensional embedding & Embedding \\
    D2 & Label\_Volume & Data volume of the product (size or quantity) & Extract the data volume label of the product & Numeric \\
    D3 & Label\_History & Historical time span covered by the product (including access to past versions) & Extract the number of accessible historical versions or the number of days covered & Numeric \\
    D4 & Label\_Future & Whether future versions of the product are accessible & 1 if accessible, 0 otherwise & Boolean \\
    D5 & Label\_Sensitive & Level of sensitive information contained in the product & 2 indicates personal sensitive information, 1 indicates partially sensitive information, 0 indicates no sensitive information & Numeric \\
    D6 & Label\_Coverage & Geographic coverage of the product & Number of countries covered by the product & Numeric \\
    D7 & Label\_Data Sample & Whether the product provides a data sample & 1 if provided, 0 otherwise & Boolean \\
    D8 & Support Email & Whether contact information for technical support is provided & 1 if provided, 0 otherwise & Boolean \\
    D9 & Support URL & Whether an online support link is provided & 1 if provided, 0 otherwise & Boolean \\
    D10 & Label\_Format & Supported data formats of the product & Encode supported formats (e.g., bin, sql, csv, xml, etc.) into categorical labels & Categorical \\
\hline
\end{tabular}
}
\end{table}

\begin{table}[htbp]
\centering
\TableSpaced
\caption{Features of data provider nodes}
\label{tab:A2}
\resizebox{\textwidth}{!}{
\begin{tabular}{llp{6cm}p{6cm}l}
\hline
No. & Variable & Definition & Calculation Method & Feature Type \\
\hline
P & Provider name & Identifies different data providers & Unique name of the data provider & Node identifier \\
P1 & Provider type & Type of the data provider (e.g., enterprise, government, others) & Labels obtained via DeepSeek & One-hot \\
P2 & Public listing status & Indicates whether the data provider is publicly listed & 1 for listed, 0 for non-listed & Boolean \\
\hline
\end{tabular}
}
\end{table}

\begin{table}[!htbp]
\centering
\TableSpaced
\caption{Features of industry nodes}
\label{tab:A3}
\resizebox{\textwidth}{!}{
\begin{tabular}{llp{4cm}p{6cm}l}
\hline
No. & Variable & Definition & Calculation Method & Feature Type \\
\hline
I & Industry type & Identifies different industries & Industry classification (e.g., finance, healthcare, manufacturing) & Node identifier \\
I1 & Industry scale 1 & Total number of data products in the industry & In-degree of data products & Numerical \\
I2 & Industry scale 2 & Total number of data providers in the industry & In-degree of data providers & Numerical \\
I3 & Industry growth rate & Measures the growth rate of the industry & Annual growth rate of industry economic activity (\%) & Numerical \\
I4 & Data sensitivity & Measures the privacy value of industry data & PrivacyScore = (Sensitivity + Identifiability + Regulatory Intensity) / 3 & Numerical \\
\hline
\end{tabular}
}

\begin{minipage}{\textwidth}
\raggedright
\footnotesize
\textit{Notes:} The industry growth rate data are obtained from \url{https://unstats.un.org/UNSDWebsite/}. Data sensitivity is evaluated based on three dimensions: data sensitivity, identifiability, and regulatory intensity. Each dimension is scored on a scale from 1 to 5 using DeepSeek, and the final score is computed as their average.
\end{minipage}

\end{table}
\begin{table}[htbp]
\centering
\TableSpaced
\caption{Features of market nodes}
\label{tab:A4}
\resizebox{\textwidth}{!}{
\begin{tabular}{llp{6cm}p{4cm}l}
\hline
No. & Variable & Definition & Calculation Method & Feature Type \\
\hline
M  & Market Identifier & Identifies different markets & Market name & Node identifier \\
M1 & Market type & Indicates whether the market has global service capability & 1 for global market, 0 for local market & Boolean \\
M2 & Market scale 1 & Measures the scale of products in the market & Number of data products in the market & Numerical \\
M3 & Market scale 2 & Measures the scale of partners in the market & Number of data providers in the market & Numerical \\
\hline
\end{tabular}
}
\end{table}

\clearpage
\subsection{Baseline Hyperparameters}

\begin{table}[!htbp]
\centering
\TableSpaced
\caption{Key hyper-parameters of the GNN baselines. The downstream MLP head shared by all four
methods uses hidden size $128$, Adam optimiser with learning rate
$10^{-3}$ and weight decay $10^{-3}$, $500$ epochs.}
\label{tab:gnn-baseline-hp}
\small
\begin{tabular}{llc}
\toprule
Model & Hyper-parameter & Value \\
\midrule
\multirow{3}{*}{DeepWalk}
  & Embedding dimension     & $128$ \\
  & Walk length             & $8$ \\
  & Walks per node          & $3$ \\
\midrule
\multirow{3}{*}{Node2Vec}
  & Return parameter $p$    & $1.0$ \\
  & In-out parameter $q$    & $2.0$ \\
  & Walk length             & $8$ \\
\midrule
\multirow{3}{*}{GraphSAGE}
  & Embedding dimension     & $64$ \\
  & Number of layers        & $1$ \\
  & Dropout                 & $0.7$ \\
\midrule
\multirow{3}{*}{GAT}
  & Embedding dimension     & $64$ \\
  & Attention heads         & $4$ \\
  & Dropout                 & $0.7$ \\
\bottomrule
\end{tabular}
\end{table}

\begin{table}[!htbp]
\centering
\TableSpaced
\caption{Key hyper-parameters of the proposed MPGNN.}
\label{tab:hyperparams}
\small
\begin{tabular}{lll}
\toprule
Group & Hyper-parameter & Value \\
\midrule
\multirow{3}{*}{Pre-processing}
  & Price transform                                      & $\log(1+x)$ \\
  & Text-embedding PCA dimension                         & $128$ \\
\midrule
\multirow{2}{*}{Meta-path sampling}
  & Max neighbours per hop    & $20$ \\
  & Top-$k$ destinations per source                       & $5$ \\
\midrule
\multirow{4}{*}{HGT encoder}
  & Hidden dimension $D_h$                               & $256$ \\
  & Attention heads                & $4$ \\
  & Price-subspace dimension $D_{\text{price}}$          & $100$ \\
\midrule
\multirow{4}{*}{Path-weight gate}
  & Query / key dimension $d_{\text{attn}}$              & $64$ \\
  & LASSO weight $\lambda_{L_1}$                         & $10^{-3}$ \\
  & Entropy weight $\lambda_{H}$                         & $0.05$ \\
  & Softmax temperature $\tau$                           & $5.0$ \\
  & Attention dimension $d_{\text{attn}}$                & $64$ \\
\midrule
\multirow{3}{*}{Loss weights}
  & Contrastive $\lambda_{\text{cl}}$                    & $0.3$ \\
  & Regression $\lambda_{\text{reg}}$                    & $0.3$ \\
  & Classification $\lambda_{\text{cls}}$                & $0.8$ \\
  & Local expert supervision $\lambda_{\text{local}}$    & $0.5$ \\
\midrule
Contrastive learning
  & Negatives per positive pair                          & $10$ \\
\midrule
\multirow{3}{*}{Price-tier head}
  & Number of tiers $C$                                  & $5$ \\
  & Quantile thresholds                                  & $[0,0.2,0.4,0.6,0.8]$ \\
  & Class-weight cap                                     & $10$ \\
\midrule
\multirow{4}{*}{Optimisation}
  & Optimiser                                            & Adam \\
  & Learning rate                                        & $1\times10^{-3}$ \\
  & Weight decay                                         & $1\times10^{-3}$ \\
  & Max epochs                               & $500$ \\
\midrule
Splits
  & Train\,:\,test ratios                                & $0.7{:}0.3,\,0.8{:}0.2,\,0.9{:}0.1$ \\
\midrule
\multirow{3}{*}{Two-stage architecture}
  & Number of experts $C$                                & $5$ \\
  & Expert hidden dimension                              & $50$ \\
  & Routing mode                                         & soft \\
\bottomrule
\end{tabular}
\end{table}

\begin{table}[!htbp]
\centering
\TableSpaced
\caption{Key hyper-parameters of the basic baselines.}
\label{tab:basic-baseline-hp}
\small
\begin{tabular}{llc}
\toprule
Model & Hyper-parameter & Value \\
\midrule
\multirow{3}{*}{Linear / Logistic Regression (LR)}
  & Max iterations (LogReg)   & $500$ \\
  & Class weighting (LogReg)  & balanced \\
  & Regularisation            & $L_{2}$ \\
\midrule
\multirow{3}{*}{ANN (MLP)}
  & Hidden layer sizes        & $(128, 64)$ \\
  & Activation / Solver       & ReLU / Adam \\
  & Max iterations            & $500$ \\
\midrule
\multirow{3}{*}{Decision Tree (DT)}
  & Max depth                 & unrestricted \\
  & Splitting criterion       & Gini \\
  & Class weighting (clf.)    & balanced \\
\midrule
\multirow{3}{*}{XGBoost}
  & Number of estimators      & $120$ \\
  & Max tree depth            & $3$ \\
  & Learning rate             & $0.035$ \\
\midrule
\multirow{3}{*}{LSTM}
  & Hidden size               & $64$ \\
  & Training epochs           & $50$ \\
  & Batch size / LR           & $32\ /\ 10^{-3}$ \\
\bottomrule
\end{tabular}
\end{table}

\end{spacing}

\end{document}

%% file: preamble.tex
\usepackage{amssymb}
\usepackage[top=1.2in, bottom=1.2in, left=1.1in, right=1.1in]{geometry}
\usepackage{setspace}
\usepackage{fixmath}
\usepackage{indentfirst}
\PassOptionsToPackage{hyphens}{url}
\usepackage[hypertexnames=false]{hyperref} 
\usepackage{bbm}
\usepackage{amsmath}
\usepackage{amsthm}
\usepackage{multicol}
\usepackage{multirow}
\usepackage{multibib}
\usepackage{tikz} 

\usetikzlibrary{arrows.meta}
\newcommand*\tikztimelineset{\pgfqkeys{/tikz/timeline}}
\tikztimelineset{
    segment width/.initial=4cm, tick length/.initial=6pt,
    topbot/.style={
    align=flush left,
    font= \footnotesize,
    text width=\pgfkeysvalueof{/tikz/timeline/segment width}
            -2*(\pgfkeysvalueof{/pgf/inner xsep})},
    top/.initial=, above/.initial=, below/.initial=, bottom/.initial=}
\newcommand*\tikztimeline[2][]{
\begin{tikzpicture}[timeline/every timeline/.try,timeline/.cd,#1]
\foreach[count=\CNT] \elem in {#2}{
    \tikztimelineset{/tikz/style/.expand once=\elem}
    \tikzset{xshift=\CNT*(\pgfkeysvalueof{/tikz/timeline/segment width})}
    \unless\ifnum\CNT=1\relax
    \draw (down:\pgfkeysvalueof{/tikz/timeline/tick length}) -- coordinate (lasttick)
        (  up:\pgfkeysvalueof{/tikz/timeline/tick length});
    \fi
    \node[above] (@above) at (right:{.5*(\pgfkeysvalueof{/tikz/timeline/segment width})})
    {\strut$\pgfkeysvalueof{/tikz/timeline/above}$};
    \node[above, timeline/topbot] at (@above.north) {\pgfkeysvalueof{/tikz/timeline/top}};
    \node[below] (@below) at (@above.south) {\strut\pgfkeysvalueof{/tikz/timeline/below}};
    \node[below, timeline/topbot] at (@below.south) {\pgfkeysvalueof{/tikz/timeline/bottom}};
}
\draw[->] (right:\pgfkeysvalueof{/tikz/timeline/segment width})
        -- ([xshift=\pgfkeysvalueof{/tikz/timeline/segment width}] lasttick) node[above=6pt] {{\it t+1}};
\end{tikzpicture}}
\tikztimelineset{every timeline/.append style={>=Latex}}

\usepackage[labelsep=period]{caption}
\usepackage{multirow}
\usetikzlibrary{decorations.pathmorphing}
\usepackage{dsfont}
\usepackage{tabularray}
\usepackage{mathtools}
\usepackage[fleqn]{nccmath}
\usepackage{footnote}
\makeatletter
\newcommand{\symbolfootnote}[2][1]{\renewcommand{\thefootnote}{\fnsymbol{footnote}}\footnote[#1]{#2}}
\makeatother
\usepackage[flushleft]{threeparttable}
\usepackage{graphicx}
\usepackage{rotating}
\usepackage{pdflscape}
\usepackage{tabularx, booktabs, multirow}
\usepackage{booktabs,tabulary,natbib}
\usepackage{siunitx}
\usepackage{adjustbox}
\usepackage{hyperref} 
\usepackage{enumerate}
\usepackage{subcaption}
\usepackage{cleveref}

\makeatletter
\newcommand*{\hyperlinkcite}[1]{\hyper@link{cite}{cite.#1}}
\makeatother
\usepackage[T1]{fontenc}
\usepackage{libertine}
\usepackage{amsmath,amsthm}
\usepackage[scaled=1.00,libertine,vvarbb]{newtxmath}

\newcolumntype{L}[1]{>{\raggedright\arraybackslash}p{#1}}
\newcolumntype{C}[1]{>{\centering\arraybackslash}p{#1}}
\newcolumntype{R}[1]{>{\raggedleft\arraybackslash}p{#1}}

\newtheorem*{remark*}{Remarks}

\usepackage{tabu}
\usepackage{graphics,graphicx,pgfplots,rotating,multirow}
\usepackage{xcolor}
\usepackage{amssymb,amsfonts,amsmath}
\usepackage{color,colortbl}
\usepackage{caption}
\usepackage{subcaption}
\usepackage{lipsum}
\usepackage{longtable,calc,array}
\newlength\mycolwidth

\usepackage{afterpage}

\usepackage{titlesec}

\makeatletter
\titleformat{\section}
 {\fontsize{16}{10}\bfseries}{\@seccntformat{section}}{0.0001em}{}
\makeatother

\titleformat{\subsection}
       {\fontsize{14}{5}\bfseries}{\thesubsection}{1em}{}
\titleformat{\subsubsection}
       {\fontsize{14}{5}\bfseries}{\thesubsubsection}{1em}{}

\definecolor{lightgray}{gray}{0.7}

\usepackage{etoolbox}
\patchcmd\abstract{\small}{\footnotesize}{}{}

\hypersetup{colorlinks,%
citecolor=blue,%
filecolor=black,%
linkcolor=blue,%
urlcolor=black
}


\pgfplotsset{compat=1.15}

\usepackage{todonotes}

\usepackage{versions,marginnote}
 \excludeversion{notes}		
\includeversion{links}          

\makeatletter\let\chapter\@undefined\makeatother 

\newcommand{\msfn}[2][]{
\renewcommand{\marginpar}{\marginnote}%
\renewcommand{\marginpar}{\oldmarginpar}}

\makeatletter
\newcommand\runinsubsubsection{\@startsection {subsubsection}{1}{\z@}%
                               {3.5ex \@plus 1ex \@minus .2ex}%
                               {-1em}%
                               {\normalfont\normalsize\bfseries}}
\makeatother

\usepackage{microtype}          
\usepackage{algorithm}
\usepackage{algorithmic}        

\newcommand{\TableSpaced}{\singlespacing}
\newcommand{\SingleSpacedXI}{\singlespacing\normalsize}